\documentclass[prb,twocolumn,superscriptaddress]{revtex4-1}

\usepackage{lipsum,graphicx}
\usepackage[dvipsnames,table,xcdraw]{xcolor}
\usepackage[english]{babel}
\usepackage[T1]{fontenc}
\usepackage[latin9]{inputenc}
\usepackage{latexsym}
\usepackage{float}
\usepackage{amsmath}
\usepackage{graphicx}
\usepackage{times}   %% Times Roman font
\usepackage{esint}
\usepackage[unicode=true,pdfusetitle,
 bookmarks=false,colorlinks=true,citecolor=blue,urlcolor=blue,linkcolor=red]{hyperref}
\usepackage{array}
\usepackage[normalem]{ulem}
\usepackage{grffile}[=v1]
\newcolumntype{L}[1]{>{\raggedright\let\newline\\\arraybackslash\hspace{0pt}}m{#1}}
\newcolumntype{C}[1]{>{\centering\let\newline\\\arraybackslash\hspace{0pt}}m{#1}}
\newcolumntype{R}[1]{>{\raggedleft\let\newline\\\arraybackslash\hspace{0pt}}m{#1}}

\makeatletter
\@ifundefined{textcolor}{}
{%
 \definecolor{BLACK}{gray}{0}
 \definecolor{WHITE}{gray}{1}
 \definecolor{RED}{rgb}{1,0,0}
 \definecolor{GREEN}{rgb}{0,1,0}
 \definecolor{BLUE}{rgb}{0,0,1}
 \definecolor{CYAN}{cmyk}{1,0,0,0}
 \definecolor{MAGENTA}{cmyk}{0,1,0,0}
 \definecolor{YELLOW}{cmyk}{0,0,1,0}
}

\@ifundefined{date}{}{\date{}}
\AtBeginDocument{
  
}
\makeatother

\newcommand{\SAVE}[1]{}

\newcommand{\Sop}{S}
\begin{document}
\renewcommand\abstractname{}

\newcommand{\SPEDITED}[2]{{\sout{#1}}{ \textcolor{red}{#2}}}
\newcommand{\SPEDITOKAY}[2]{{}{\textcolor{black}{#2}}}
\newcommand{\SP}[1]{{\color{ForestGreen}{\bf SP: #1}}}
\newcommand{\SPHIDE}[1]{{}}

%\title{Quantum antiferromagnets in the $XY$ limit}
\title{Colorful points in the $XY$ regime of $XXZ$ quantum magnets}

\author{Santanu Pal}
\affiliation{Department of Physics, Indian Institute of Technology Bombay, Mumbai, Maharashtra, 400076, India}
\author{Prakash Sharma}
\affiliation{Department of Physics, Florida State University, Tallahassee, Florida 32306, USA}
\affiliation{National High Magnetic Field Laboratory, Tallahassee, Florida 32310, USA}
\author{Hitesh J. Changlani}
\affiliation{Department of Physics, Florida State University, Tallahassee, Florida 32306, USA}
\affiliation{National High Magnetic Field Laboratory, Tallahassee, Florida 32310, USA}
\author{Sumiran Pujari}
\email{sumiran@phy.iitb.ac.in}
\affiliation{Department of Physics, Indian Institute of Technology Bombay, Mumbai, Maharashtra, 400076, India}
\date{\today}

\begin{abstract}

In the $XY$ regime of the $XXZ$ Heisenberg model phase diagram, we demonstrate that
the origin of magnetically ordered phases is influenced by the presence of solvable points 
with exact quantum coloring ground states featuring a quantum-classical correspondence. 
Using exact diagonalization and density matrix renormalization group calculations, 
for both 
the square and the triangular lattice magnets, we show that the 
ordered physics of the solvable points
in the extreme $XY$ regime, at $\frac{J_z}{J_\perp}=-1$ 
and $\frac{J_z}{J_\perp}=-\frac{1}{2}$ respectively with $J_\perp > 0$, 
adiabatically extends to
the more isotropic regime $\frac{J_z}{J_\perp} \sim 1$.
We highlight the projective structure of the coloring ground states
to compute the correlators in fixed magnetization sectors which enables
an understanding of the features in the static spin structure factors and
correlation ratios.
These findings are contrasted with an anisotropic generalization of the celebrated one-dimensional 
Majumdar-Ghosh model, which is also found to be (ground state) solvable. 
For this model, both exact dimer and three-coloring ground states exist at 
$\frac{J_z}{J_\perp}=-\frac{1}{2}$ but only the two dimer ground states survive for any 
$\frac{J_z}{J_\perp} >-\frac{1}{2}$.
\end{abstract}

\maketitle

%===============================================================

\section{Introduction}
\label{sec:intro}

The question of magnetic long-range ordering (LRO), or lack thereof, 
in quantum antiferromagnetic insulators in low dimensions has been 
of prime interest in the field of quantum magnetism. 
One of the hallmark results is the absence of 
true LRO for the quantum Heisenberg spin chain  
owing to strong quantum mechanical 
fluctuations in one dimension and the associated fractional spinon excitations.~\cite{Giamarchi_book} LRO does 
exist in two dimensions, but only at zero temperature,  
for the square lattice Heisenberg model, as well as other (unfrustrated) bipartite 
lattices.~\cite{Chakravarty_Halperin_Nelson_1989}
In three dimensions, LRO exists at finite temperatures as well.
Compounding this issue is the ingredient of frustration; it was initially suggested 
as a possible mechanism to suppress LRO in the 
triangular lattice Heisenberg antiferromagnet. 
Theoretical~\cite{Huse_Elser,Jolicoeur1989,Singh1992,Bernu1992,Capriotti1999}
and experimental studies~\cite{Kadowaki1995,Ishii2011,Shirata2012} 
have revealed that LRO indeed survives in
the triangular lattice geometry, however, other frustrated geometries and interactions
have continued to be the subject of intense study.

Given the complexity of such problems, exactly solvable Hamiltonians form important cornerstones
in our understanding of quantum magnetism and, more generally, 
quantum matter in its vast variety.
Bethe's solution of the one-dimensional ($1d$) 
Heisenberg chain\cite{Bethe_ansatz} has led to an entire field of activity
\cite{Guan_Bethe_ansatz_review, Karabach_Bethe_ansatz_review1,Karabach_Bethe_ansatz_review2,
Fedor_Bethe_ansatz_review} with Bethe ansatz methods applied
to a host of $1d$ models including the spin-$\frac{1}{2}$ $XXZ$ model.
Additionally, the ground state solvable one-dimensional 
Majumdar-Ghosh model \cite{Majumdar_Ghosh}, a precursor to the $S=1$ AKLT chain \cite{AKLT}, 
has led to many insights into the valence bond physics of $1d$ frustrated systems.
In higher dimensions, however, there are fewer solvable examples
for both unfrustrated and frustrated quantum magnets, notably 
the Shastry-Sutherland model \cite{Shastry_Sutherland}
and Kitaev honeycomb model.\cite{Kitaev_anyons}
In this spirit, this work will show the influence of exactly solvable points
in the $XXZ$ parameter space,
\begin{equation}
H_{XXZ} = J_\perp \sum_{\langle i,j \rangle} \left( S^x_i S^x_j + S^y_i S^y_j \right)
+ J_z \sum_{\langle i,j \rangle} S^z_i S^z_j
\label{eq:XXZdefn}
\end{equation} 
on magnetic LRO in
two dimensional quantum antiferromagnets (QAFM). We will set $J_\perp=1$
throughout this paper.

The motivation for our present work stems from a recent finding of a higher-dimensional example
of ground state solvable frustrated quantum magnets described by $H_{3c} \equiv H_{XXZ}[J_z=-1/2]$
\cite{Changlani_PRL} (previously referred to as ``$XXZ0$'') on any lattice composed of triangular motifs that allow for a consistent ``three-coloring"
of the lattice sites where no two sites connected by a bond share the same color. 
Even though this work was situated in the context of the Kagome antiferromagnet,
the general principle applies to a host of lattices including
the triangular lattice.
At this solvable $3c$ point, there is 
a one-to-one correspondence between the classical and quantum ground states.
Adding realistic perturbations away from such a point in parameter space 
thus potentially offers a new way of understanding the phases 
that are stabilized by quantum fluctuations.

The coloring states remain exact ground states when
projected to a specific magnetization sector due to $U(1)$ symmetry
of the $XXZ$ model.
This general projection structure of the exact ground 
state is an important feature of the solvable point, e.g. it was 
utilized~\cite{changlani2019resonating} to explain the magnon crystal 
associated with the 
$\frac{m}{m_s}=\frac{7}{9}$ high-magnetization plateau state
on the Kagome lattice~\cite{Richter_lectures,okuma2019series,Schnack2020magnon}
where $m$ is the magnetization, and $m_s$ is its saturation value.
This was achieved by an exact mapping of three-colorings to localized magnons
using the projection structure (also see the recent Ref.~\onlinecite{Derzhko2020} for 
analogous mappings on the sawtooth lattice).  
The unprojected exact solution in the context of the 
triangular lattice has previously been noted by Ref.~\onlinecite{Momoi}.

\begin{figure}
	\includegraphics[width=\linewidth,trim=0 30mm 0 0,clip=true]{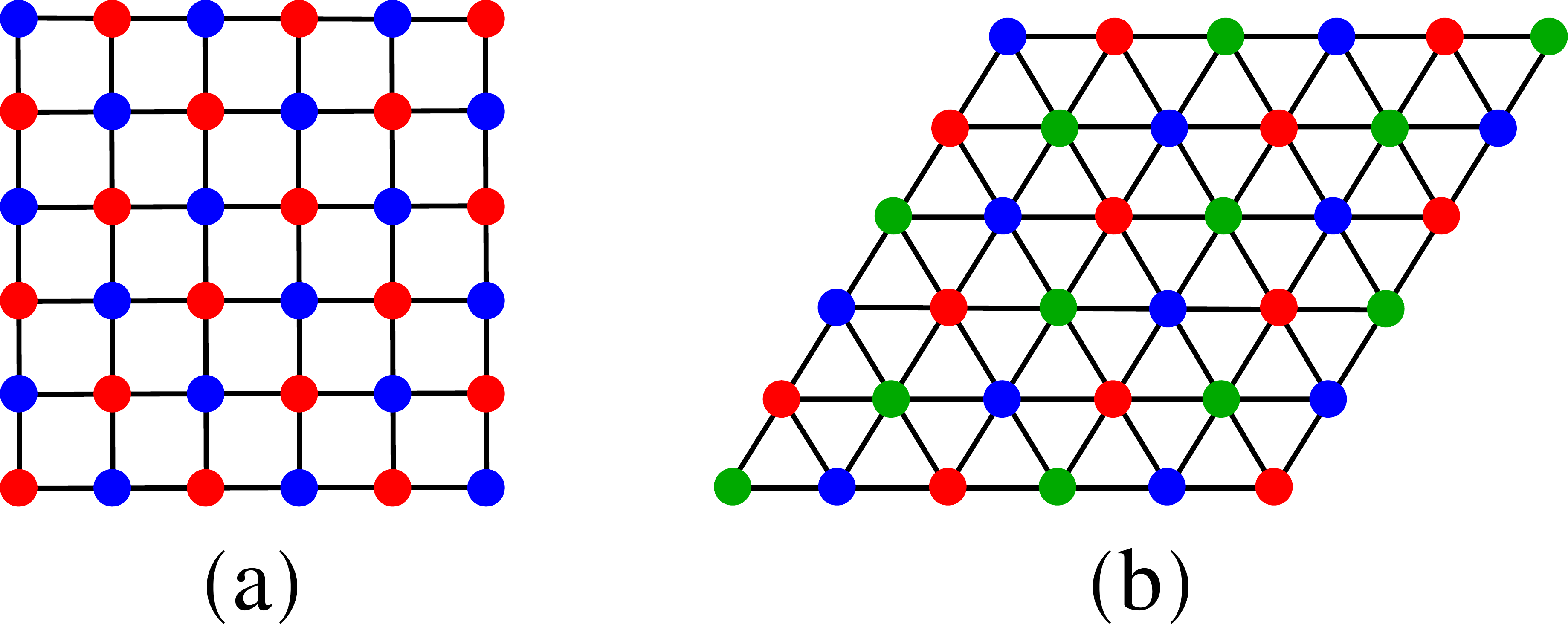}
	\caption{\label{fig:cartoon} Illustration of the unique two-coloring on the square lattice,
	and one of the two three-colorings on the triangular lattice. The colorings 
	directly correspond to magnetically ordered states. 
	}
\end{figure}

Here we address two cases with magnetic LRO - 
the frustrated triangular and unfrustrated square lattice which admit
coloring ground states as shown in Fig.~\ref{fig:cartoon}. For the unfrustrated case,
the exact ground state corresponds to a two-coloring which is
applicable for any bipartite lattice in any dimension and occurs for $J_z = -1$.
Because of the projection structure of these states, we work in
fixed magnetization sectors of choice. 
For the square lattice case, we focus on the zero magnetization sector and for the triangular case on the 
$m=0$ and $\frac{m}{m_s}=\frac{1}{3}$ sectors, the latter 
being a known plateau state at the Heisenberg point.~\cite{Chubukov_jpcm1991,Pal_jpcm2020} 
For these projected coloring states, we establish the presence of magnetic LRO 
by calculating two-point correlators.
Since these points in parameter space do not have the full $SU(2)$
but only $U(1)$ symmetry, the corresponding ground state in the zero magnetization sectors are 
AFM ordered in the $XY$ plane.

We next investigate, using exact diagonalization (ED) and density matrix renormalization group (DMRG) 
\cite{White1992,ITensor}
calculations, how these coloring ground states are
connected to the more isotropic regime $J_z \sim 1$. Using various measures,
we provide evidence for the emergence of magnetic LRO in the square and triangular
Heisenberg magnets from the solvable points. 
Interestingly, both the three-coloring and two-coloring solvable points
sit at the quantum critical point between the $XY$ N\'eel LRO and ferromagnetic
ground state. Thus, the exact ground states contain the seeds for both AFM
and FM ordering.
This basic structure of the $XXZ$ phase diagram for magnetically LRO magnets
is the central result of this paper.

However, the presence of an exactly solvable point with 
quantum coloring ground states in the extreme anisotropic limit 
does not necessarily guarantee the existence of LRO away from it. 
We demonstrate this in the context of the anisotropic generalization of the celebrated 
Majumdar-Ghosh model where both coloring and dimer ground states are exact solutions 
at $J_z=-1/2$.  
The model is characterized by competing coloring and dimerized (valence bond) 
ground states; perturbing towards the isotropic point 
favors the dimer solutions 
rather than the coloring solutions.

The paper is organized as follows. In Sec. \ref{sec:square_zeromag}, 
we discuss the case of the $m=0$ sector of the 
square lattice AFM in the context
of the $J_z=-1$ point. 
In Secs. \ref{sec:triangle_zeromag} and \ref{sec:triangle_1by3},
 we present our findings for the $m=0$ and $\frac{m}{m_s}=\frac{1}{3}$ sectors 
of the triangular AFM. As mentioned above, we contrast these findings with that for an anisotropic generalization 
of the Majumdar-Ghosh model in Sec. \ref{sec:dimers}. 
In the appendices Apps. \ref{app:2c_mat_elem}-\ref{app:tables}, 
we provide derivations for correlations and structure factors
induced by projection, 
applicable on any lattice and some additional useful information.

%===============================================================

\section{The square lattice antiferromagnet}
\label{sec:square_zeromag}

We consider the case of the Hamiltonian $H_{2c} \equiv H_{XXZ}[J_z=-1]$, 
where an exact ground state solution is guaranteed 
on any bipartite lattice in any dimension. We focus on the $2d$ square lattice where the existence of N\'eel LRO
at the Heisenberg point $J_z=1$ is well established\cite{Richter_lectures}, in comparison to the $1d$ chain 
which has only quasi-LRO with polynomially decaying spin-spin correlations. 
The ground state of $H_{2c}$ corresponds to a \emph{unique} two-coloring of the bipartite lattice.

Let the two colors, denoted by red ($|r \rangle$) and blue ($|b \rangle$) labels, 
represent the $\Sop^x$ eigenstates on a single site,
\begin{eqnarray}
|r\rangle \equiv \frac{1}{\sqrt{2}}(|\uparrow\rangle + |\downarrow\rangle)~~~~~~~~~~ 
|b\rangle \equiv \frac{1}{\sqrt{2}}(|\uparrow\rangle - |\downarrow\rangle).  
\label{eq:rb}
\end{eqnarray}
The 
%\textcolor{red}{(unprojected)} 
(unprojected) ground state at $H[J_z=-1]$ is
\begin{equation}
|C\rangle \equiv \left( \prod_{i\in A} \otimes_i |r\rangle_i \prod_{j\in B} \otimes_j |b\rangle_j \right)
\label{gswf}
\end{equation}
where $A, B$ are the two sublattices of any
bipartite lattice, for example, in 1D: chain, ladders; 2D: square, honeycomb; 3D: cube, hyper-honeycomb etc. 

To show the ground state property of Eq.~\ref{gswf}, we write $H_{2c}$ as a sum of bond Hamiltonians $H_{ij} \equiv \Sop_i^x\Sop_j^x+\Sop_i^y\Sop_j^y-\Sop_i^z\Sop_j^z$. 
On a given bond, the eigensystem of $H_{ij}[J_z=-1]$ consists of 
the polarized states $|\uparrow\uparrow\rangle$, $|\downarrow\downarrow\rangle$, 
and the bond singlet $|0;0\rangle\equiv (|\uparrow\downarrow\rangle-|\downarrow\uparrow\rangle)/\sqrt{2}$ as ground states with energy $-1/4$,
while the state $|1;0\rangle\equiv (|\uparrow\downarrow\rangle+|\downarrow\uparrow\rangle)/\sqrt{2}$ 
is an excited state with energy $+3/4$. 
Then, $H_{ij} = \sum_{k=1}^{4} E_k |\psi_k\rangle \langle \psi_k|$ where $E_k$ are the four eigenenergies 
of the bond, and $|\psi_k\rangle$ are the corresponding eigenvectors. Using the identity $1= \sum_{k=1}^{4} |\psi_k\rangle \langle\psi_k|$, 
$H_{2c} = \sum_{\langle ij \rangle} H_{ij}$ is recast purely in terms of the bond projectors $P_{ij} \equiv |1;0\rangle\langle1;0|$,
\begin{eqnarray}
	H_{2c} &=& \sum_{\langle ij\rangle}P_{ij}-\frac{1}{4}N_{\text{bonds}}
\end{eqnarray}
Since the coefficient in front of the projectors is positive, any wavefunction that simultaneously zeros out the projector on each bond is a ground state. 
Zeroing out a projector requires that only components orthogonal to $|1;0\rangle$ enter 
the many-body wavefunction. This is indeed achieved by $|C\rangle$.
(Expanding out the product state for one $|r\rangle$ and one $|b\rangle$ 
gives $|\uparrow\uparrow\rangle - |\downarrow \downarrow \rangle - \sqrt{2}|0;0\rangle$, 
with each individual term being orthogonal to $|1;0\rangle$.)

One can see, at the level of a single bond, 
that there is inherent competition between FM 
($|\uparrow\uparrow\rangle$,$|\downarrow\downarrow\rangle$) 
and AFM ($|0;0\rangle$) correlations, and the two states become exactly degenerate at $J_z=-1$. 
This, in turn, 
results in $J_z=-1$ being a critical point in the $XXZ$ phase diagram.
Since total $S^z$ is conserved, the projected coloring state 
%\textcolor{red}{
\begin{equation}
|C_{S^{z}}\rangle \equiv P_{S^{z}}|C\rangle
\label{eq:projection}
\end{equation}
%}
%$|C_{S^{z}}\rangle \equiv P_{S^{z}}|C\rangle$
is also the exact ground state in every $S^z$ sector, where 
$P_{\Sop^z}$ is the projection to a given total $\Sop^z$ sector.~\cite{projector} 
This construction gives a 
unique ground state in each total $\Sop^z$ sector. 
For a lattice with $N$ sites, there are thus $(N + 1)$ degenerate ground states
which is readily verifiable in Exact diagonalization (ED) for accessible systems, 
as well as their ground state energy value 
(Table~\ref{tab:ED_comp} in App.~\ref{app:tables}). 

\begin{figure*}[]
  \includegraphics[width=.47\linewidth]{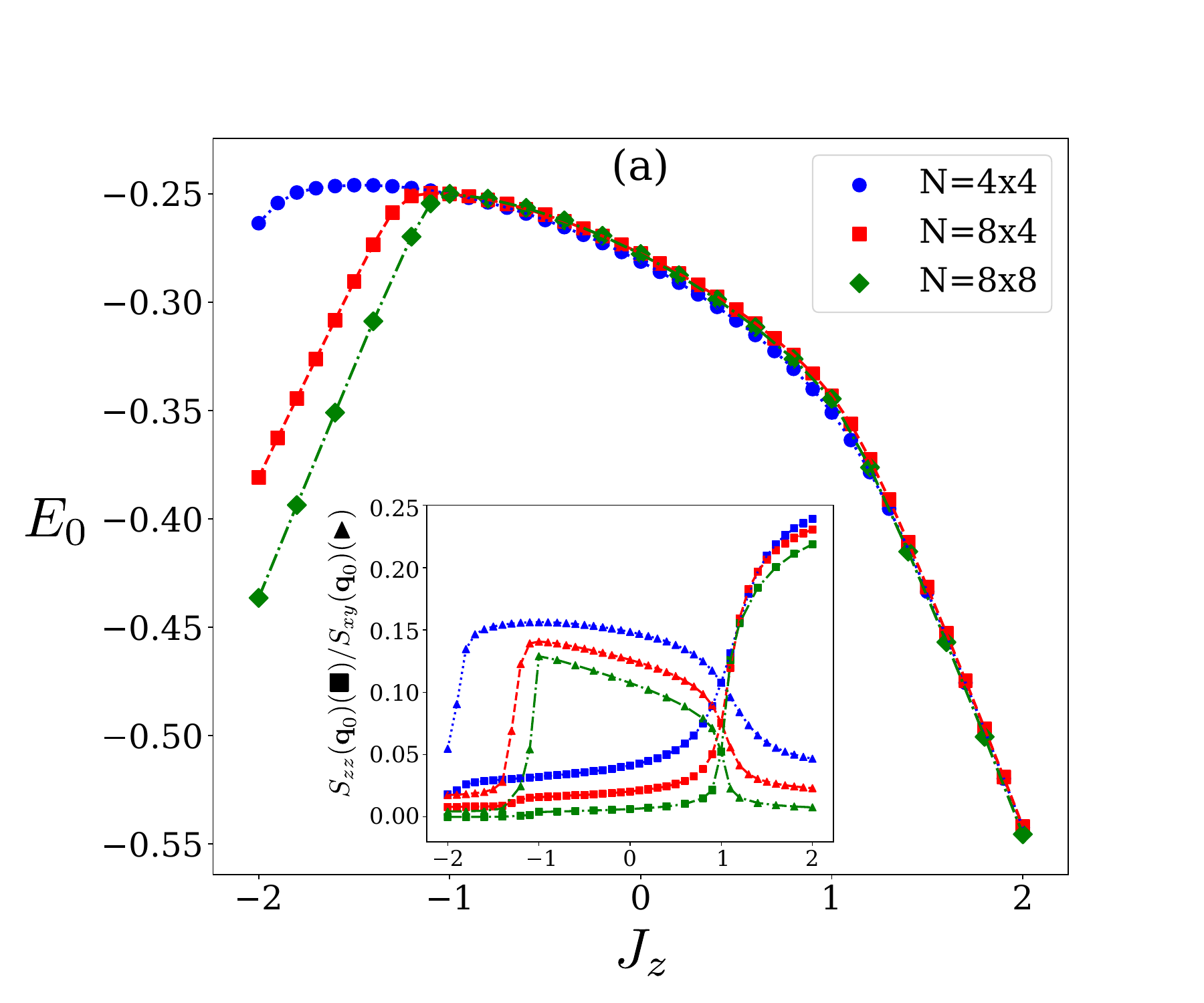}
\includegraphics[width=.47\linewidth]{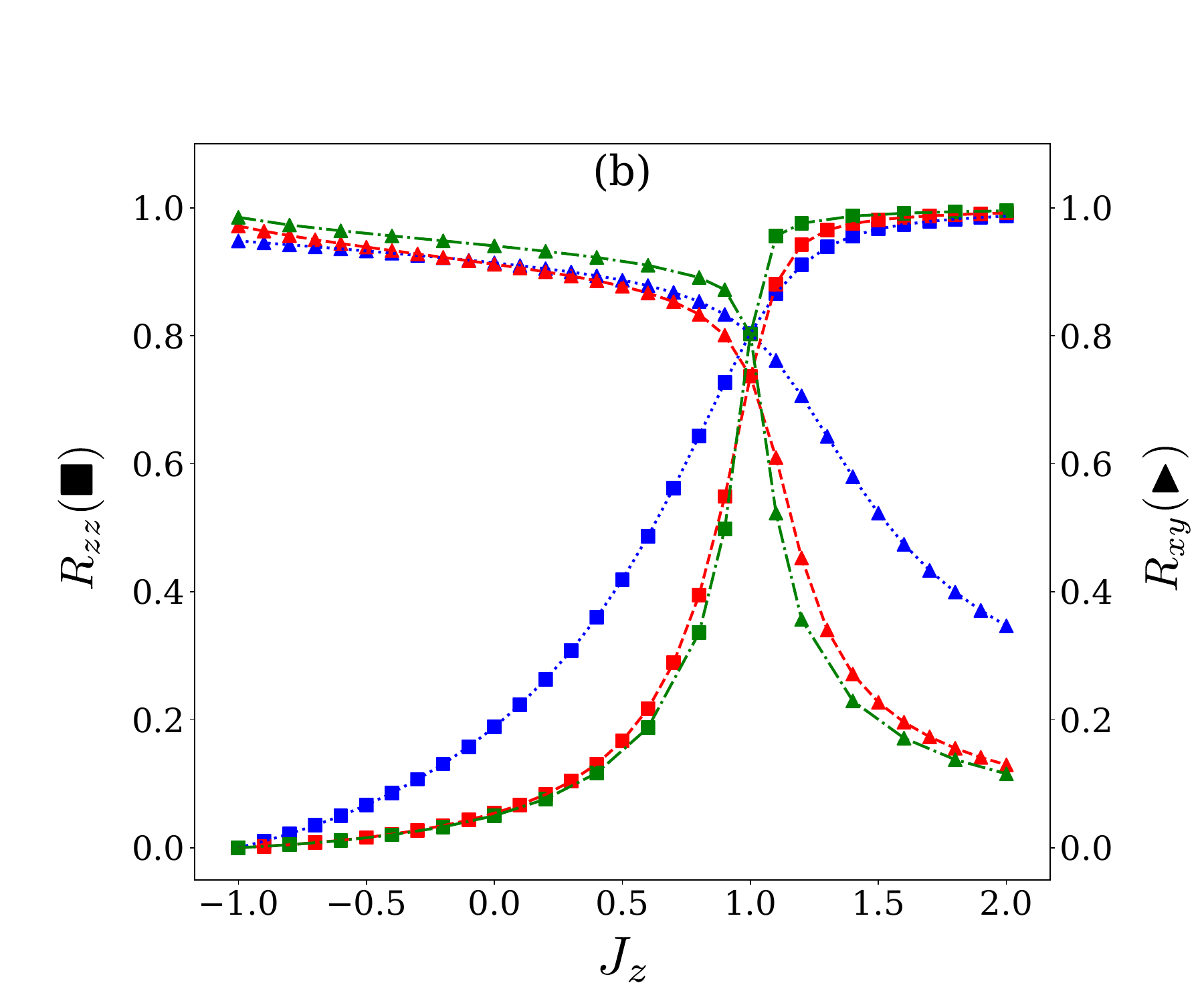}
  \caption{\label{fig:square_ED}
	For the square lattice $XXZ$ Hamiltonian in the $m=0$ sector, the left panel shows the ground state energy per bond ($E_0$) vs. $J_z$. 
	The inset of the left panel shows the evolution of the structure factors ($\frac{S_{zz}(\mathbf{q}_0)}{N}$ and $\frac{S_{xy}(\mathbf{q}_0)}{N}$ such that they are intensive), 
	calculated at the ordering vector $\mathbf{q}_0=(\pi,\pi)$.
	The right panel shows the correlation ratio, as defined in the text (Eq.~\ref{eqn:CR}), vs. $J_z$ 
	at the ordering wave-vector $\mathbf{q}_0=(\pi,\pi)$ for the representative case of 
	$\Delta \mathbf{q}=(\frac{2\pi}{L_x},0)$. 
	}
\end{figure*}

We note that the choice of the two colors in Eq.~\ref{eq:rb} has a (global) gauge freedom. 
The present choice is along the $X$ direction in $XY$ plane of the Bloch sphere. 
They can be chosen to be in any direction in the $XY$ plane owing to the $U(1)$ 
symmetry of $H[J_z]$. This is also seen at a classical level through a
Luttinger-Tisza analysis of $H_{XXZ}$ which leads to a classical phase transition 
at $J_z = -1$, with a FM solution along the $Z$ axis for $J_z < -1$, and an 
AFM solution in the $XY$ plane for $J_z > -1$. This 
freedom of choice in direction in the $XY$ plane is the classical counterpart 
of the global gauge freedom seen in the quantum mechanical case. 
A similar classical-quantum correspondence works for
$H_{3c}$.\cite{changlani2019resonating}
In the classical case, the $XY$-plane AFM solution holds true only up to $J_z < 1$, 
after which the AFM solution lies along the $Z$ axis. 
At $J_z = 1$, the AFM solution can lie in any direction. 
In the quantum case, this translates to full $SU(2)$ symmetry at the Heisenberg point. 
We can thus anticipate that the $U(1)$ symmetric $XY$ N\'eel state as in Eq. \ref{gswf} 
evolves in an adiabatic fashion to a $SU(2)$ symmetric N\'eel state, 
since both are essentially N\'eel-ordered states in the same total $\Sop^z$ sector.
%-------------------------------------------------------------------------------

The above can also be understood as a consequence of a ``superspin" 
with length $\frac{N}{2}$ with a degeneracy of 
$2\left(\frac{N}{2}\right)+1$, even though $H_{2c}$ is not $SU(2)$ 
symmetric and is short-ranged.
A more familiar and direct example of such a superspin 
is rather the long-range all-to-all coupled $SU(2)$-symmetric
Hamiltonian~\cite{Lieb_Mattis_1962}
$J \sum_{i \in A,j \in B} \left( S^x_i S^x_j + S^y_i S^y_j + S^z_i S^z_j \right)
= J (\mathbf{S}_A \cdot \mathbf{S}_B)$ where $\mathbf{S}_A$/$\mathbf{S}_B$ are superspins
with length $\frac{N}{2}$. To see how this arises in $H_{2c}$, it is useful
to compare its solution with that of the \emph{ferromagnetic} 
Hamiltonian $H_{\text{FM}}=- H_{XXZ}[J_\perp = J_z]$
through a projector point of view.
Recasting this ferromagnetic Hamiltonian as a sum of projectors,
$H_{\text{FM}} = \sum_{\langle i,j \rangle} Q_{ij}$ after a trivial shift
of $0.25$ per bond, where $Q_{ij}$'s are non-commuting semi-definite projectors to the
singlet state $|0;0\rangle$ on bonds $\langle i,j \rangle$ respectively.
The familiar eigensystem
here is $|\uparrow \uparrow\rangle$, $|\downarrow \downarrow\rangle$
and $|1;0\rangle$
as ground states (with energy $-1/4$), and $|0;0\rangle$
as an excited state (with energy $+3/4$).
The unprojected ground state is now achieved by 
%\textcolor{red}{$\prod_{i \in \{A,B\}} \otimes_i |r\rangle_i$},
$\prod_{i \in \{A,B\}} \otimes_i |r\rangle_i$,
and projection to desired total $S^z$ sectors may again
be done as before. Since total $\mathbf{S}^2$ is also conserved
for $H_{\text{FM}}$, this projection gives rise to the
usual multiplet structure expected for $SU(2)$ symmetry, i.e. 
the $2 \left(\frac{N}{2}\right)+1 = N+1$ degeneracy due to a superspin structure.
Also, since we project out the singlet on each bond, we only
get FM correlations here as expected.
In contrast, for $H_{2c}$ there is no $SU(2)$ symmetry,
and therefore the ground state in any total $S^z$ sector
is a superposition of various total $\mathbf{S}^2$ sectors.
Nonetheless, we see that the superspin structure of the ground state of $H_{\text{FM}}$
gets exactly mirrored in the ground state of $H_{\text{2c}}$
because of the close relation of the two ground states,
$|C\rangle$ and 
%\textcolor{red}{$\prod_{i \in \{A,B\}} \otimes_i |r\rangle_i$}.
$\prod_{i \in \{A,B\}} \otimes_i |r\rangle_i$.
A phase change of $(-1)^{\#\downarrow_B}$ where $\#\downarrow_B$ are the number
of down-spins 
on the B sublattice to the wavefunction coefficients (in the $S^z$ basis) 
 maps uniquely 
%\textcolor{red}{$|C\rangle$ to $\prod_{i \in \{A,B\}} \otimes_i |r\rangle_i$ and vice versa},
$|C\rangle$ to $\prod_{i \in \{A,B\}} \otimes_i |r\rangle_i$ and vice versa,
% the two wavefunctions unto each other
and this mapping carries
over under projection $P_{S^z}$ as well.

\begin{figure*}[]
	\includegraphics[width=.47\linewidth]{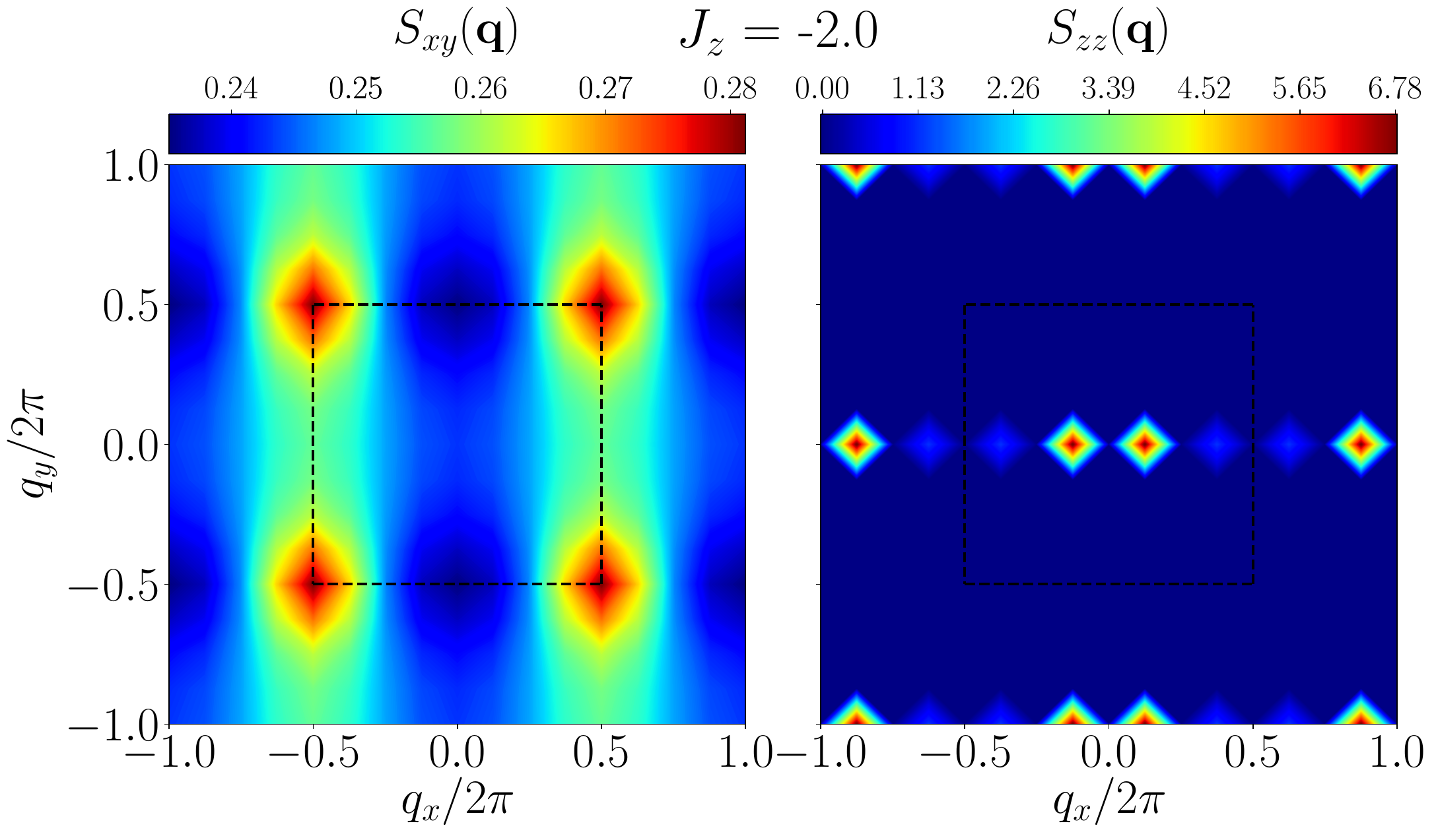}
	\includegraphics[width=.47\linewidth]{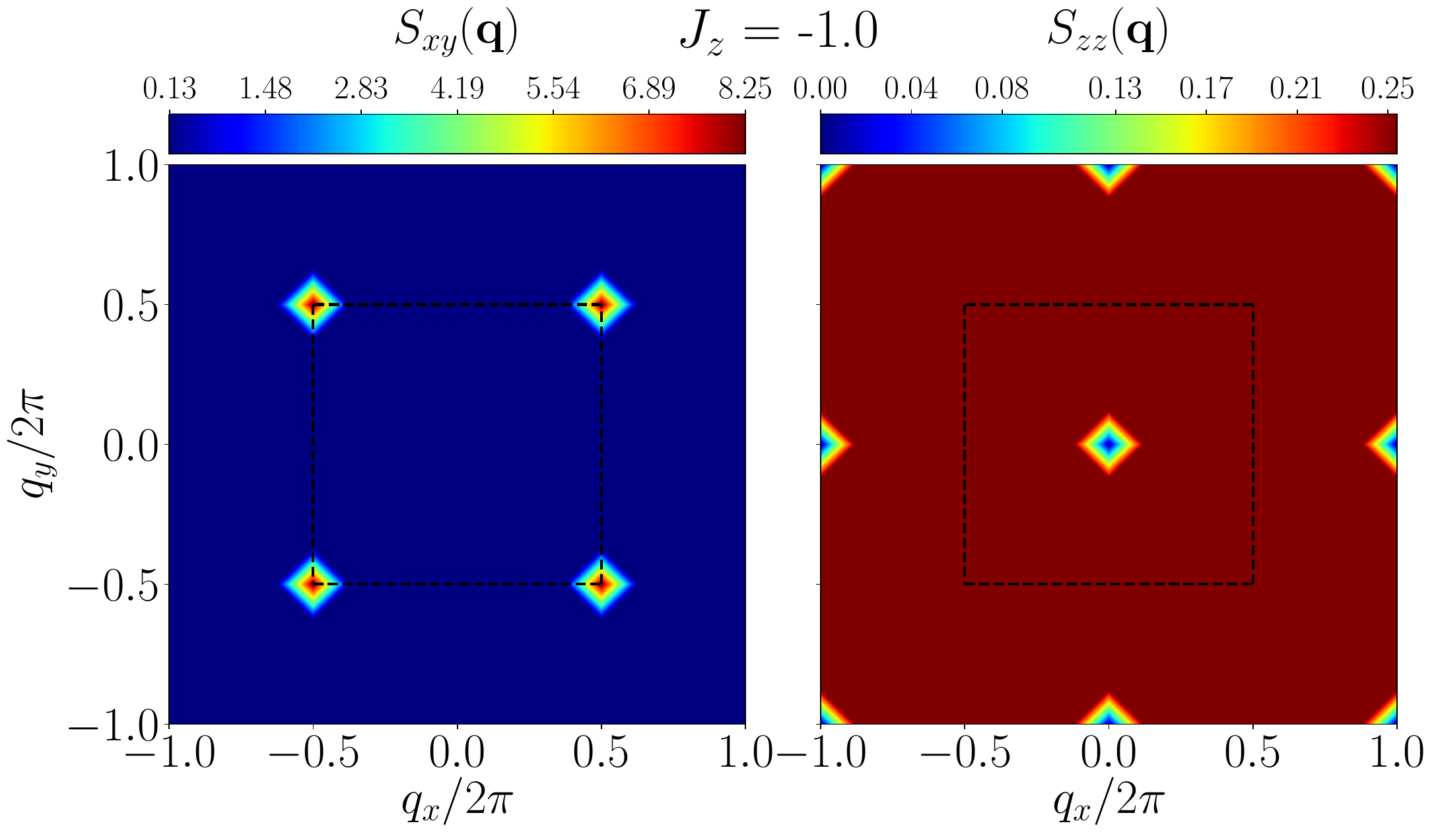} \\
	\vspace{2mm}
	\includegraphics[width=.47\linewidth]{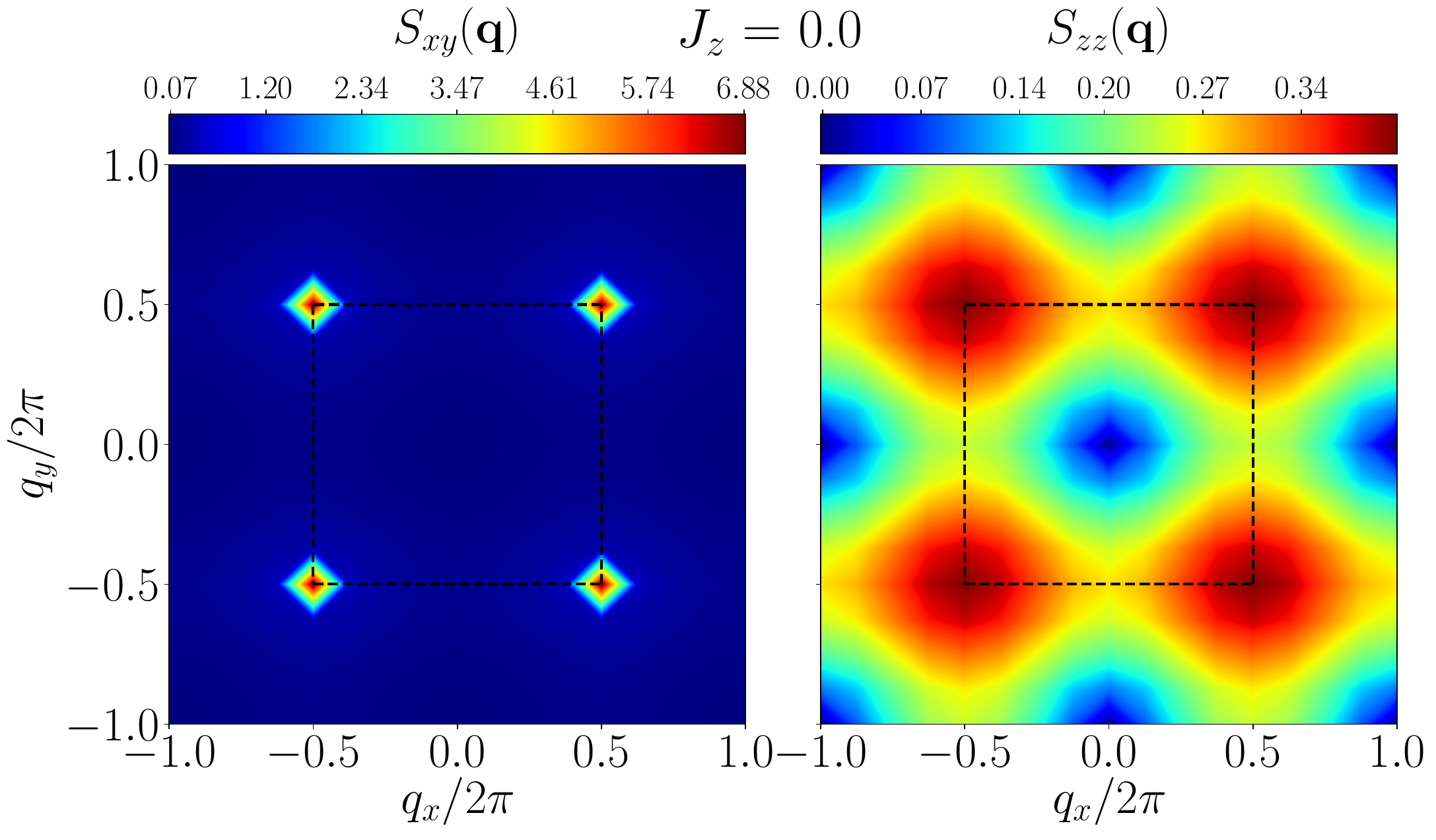}
	\includegraphics[width=.47\linewidth]{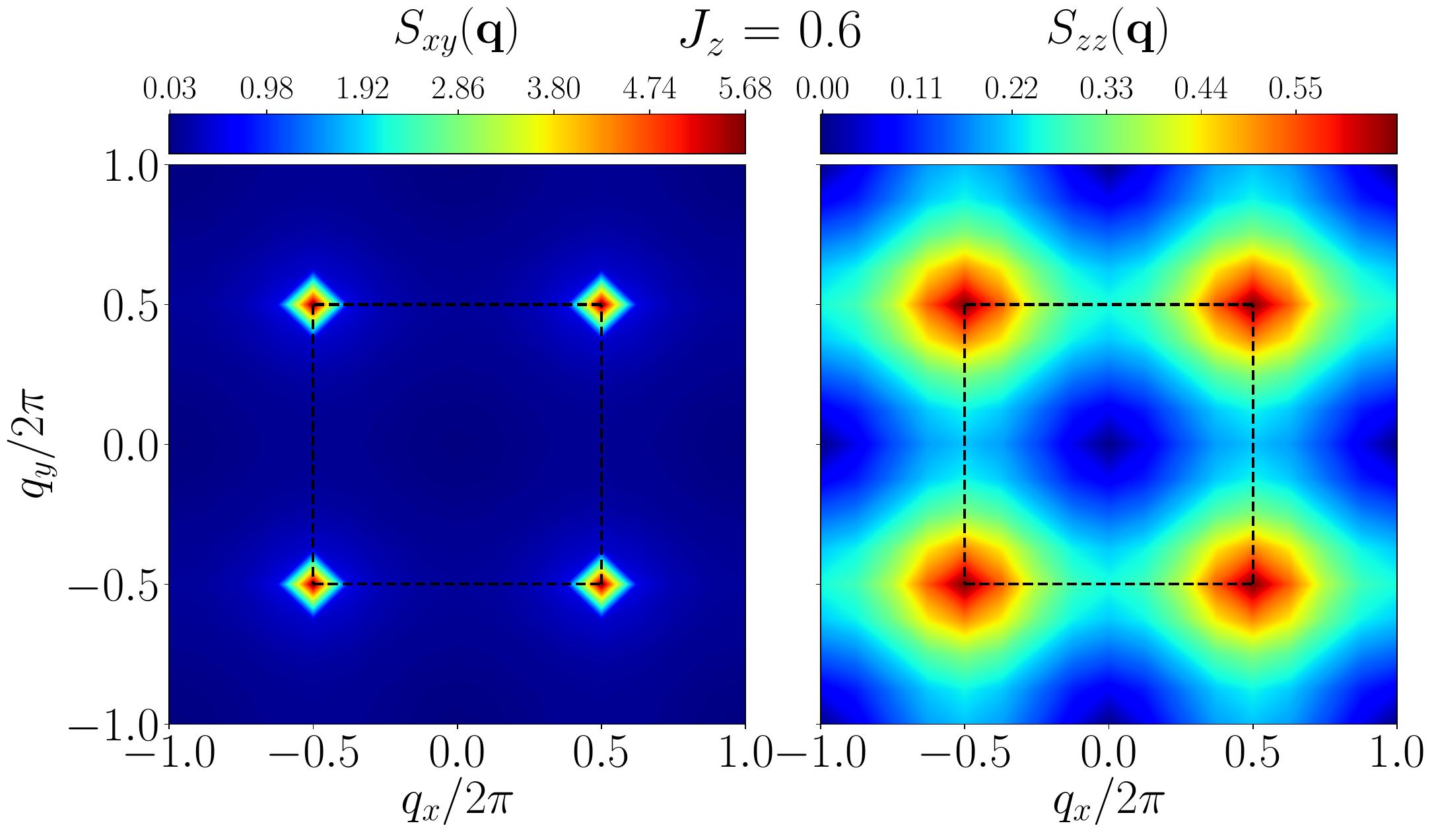} \\
	\vspace{2mm}
	\includegraphics[width=.47\linewidth]{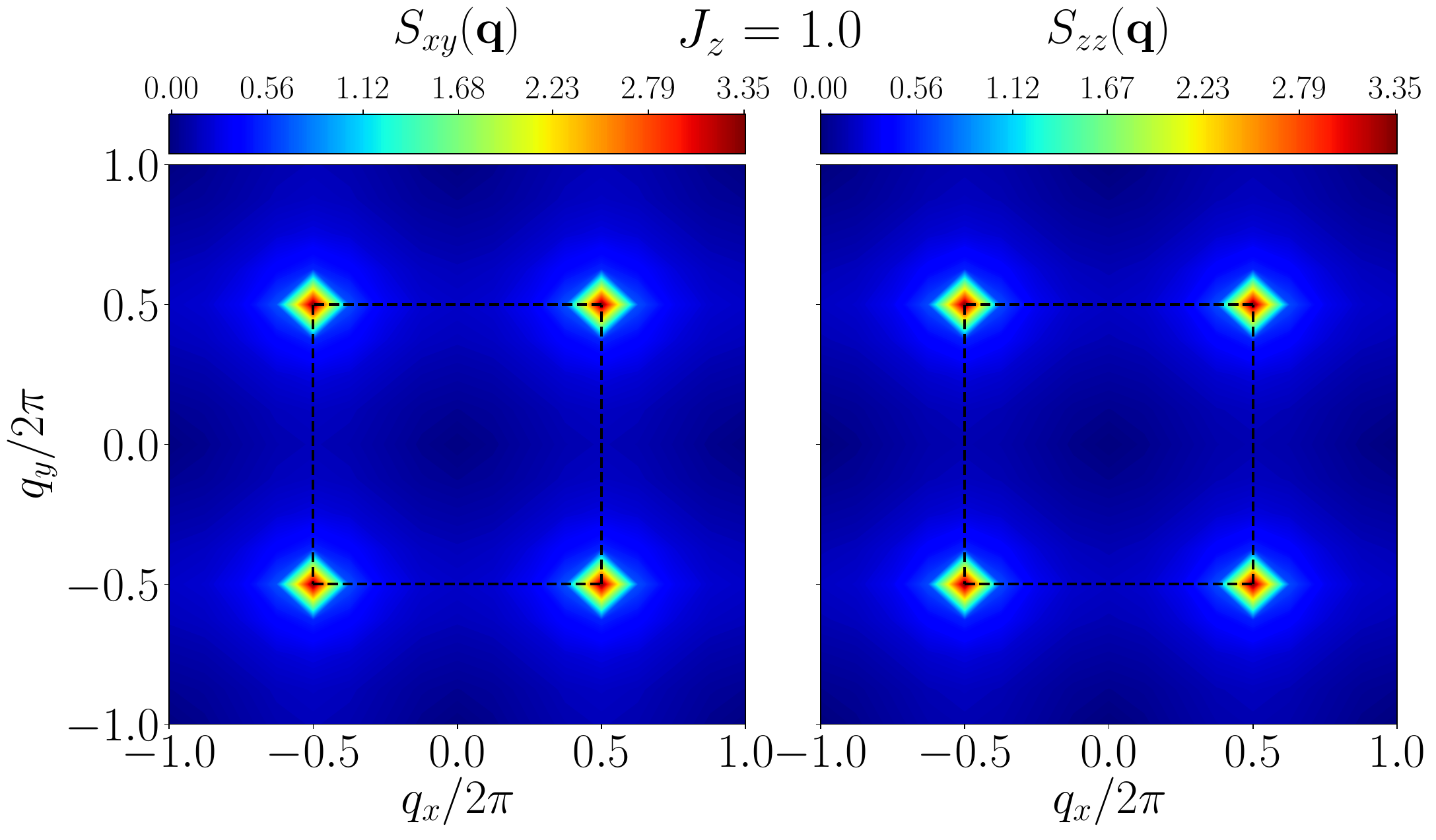}
	\includegraphics[width=.47\linewidth]{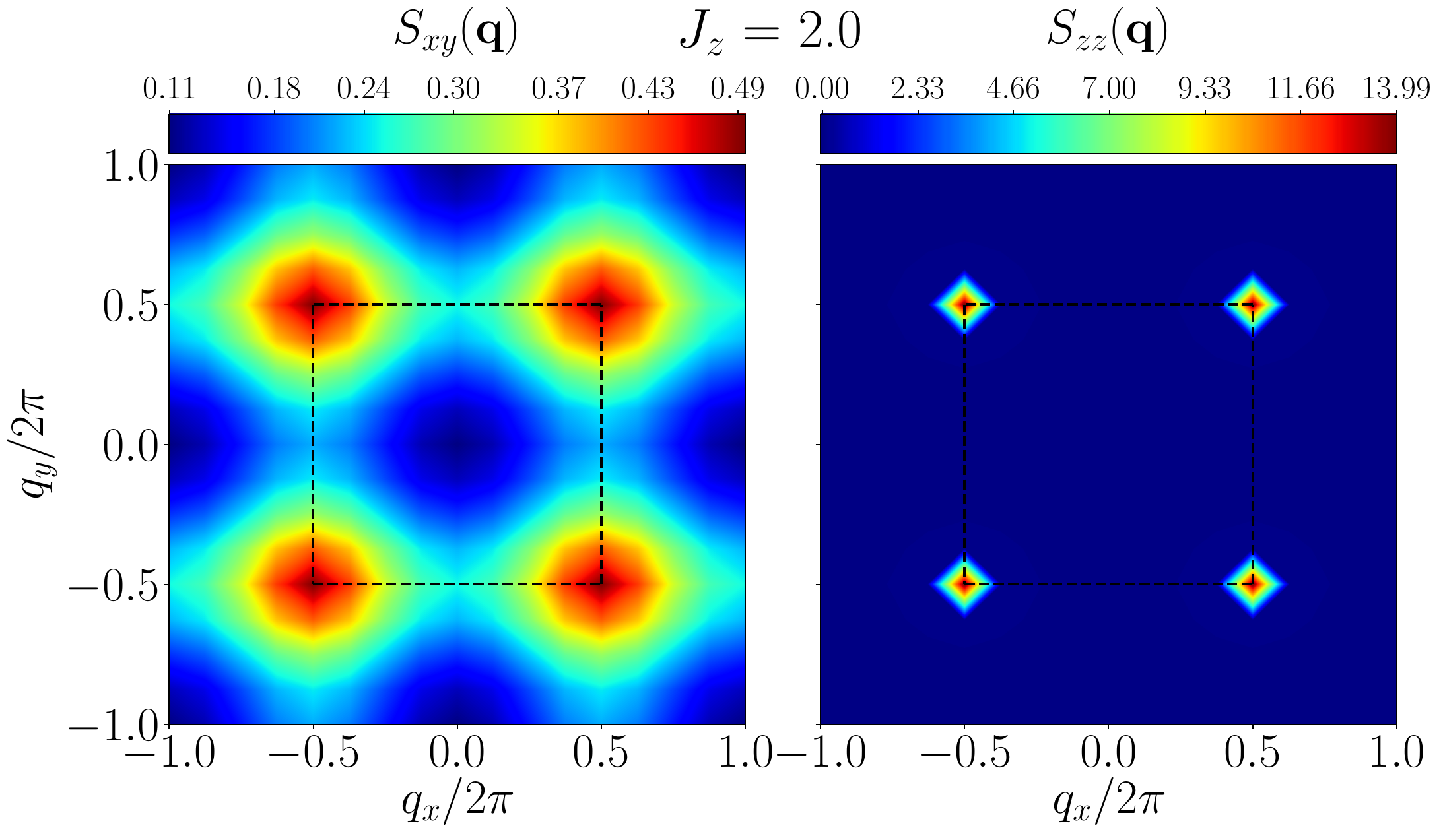} \\
	\caption{\label{fig:S(q)_square_Sz=0} The panels show $S_{xy}(\mathbf{q})$ and $S_{zz}(\mathbf{q})$ 
	for the square lattice, in the first Brillouin zone (highlighted by the dashed black lines) 
	for various representative $J_z$ in the $m=0$ magnetization sector for
	the $N=8 \times 8$ cylinder. 
 }
\end{figure*}

We now calculate the spin correlations in the state $|C_{S^z=0}\rangle$.
Following Ref.~\onlinecite{Changlani_PRL}'s supplementary, 
we have
\begin{align}
\langle C_{S^z} | C_{S^z} \rangle
\propto 
\frac{1}{N+1} \sum_p \prod_j \sum_{s_j}
e^{i p s_j} \langle c_j | s_j \rangle \langle s_j | c_j \rangle
	e^{-i p S^*_z}
\label{eq:mat_elem_def}
\end{align}
up to an overall normalization,
where $p$ runs from $0$ to $\frac{2 \pi N}{N+1}$ in steps
of $\frac{2 \pi}{N+1}$, and $S^*_z$ is desired total $\Sop^z$ sector.
We work with even $N$ to ensure that $S^*_z=0$.
$|c_j\rangle$ refers to the coloring
of the site $j$ in $|C\rangle$, i.e. $|c_j\rangle=|r\rangle$ or $|b\rangle$ for
$A$/$B$ sublattices respectively, and $|s_j\rangle$ are Ising states
$|\pm \frac{1}{2}\rangle$.
Taking into account
the number of states in the $S^*_z=0$ sector compared to the full Hilbert space,
we are guaranteed that $\langle C_{S^z=0} | C_{S^z=0} \rangle = 
\frac{1}{N+1} \frac{2^N}{^N C_{N/2}}\sum_p \left( \cos \frac{p}{2} \right)^N = 1$
as expected. \textcolor{black}{$^N C_M$ stands for the 
``$N$ choose $M$" combinatorial function
everywhere in this paper, i.e. $^N C_M = N!/(M!(N-M)!)$.} 
For the two-point correlators, we perform similar calculations and arrive at
\begin{eqnarray}
\langle C_{S^z=0} | \Sop^z_m \Sop^z_n | C_{S^z=0} \rangle & = & -\frac{1}{4} \frac{1}{N-1}  
\label{eq:mat_elem_SzSz} \\
\langle C_{S^z=0} | \Sop^x_m \Sop^x_n | C_{S^z=0} \rangle & = & \langle C_{S^z=0} |\Sop^y_m \Sop^y_n | C_{S^z=0} \rangle \nonumber \\
							  & = & \frac{\epsilon_{mn}}{8} \frac{N}{N-1} 
\label{eq:mat_elem_SxSx}
\end{eqnarray}
where $\epsilon_{mn}=-1$ for a pair of sites $\{m,n\}$ with different colors,
and $\epsilon_{mn}=1$ for $\{m,n\}$ with the same color.
These exact expressions are readily verifiable 
by performing ED on small systems (see Table~\ref{tab:ED_comp}).
Details of the derivations are given in App.~\ref{app:2c_mat_elem}.
We see from the above that projection to $S^z=0$ sector
introduces only sub-dominant corrections of $O(1/N)$ in the LRO correlations of 
$|C_{S^z=0}\rangle$ when compared to the unprojected state $|C\rangle$
which is another generic feature of the quantum-classical correspondence
in all the examples considered in this paper, and are to be
expected in other ordered cases as well.

We now show that the unprojected state $|C\rangle$ is a gapless ground state  
in the thermodynamic limit. Consider the following %\sout{variational} 
(unprojected) state 
%\textcolor{red}{
$|C'\rangle = \left( \prod_{i \in A} \otimes_i e^{i \Sop_i^z \delta_i} 
|r\rangle_i
\prod_{j \in B} \otimes_j e^{i \Sop_j^z \delta_j} |b\rangle_j \right)$ 
%}
built by modulating the two-colorings of $|C\rangle$ 
in the $XY$ plane of the Bloch sphere
by a small angle $\delta_i$ that oscillates at a \emph{non-zero} 
wavevector $\mathbf{q} \rightarrow 0$,
e.g. $\delta_i = \delta \sin(\mathbf{q} \cdot \vec{r}_i)$ with a small $\delta$.
This variational state is like a Goldstone mode associated with $U(1)$-symmetry breaking,
\textcolor{black}{however it is not orthogonal to $|C\rangle$.
Thus, for an excited state, we consider the following variational
state $|\psi \rangle \propto |C'\rangle - \langle C | C' \rangle |C\rangle$.
$|\psi\rangle$ is orthogonal to $|C\rangle$ by construction.
}
%\sout{$|C'\rangle$ satisfies
%$\langle C' | C \rangle \rightarrow 0$ as $N \rightarrow \infty$.
%\cite{zero_overlap}
%Also, $\langle C' | H | C' \rangle \rightarrow \langle C | H | C \rangle$
%as $\mathbf{q} \rightarrow 0$, which implies
% gapless excitations.\cite{finite_N_gap}}
\textcolor{black}{The variational estimate for the excitation energy
$\Delta E \equiv \langle \psi | H | \psi \rangle / \langle \psi | \psi \rangle
- \langle C | H | C \rangle$
scales to zero as $N \rightarrow \infty$,
provided the variational parameter $\delta$ is chosen to scale 
as $N^\alpha$ with $-1/2 < \alpha < 0$.
The details are given in App.~\ref{app:gapless}.
%, and we just note
%here that $\langle C | C' \rangle$ scales to zero with (finite) $N$ only in this
%range of $\alpha$ or above whence $|C'\rangle \sim |\psi\rangle$ effectively,
%and not when $\alpha < -1/2$ (App.~\ref{app:gapless}).
%Again because of Eq.~\ref{eq:projection} and Ref.~\onlinecite{projector}, 
%a gapless spectrum at an unprojected level implies gaplessness at a projected
%level as well.
}
The foregoing discussions are thus highly suggestive of $|C_{S^z=0}\rangle$
(and $|C\rangle$)
being adiabatically connected to the $SU(2)$ symmetric
N\'eel ground state,
which we numerically demonstrate next for the case of two dimensions
and expect to hold for higher dimensions.

%-------------------------------------------------------------------------
To analyze the magnetic structure and adiabaticity of the $XY$ LRO upto the $SU(2)$ symmetric Heisenberg point,
we calculate the structure factors  
defined by 
\begin{eqnarray}
S_{zz}(\mathbf{q}) &=& \frac{1}{N}\sum_{m,n}e^{-i\mathbf{q} \cdot (\textbf{r}_m-\textbf{r}_n)}\langle S^z_mS^z_n\rangle\nonumber \\
S_{xy}(\mathbf{q}) &=& \frac{1}{N}\sum_{m,n}e^{-i\mathbf{q} \cdot (\textbf{r}_m-\textbf{r}_n)} \frac{\langle S^x_mS^x_n\rangle
+ \langle S^y_mS^y_n\rangle}{2}
\label{eqn:SF}
\end{eqnarray}
where we set $L_x$, $L_y$ as the 
number of unit cells along the primitive lattice directions such that
the number of sites $N = L_x \times L_y$, and
$\mathbf{r}_m$ is the Bravais lattice vector for site $m$,
while $\mathbf{q}$ is a reciprocal lattice vector in the (first) Brillouin zone. 
We also calculate correlation ratios  defined as 
\begin{eqnarray}
R_{\alpha}=1-\frac{S_{\alpha}(\mathbf{q}_0-\Delta \mathbf{q})}{S_{\alpha}(\mathbf{q}_0)}~, 
\label{eqn:CR}
\end{eqnarray}
where $\alpha \in (zz,xy)$, $\mathbf{q}_0$ is a chosen wave vector in the Brillouin zone,
 and $\mathbf{q}_0 - \Delta \mathbf{q}$ represents one choice of the nearest wave vectors
allowed on the discrete lattice.
This quantity scales to $1$ if there is a Bragg peak at $\mathbf{q}_0$
implying ordering at that wavevector in the $\alpha$ channel, and scales to $0$ if there
is no such ordering. 
This ratio is designed to approach unity independent of the strength of quantum fluctuations
as long as there is spin LRO at the chosen wave vector.

For the square lattice, the two structure factors $S_{zz}(\mathbf{q_0})$ 
and $S_{xy}(\mathbf{q_0})$ at $\mathbf{q}_0=(\pi,\pi)$ 
are useful order parameters to measure the diagonal or Ising AFM ordering, 
and off-diagonal or $U(1)$/$XY$ AFM ordering respectively.
In Fig. \ref{fig:square_ED}, we show the ground state energy per 
bond, and the above order parameters
and corresponding correlation ratios at the AFM ordering wave 
vector $\mathbf{q}_0=(\pi,\pi)$ computed using ED and DMRG
(with bond dimension $8000$) in the zero magnetization sector. 
From Fig. \ref{fig:square_ED}(a), we see a monotonic behavior in the
ground state energy in the $XY$ regime extending up to $J_z=-1$ on one side and $J_z=1$ 
on the other side. This is the first piece of data that signals that 
%the regime $J_z \in (-1,1)$ is a single phase 
a single phase encompasses the regime $J_z \in (-1,1)$
of the $XXZ$ phase diagram on the square lattice. 
At the end points of this regime, we observe kinks in the ground 
state energy curves.\cite{bishop2017spin}
At $J_z=-1$, this kink behavior is quite pronounced, and it corresponds to development of 
ferromagnetic order. 

On the other hand, at $J_z=1$,
the kink behavior is less pronounced. However, by looking at the structure factors in the inset of Fig. \ref{fig:square_ED}(a),
we see that $S_{zz}(\mathbf{q}_0)$ dominates over $S_{xy}(\mathbf{q}_0)$ on the Ising side. 
This is consistent with the development of Ising AFM order~\cite{Cuccoli_etal_2003}, which is confirmed by
the fact that $R_{zz}$ is essentially one on this side in Fig. \ref{fig:square_ED}(b).
Germane to the solvable point $2c$ and as is also seen from Eq.~\ref{eqn:SF}, 
we observe that $R_{xy}$ tends to one strongly in the whole regime $J_z \in (-1,1)$.
This second piece of data convincingly establishes
that the $U(1)$ AFM LRO state at $J_z=-1$ is adiabatic all the way
to the $SU(2)$-symmetric point. Our results show that the $XY$ regime of the square lattice unfrustrated magnet,
and likely other unfrustrated magnets, 
has a ground state whose essential properties are captured by the correlations of the
exact ground state $|C_{S^z=0}\rangle$.
We finally note that the FM-AFM phase transition at $H_{2c}$ is a first-order level-crossing transition as 
can be seen in Fig. \ref{fig:square_ED}(a).

Our findings in Fig.~\ref{fig:square_ED} are further substantiated 
in Fig.~\ref{fig:S(q)_square_Sz=0} where we have plotted 
the full structure factor as a function of $\bf{q}$ for the $8 (L_x) \times 8 (L_y)$ cylinder. 
At $J_z=-2$ (top left), there is ferromagnetic order in the system. 
Imposing the constraint of $S_z=0$ in our DMRG calculations leads to a state with
two domains arranged along the length of the cylinder, 
preserving the translational invariance along 
the $y$-direction due to periodic boundary conditions imposed in the $y$-direction. 
As a result of this modulation in the $x$-direction, 
the ordering wavevector in the $zz$ channel %for the $ZZ$ correlations
is not $(0,0)$. 
Instead, peaks occur at the smallest allowable nonzero $|q_x|=\frac{2\pi}{L_x}$ and $q_y=0$. 
The contributions to $S_{xx}(\mathbf{q})$ throughout the entire Brillouin zone 
are significantly smaller and arise purely near the domain wall due to transverse spin fluctuations 
(see App.~\ref{app:real_space_corr} for real space plots of the 
spin-spin correlations for further discussion). 
Moving on to $J_z=-1$ (top right), DMRG correctly captures the exact two-coloring ground 
state; for this state the Fourier transform of the real space spin-spin correlators 
corresponding to the two-coloring wavefunction (Eq.~\ref{eq:mat_elem_SzSz} 
and Eq.~\ref{eq:mat_elem_SxSx}) 
can be computed analytically (App.~\ref{app:exact_struc_fac}). 
$S_{zz}(\mathbf{q})$ is precisely $1/4$ at all points in the first Brillouin zone 
except for $\mathbf{q}=(0,0),$ where its 
value is exactly zero. This is a direct consequence of the sum-rule 
$S_{zz}(0,0) = \frac{1}{N} \langle(\sum_{i=1}^{N}{S^{i}_z})^2 \rangle = 0$ where $N=L_x L_y$. 
$S_{xy}(\mathbf{q})$ has a Bragg peak at $\mathbf{q}_0=(\pi,\pi)$ 
and no peaks elsewhere as might be expected from the quantum-classical correspondence mentioned 
previously.

These features associated with perfect co-planar N\'eel order in $S_{xy}(\mathbf{q})$ 
are quantitatively modified on moving towards the Heisenberg point. 
For $S_{zz}(\mathbf{q})$, there is also a qualitative reorganization of spectral weight.
The featureless $S_{zz}(\mathbf{q})$ at $J_z=-1$ now starts to develop a maximum at $\mathbf{q_0}$.  
For example, at $J_z=0.0$ (middle left) the dip at $(0,0)$ has broadened out significantly. 
As $J_z$ keeps increasing, the maxima at $\mathbf{q_0}$ also acquire appreciable weight
as shown for $J_z=0.6$ (middle right). These features are further enhanced as one approaches the Heisenberg regime, 
and at exactly $J_z=1$ (bottom left) both correlators become identical due to $SU(2)$ symmetry.
Beyond $J_z>1$ (bottom right),  
the dominant correlations are now present in the $zz$ channel 
seen clearly as a Bragg peak at $\mathbf{q}_0$
reflecting Ising LRO, while there are no peaks in the $XY$ channel
but only a broad maximum at $\mathbf{q}_0$ in agreement with lack of 
$U(1)$ AFM LRO as surmised from $R_{xy}$ on the Ising side in Fig.~\ref{fig:square_ED}. 

%===============================================================
\section{The triangular lattice antiferromagnet}
%===============================================================
\subsection{Zero magnetization sector}
\label{sec:triangle_zeromag}

\begin{figure*}[t]
	\includegraphics[width=.47\linewidth]{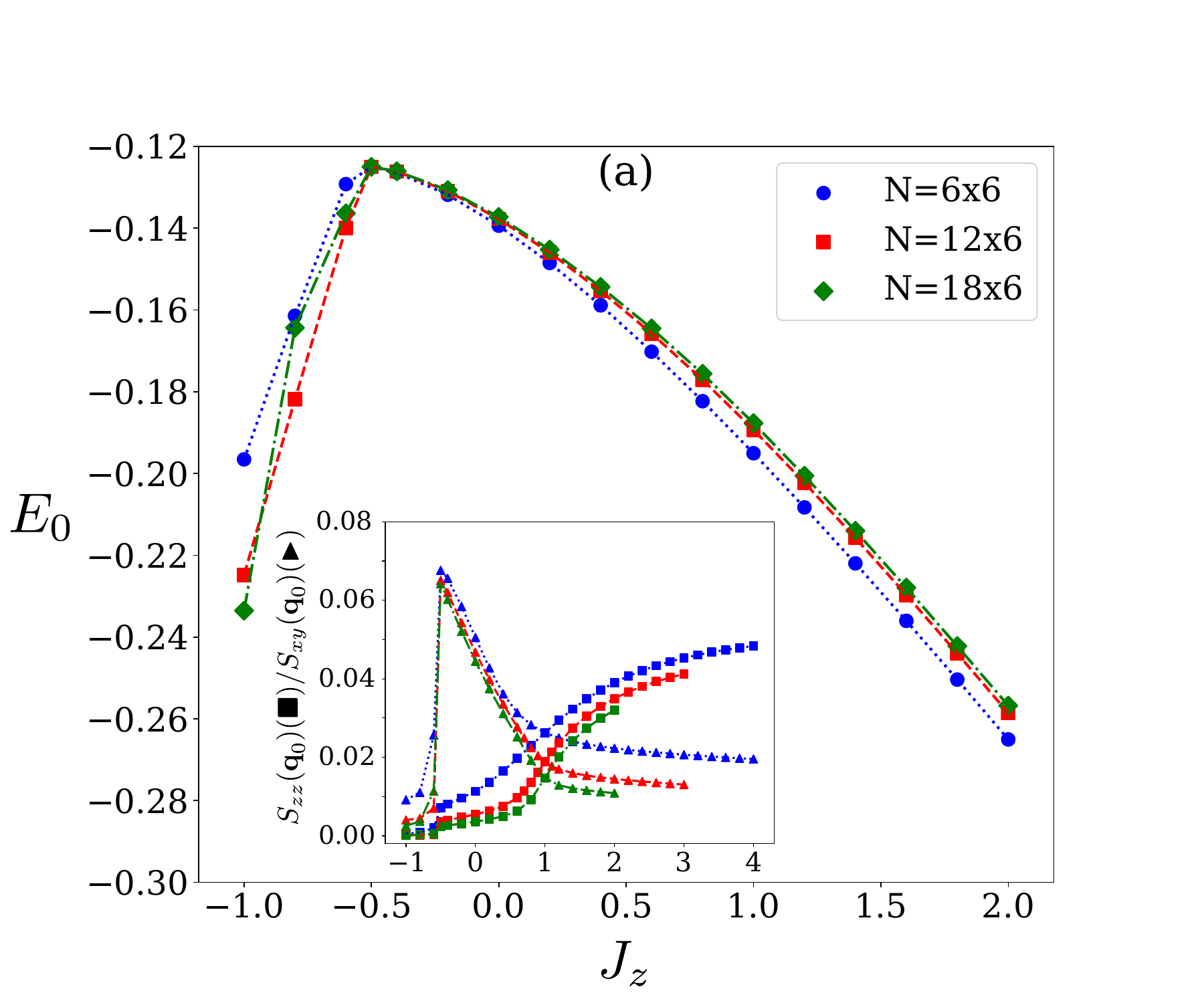}
	\includegraphics[width=.47\linewidth]{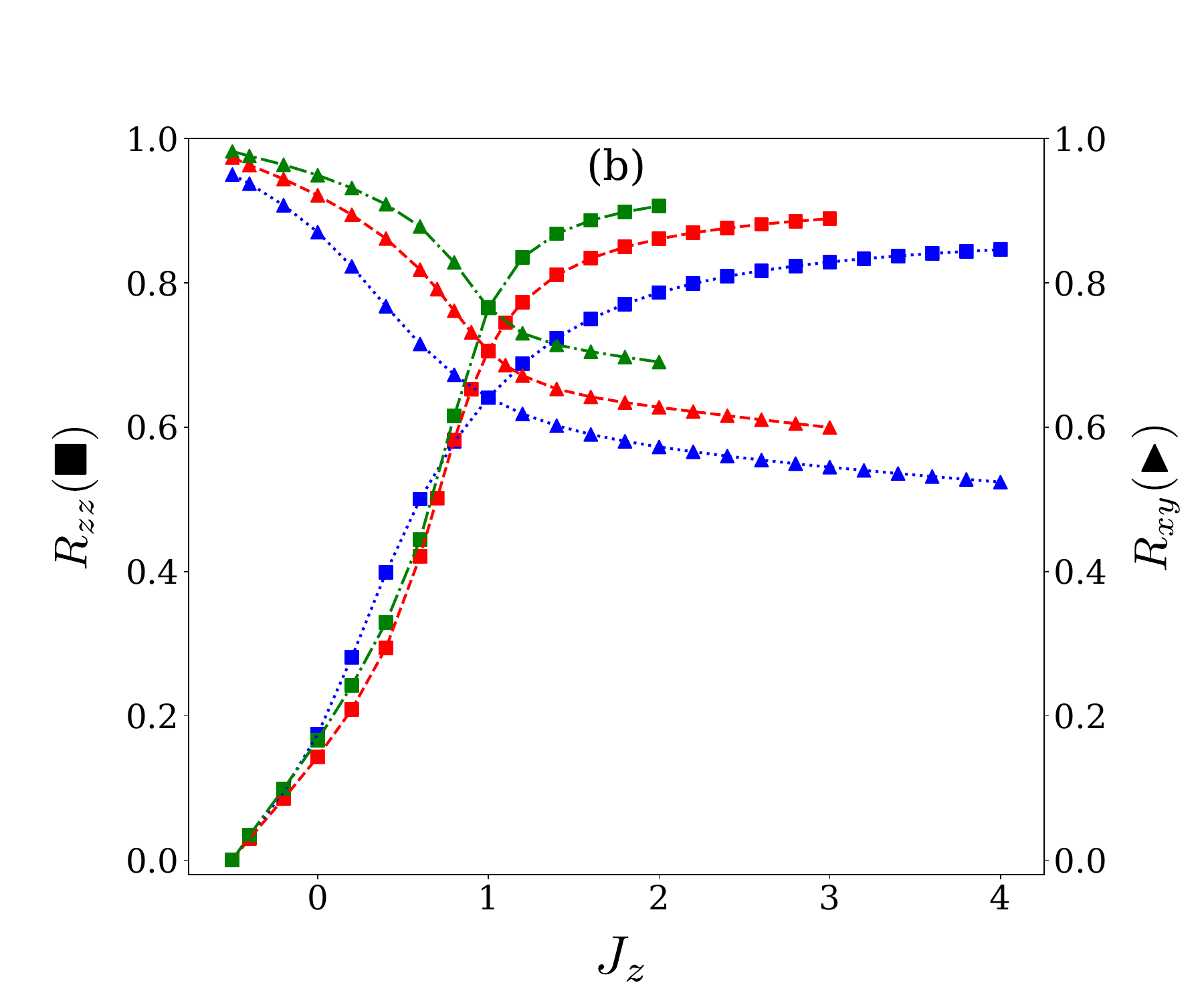}
  \caption{\label{fig:triangular_Sz=0} 
	For the triangular lattice $XXZ$ Hamiltonian in the $m=0$ sector, the left panel shows the ground state energy per bond ($E_0$) vs. $J_z$
	The inset of the left panel shows the evolution of the structure factors ($\frac{S_{zz}(\mathbf{q}_0)}{N}$ and $\frac{S_{xy}(\mathbf{q}_0)}{N}$),
        calculated at the ordering vector $\mathbf{q}_0=(4\pi/3,0)$.
	The right panel shows the correlation ratios ($R_{zz}$ and $R_{xy}$), as defined in the text (Eq.~\ref{eqn:CR}), vs. $J_z$
	at the ordering wave-vector $\mathbf{q}_0$ for the representative case of
	        $\Delta \mathbf{q}=(\frac{2\pi}{L_x}, \frac{4\pi}{\sqrt{3}L_y} - \frac{2\pi}{\sqrt{3}L_x})$.
	}
\end{figure*}

In this and the next section, we turn our attention to the triangular lattice
AFM with its frustrated geometry. This geometry harbors a different solvable
point $H_{3c}$ at $H[J_z=-1/2]$ as introduced in Sec.~\ref{sec:intro} such
that the exact ground states are three-coloring states.
On the triangular lattice, there are two distinct such ground states
one which is sketched in Fig.~\ref{fig:cartoon}. Analogous to the two-coloring case, these
ground states also possess LRO in the $XY$ plane.
Based on our knowledge of the $120^{\circ}$ ordered Heisenberg point~\cite{Richter_lectures},
we expect that LRO of the solvable point $H_{3c}$ extends to the
$SU(2)$ symmetric point analogous to the situation on the square lattice.

In the zero magnetization sector, the two ground states may be written down as
%\textcolor{red}{
\begin{eqnarray}
|C^{(1)}_{Sz=0}\rangle &\equiv & P_{S^z=0} \left( \prod_{i \in A} \otimes_i |r\rangle_i 
\prod_{j \in B} \otimes_j |b\rangle_j \prod_{k \in C} \otimes_k |g\rangle_k
\right)\nonumber\\
|C^{(2)}_{S^z=0}\rangle &\equiv & P_{S^z=0} \left( \prod_{i \in A} \otimes_i |r\rangle_i 
\prod_{j \in B} \otimes_j |g\rangle_j \prod_{k \in C} \otimes_k |b\rangle_k
\right) \; \; \; \; \; \; \; \;
\label{eq:wavefn}
\end{eqnarray}
%}
where $A/B/C$ are the three sublattices, 
and $|r\rangle \equiv \frac{1}{\sqrt{2}}(|\uparrow\rangle+|\downarrow\rangle)$, 
$|b\rangle \equiv \frac{1}{\sqrt{2}}(|\uparrow\rangle+\omega|\downarrow\rangle)$ 
and $|g\rangle \equiv \frac{1}{\sqrt{2}}(|\uparrow\rangle+\omega^2|\downarrow\rangle)$. 
$\omega=e^{i 2\pi/3}$ and $\omega^2=\omega^*$ are the cube roots of unity.
$|r\rangle, |b\rangle, |g\rangle$ may be chosen to be any triad of $120^\circ$ states in the $XY$ plane of 
the Bloch sphere due to the presence of $U(1)$ symmetry, this choice being a global gauge choice.  

Based on the existence of the $J_z=-\frac{1}{2}$ point coupled with
linear spin-wave calculations, Ref.~\onlinecite{Momoi} argued 
the adiabaticity of the coloring ground states to the ground state at the
$SU(2)$ point. 
In what follows, we will work in a fixed magnetization sector 
and numerically demonstrate this adiabaticty working with projected wavefunctions
by calculating structure factors and correlation ratios.

For the $N$ site triangular lattice, the
overlap between the two three-coloring states is given by
%\textcolor{red}{
\begin{equation}
  \langle C^{(k)}_{S^z=0}|C^{(l)}_{S^z=0}\rangle =\begin{cases}
    1, & \text{for $k=l$}.\\
    \frac{{^{N/3}}C_{N/6}}{{^N}C_{N/2}}, & \text{for $k\neq l$}.
  \end{cases}
\label{eq:3c_zeromag_overlap}
\end{equation}
%}
%\begin{eqnarray}
%\langle C^{(k)}_{S^z=0}|C^{(l)}_{S^z=0}\rangle &=& 1 ~~~\text{for $k=l$}\nonumber \\ 
%	&=& \frac{{^{N/3}}C_{N/6}}{{^N}C_{N/2}}~~~\text{for $k\neq l$}~.
%\label{eq:3c_zeromag_overlap}
%\end{eqnarray}
where $k,l\in (1,2)$.
%\textcolor{red}{and the combinatorial functions 
%defined as before ${^{N}}C_{M}=\frac{N!}{(N-M)!M!}$}. 
It goes to zero for $k\neq l$ exponentially as $N \to \infty$ 
due to the macroscopic difference in the colors in the 
two wavefunctions.
Perturbing away from the $3c$ point
towards the Heisenberg point brings in matrix elements with magnitude that are 
exponentially small in $N$ between
$|C^{(1)}\rangle$ and $|C^{(2)}\rangle$ at lowest-order 
resulting in an exponentially small splitting. 
As one goes further away from the $3c$ point, non-perturbative effects result in 
a finite splitting such that
there is a unique ground state at the Heisenberg point.
Alternatively, this can be understood by starting at the Heisenberg point
which, being fully $SU(2)$-symmetric,  harbors the low-energy quasi-degenerate
Anderson tower of states whose
energy spectrum is given by $\sim \frac{S(S+1)}{N}$.~\cite{Lhuillier} 
Appropriate linear combinations
of these states are known to give symmetry broken states.~\cite{Bernu1992} 
Thus, the effect of 
$XY$ anisotropy is to break this quasi-degeneracy of the Heisenberg point
and lead to the (two) AFM ordered states. At and near the Heisenberg point, 
these states have significant quantum fluctuations~\cite{White_Chernyshev_2007}
which become effectively absent at the $3c$ point (Eq.~\ref{eq:wavefn}). 

\begin{figure*}[t]
	\includegraphics[width=.47\linewidth]{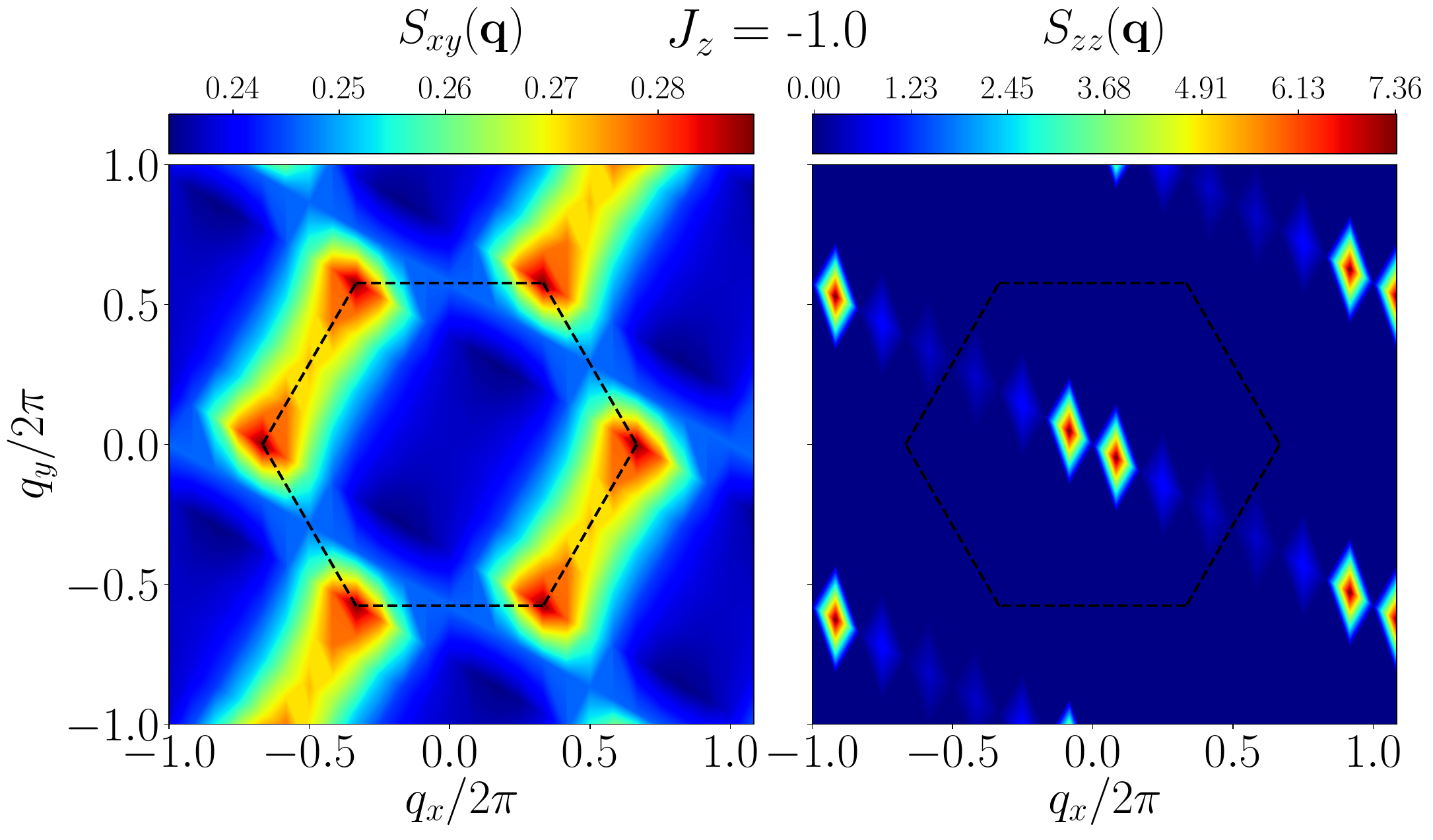}
	\includegraphics[width=.47\linewidth]{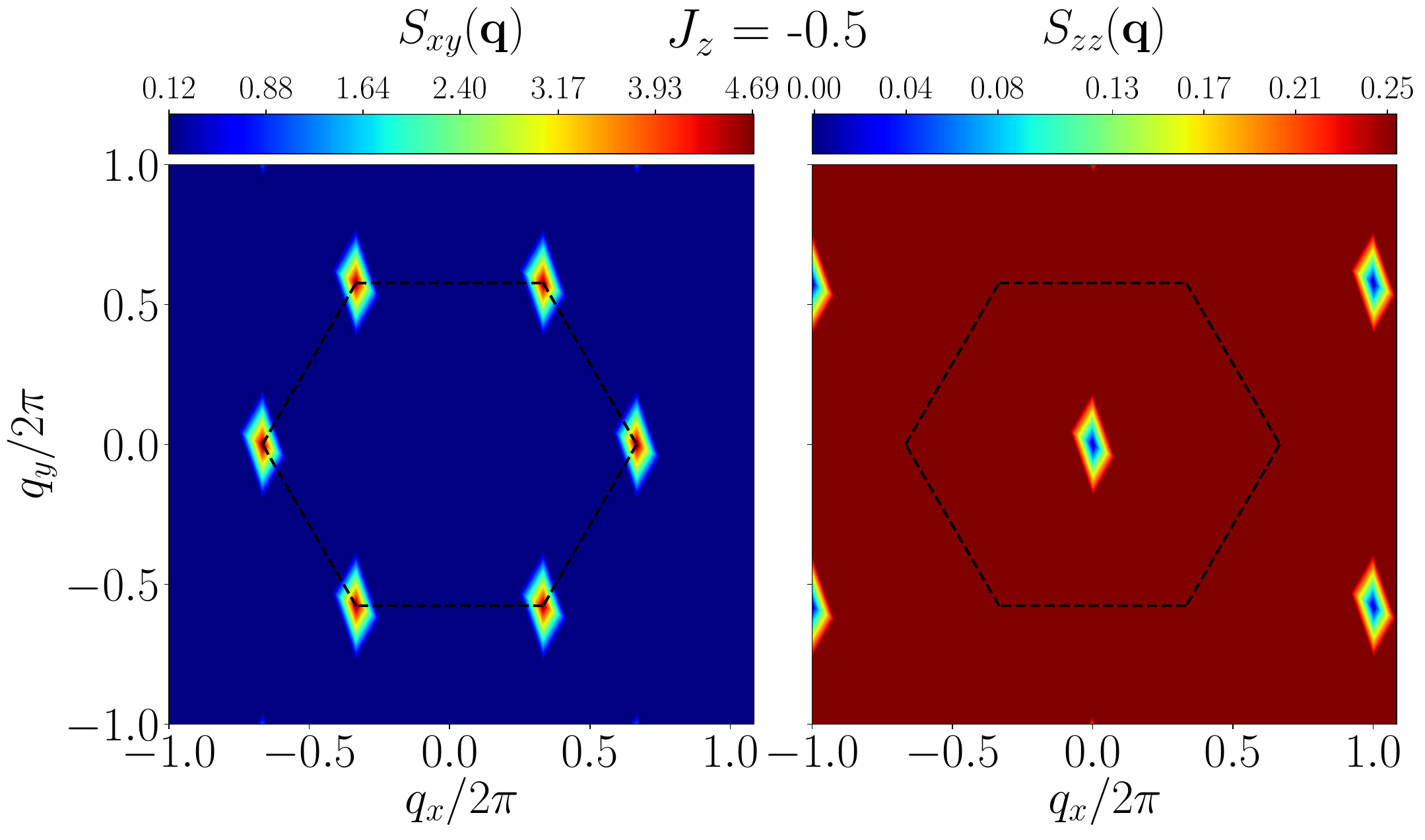} \\
	\vspace{2mm}
	\includegraphics[width=.47\linewidth]{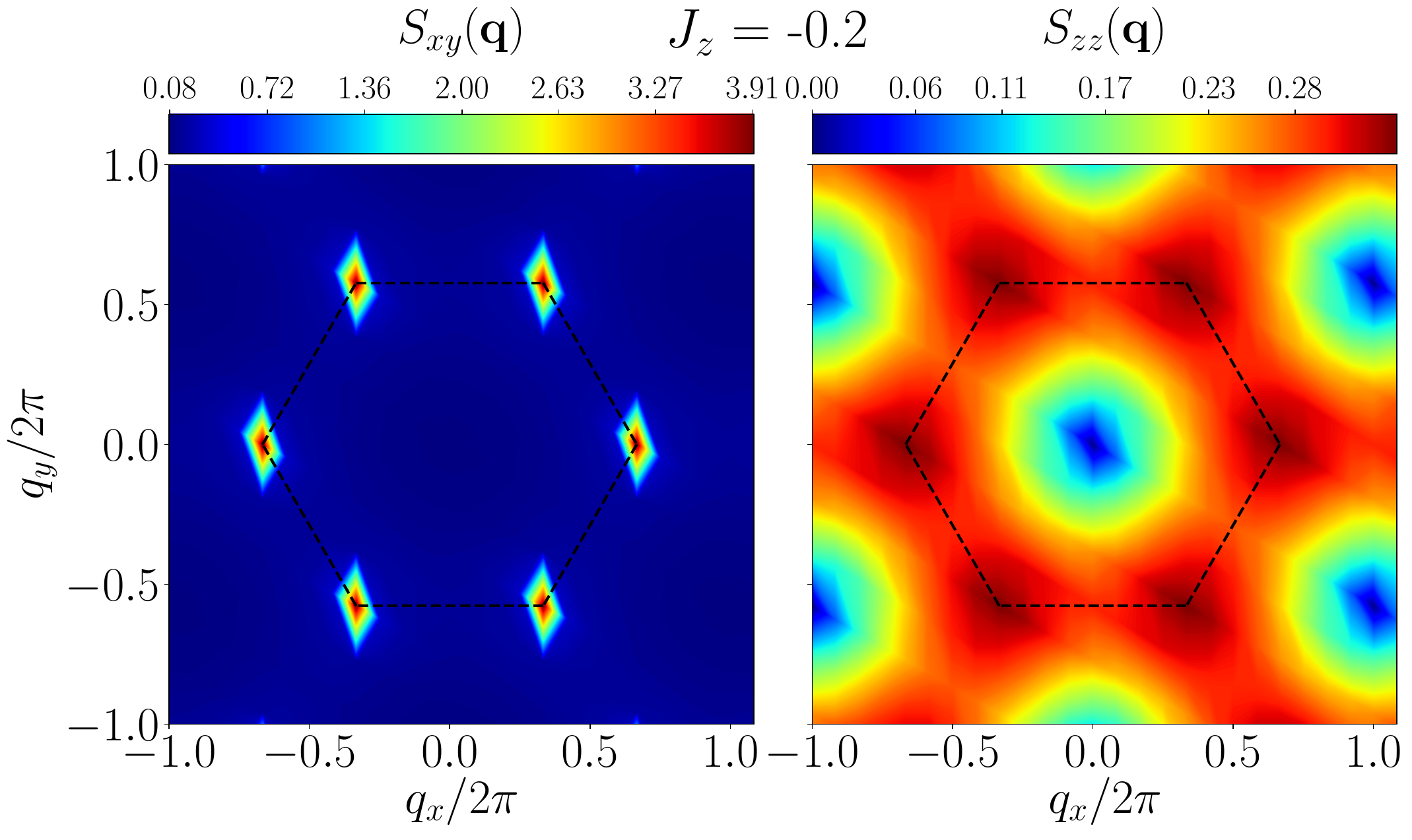}
	\includegraphics[width=.47\linewidth]{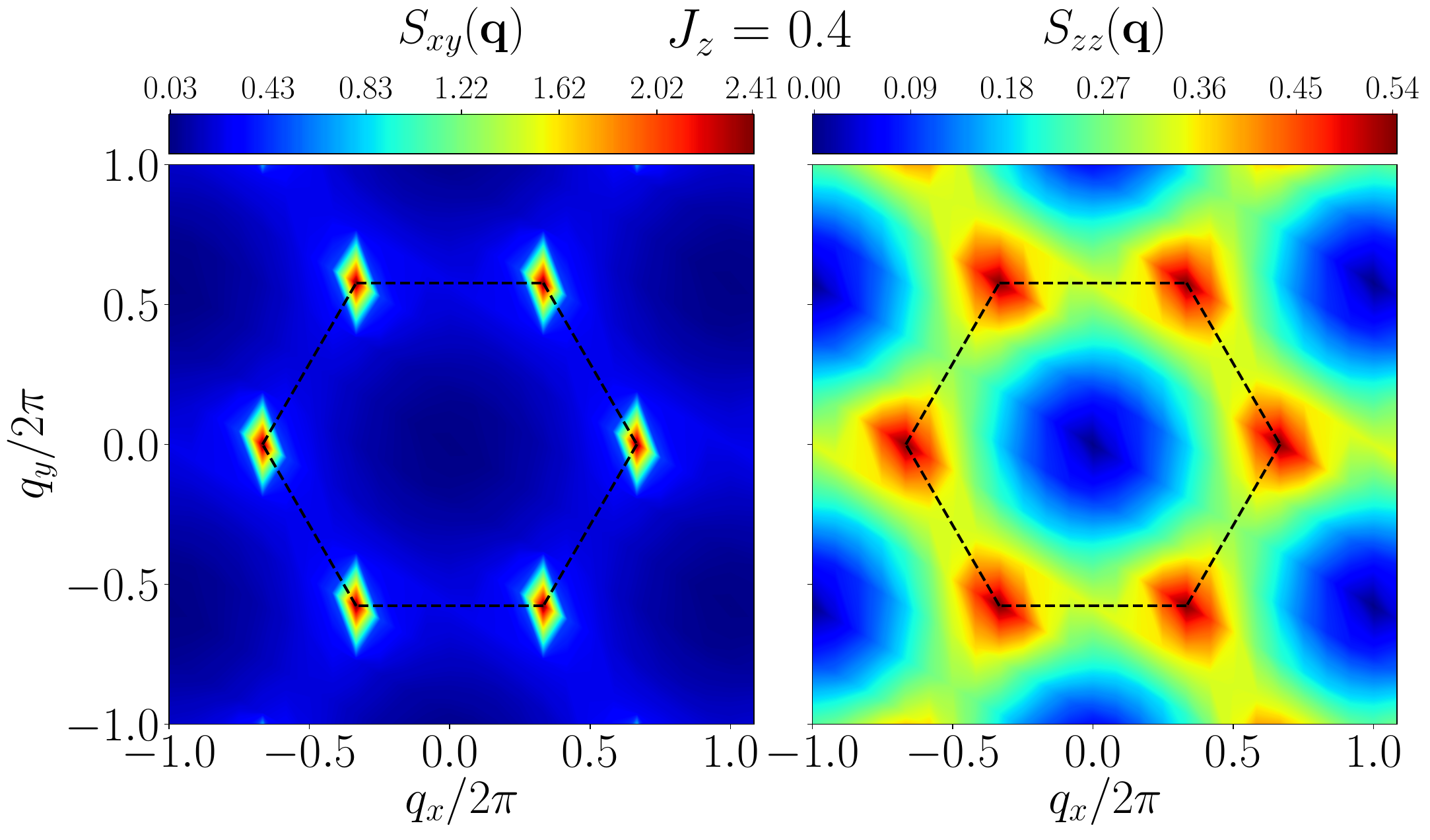} \\
	\vspace{2mm}
	\includegraphics[width=.47\linewidth]{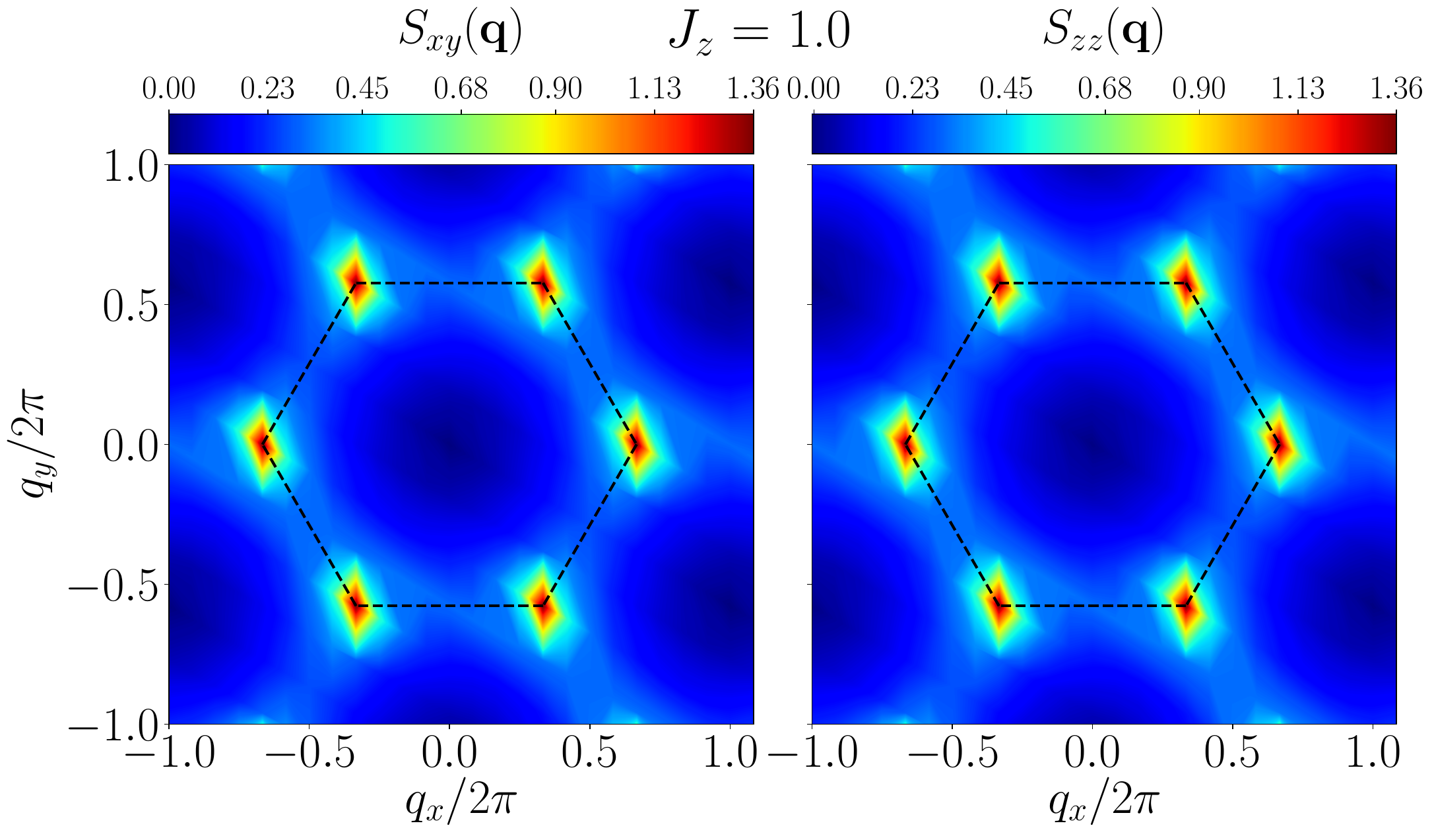}
	\includegraphics[width=.47\linewidth]{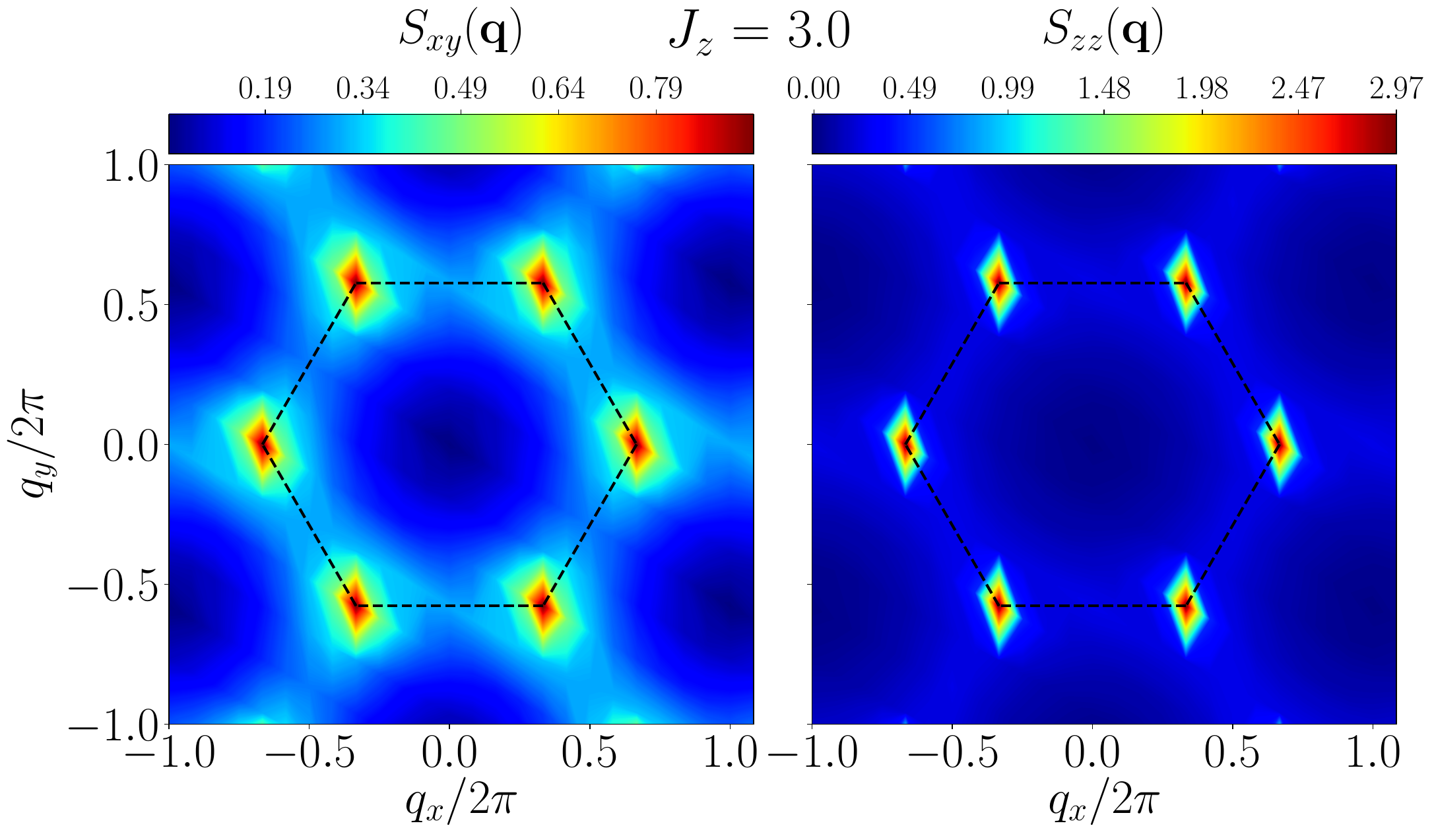}
  \caption{\label{fig:S(q)_triangular_Sz=0} The panels show $S_{xy}(\mathbf{q})$ and $S_{zz}(\mathbf{q})$ 
	for the triangular lattice in the first Brillouin zone (highlighted by the dashed black lines) 
	for various representative $J_z$ in the $m=0$ magnetization sector for the $N=12 \times 6$ cylinder. 
 }
\end{figure*}

A similar calculation (see App.~\ref{app:3c_zeromag_mat_elem}) 
for the correlators in either of the two ground states gives
\begin{eqnarray}
\langle C^{(k)}_{S^z=0} | \Sop^z_m\Sop^z_n | C^{(k)}_{S^z=0} \rangle &=& -\frac{1}{4}\frac{1}{N-1} 
\label{eq:3c_zeromag_corr_zz} \\
\langle C^{(k)}_{S^z=0} | \Sop^x_m\Sop^x_n | C^{(k)}_{S^z=0} \rangle
	&=& \langle C^{(k)}_{S^z=0} | \Sop^y_m\Sop^y_n | C^{(k)}_{S^z=0} \rangle  \nonumber \\
&=& \frac{\epsilon_{mn}}{8}\frac{N}{N-1}
\label{eq:3c_zeromag_corr}
\end{eqnarray}
where $\epsilon_{mn}=-1/2$ for a pair of sites $\{m,n\}$ with different colors,
and $\epsilon_{mn}=1$ for $\{m,n\}$ with the same color.
For $N\rightarrow\infty$,
$\langle\Sop^z_m\Sop^z_n\rangle\rightarrow 0$ 
and $\langle\Sop^x_m\Sop^x_n\rangle\propto \epsilon_{mn}$ 
reflect the three sublattice or $120^\circ$ order solely lying in the $XY$-plane.

The structure factors (Eq. \ref{eqn:SF}) $S_{zz}(\mathbf{q_0})$ 
and $S_{xy}(\mathbf{q_0})$ 
at $\mathbf{q}_0=(\pm 4\pi/3,0)$ are useful order parameters for this case. 
They quantify the presence or absence of ``diagonal" and ``off-diagonal" LRO respectively 
in terms of the mapping between $S=\frac{1}{2}$ degrees of freedom and hardcore bosons
($S^+_m \to b^\dagger_m$, $S^-_m \to b_m$ and $S^z_m \to b^\dagger_m b_m - 1/2$ on site $m$).
If $S_{zz}(\mathbf{q})$ is finite as $N \to \infty$, 
the system has a broken sublattice $\Sop^z$ 
symmetry which corresponds to the boson occupation density wave at wavevector $\mathbf{q}$, 
whereas a finite $S_{xy}(\mathbf{q})$ as $N \to \infty$ represent a 
broken $U(1)$ rotational symmetry which corresponds to superfluid ordering of the bosons.
\cite{sellmann2015phase}

In Fig.~\ref{fig:triangular_Sz=0}(a), 
we show the ground state energy per bond 
using both ED on toric and DMRG (with bond dimension $7000$) on cylindrical geometries. 
Its behavior is featureless as we scan from $H_{3c}$ 
to the Heisenberg point and beyond when compared to the corresponding
data set for the square lattice (Fig. \ref{fig:square_ED}(a)). 
In the inset of Fig.~\ref{fig:triangular_Sz=0}(a),  we show the
magnitude of structure factors at the 
ordering vector $\mathbf{q}_0=(4\pi/3,0)$.
In the range $-0.5<J_z<1$, $S_{xy}(\mathbf{q}_0)$ dominates over 
$S_{zz}(\mathbf{q}_0)$. Their finite size dependence suggests the absence of boson occupation ordering,
and the presence of three-sublattice AFM LRO
lying in the $XY$-plane tied to the $3c$ point (Eq.~\ref{eq:3c_zeromag_corr}) 
corresponding to the superfluid state in the hardcore boson 
language. In contrast, $S_{zz}(\mathbf{q}_0)$ dominates over
$S_{xy}(\mathbf{q}_0)$ for $J_z>1$. 
The finite size dependence of $S_{zz}(\mathbf{q}_0)$ clearly shows the presence of boson occupation
order in this regime. Moreover, the finite size dependence of $S_{xy}(\mathbf{q}_0)$ suggests
a coexistence of superfluid ordering in this regime, i.e. supersolid order, in agreement with
earlier studies.
\cite{wang2009extended,jiang2009supersolid,heidarian2010supersolidity}

However, inferring the thermodynamic behavior from the 
finite size dependence of order parameters can sometimes be inconclusive, especially if the
extrapolated value is small as is the case for $S_{xy}(\mathbf{q}_0)$ 
for $J_z>1$ (inset of Fig.~\ref{fig:triangular_Sz=0}(a)).
In such situations, correlation ratios as defined in Eq.~\ref{eqn:CR}, have proved
especially useful since they have been shown to be less susceptible
to finite size effects.\cite{pujari2016interaction}
Thus, we utilize them to  probe the coexistence
of density wave and superfluid LRO for $J_z>1$ which is shown in
Fig.~\ref{fig:triangular_Sz=0}(b), choosing a representative $\Delta\mathbf{q}$ for
computations.
In the $XY$ regime, we see that $R_{xy}$ tends towards unity with increasing system size, while
$R_{zz}$ decreases towards zero.
This is consistent with the presence of $120^\circ$ AFM in the $XY$ plane
or superfluid order. As we go beyond the Heisenberg point ($J_z>1$),
we see that $R_{zz}$ now increases towards one providing evidence for
boson density wave ordering. Furthermore, we see that $R_{xy}$ is quite appreciable
and evidently consistent with a non-zero value that is increasing towards unity
as we go towards the thermodynamic limit for the
system sizes studied here. This
provides strong evidence for the coexistence of superfluid
and boson density ordering in the zero magnetization sector of the triangular AFM 
on the Ising side.

Given the unusual coexistence of diagonal and off-diagonal orders presented above
unlike the square lattice case discussed in the previous section, 
we address how they are reflected in the spin structure factors. 
For the $12 \times 6$ cylinder we plot $S_{zz}(\mathbf{q})$ and $S_{xy}(\mathbf{q})$  
as a function of $J_z$ (Fig.~\ref{fig:S(q)_triangular_Sz=0}). 
Our findings bear many qualitative similarities to the square lattice case on the $XY$ side. 
At $J_z=-1$ (top left), there is ferromagnetic order in the system with domains,  
and accordingly, the peaks in $S_{zz}(\mathbf{q})$ occur at the smallest allowable 
nonzero $|q_x|=\frac{2\pi}{L_x}$ and the corresponding $q_y$. 
Then, at $J_z=-1/2$ (top right), DMRG spontaneously picks one of the two three-colorings, 
and the features seen can be matched by exact computations 
(Eqs.~\ref{eq:3c_zeromag_corr_zz}, ~\ref{eq:3c_zeromag_corr}, App.~\ref{app:exact_struc_fac}, also see App.~\ref{app:real_space_corr}). 
$S_{zz}(\mathbf{q})$ is again precisely $1/4$ at all points in the first Brillouin zone 
except for $\mathbf{q}=(0,0),$ where its 
value is exactly zero. 
$S_{xy}(\mathbf{q})$ has Bragg peaks at $\mathbf{q}_0=(\frac{4\pi}{3},0)$ and
symmetry related points in the Brillouin zone. 
For $J_z>-1/2$, the sequence of panels in Fig.~\ref{fig:S(q)_triangular_Sz=0} 
from $J_z=-0.2$ to $J_z=1.0$
confirm that the features associated with perfect coplanar $120^\circ$ order 
at $J_z=-1/2$
are only quantitatively modified on moving towards the Heisenberg point. 
Beyond $J_z>1$ (bottom right), 
the correlations are again dominated by the $zz$ channel with 
pronounced Bragg peaks seen at $\mathbf{q}_0$ signaling the diagonal LRO. 
However, the maxima in the $XY$ channel at $\mathbf{q}_0$ are 
\emph{also} Bragg peaks as confirmed through the size dependence of
correlation ratio $R_{xy}$ on the Ising side (Fig.~\ref{fig:triangular_Sz=0})
which is the expected signature of the co-existence of superfluid LRO in the structure factor, 
as opposed to the square lattice case where only
a broad maximum was present at the ordering wavevector $(\pi,\pi)$.

Our ED and DMRG results are in agreement with the previous studies that
have focused on $J_z>0$. 
Our study shows that the properties on the $XY$ side 
originate from the $3c$ point including the $120^\circ$ order at the Heisenberg point. 
Thus, for zero magnetization, we may say that the Heisenberg points 
on triangular and square lattices are ``inheriting" the long-range AFM order of 
their respective solvable points $H_{3c}$ and $H_{2c}$.
Moreover, on the Ising side past the Heisenberg point, the correlation ratio
data provides compelling evidence for the
coexistence of diagonal and off-diagonal LRO.

%===============================================================

\subsection{$\frac{m}{m_s}=\frac{1}{3}$ sector}
\label{sec:triangle_1by3}

\begin{figure*}[]
  \centering
  \includegraphics[width=.47\linewidth]{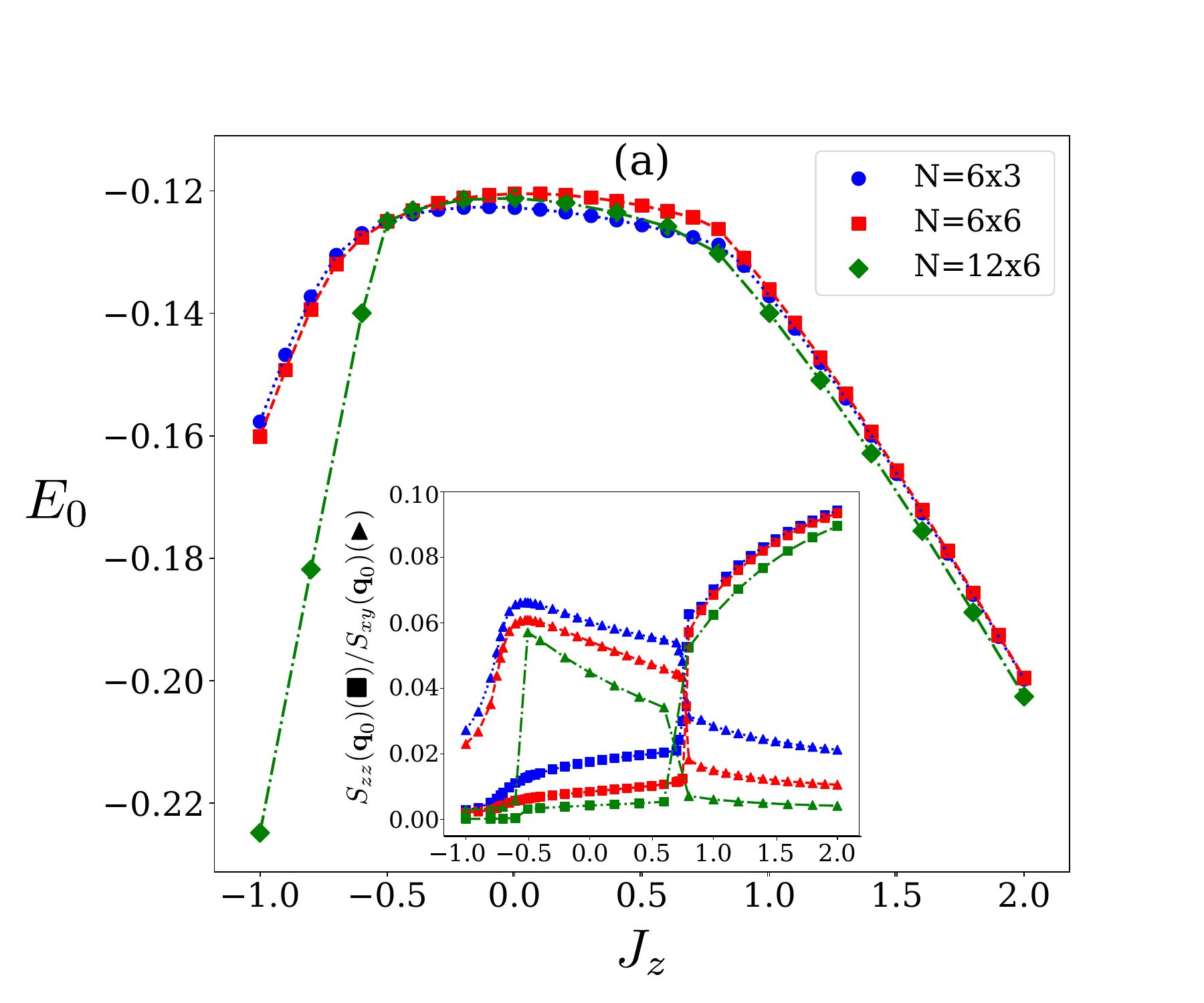}
  \includegraphics[width=.47\linewidth]{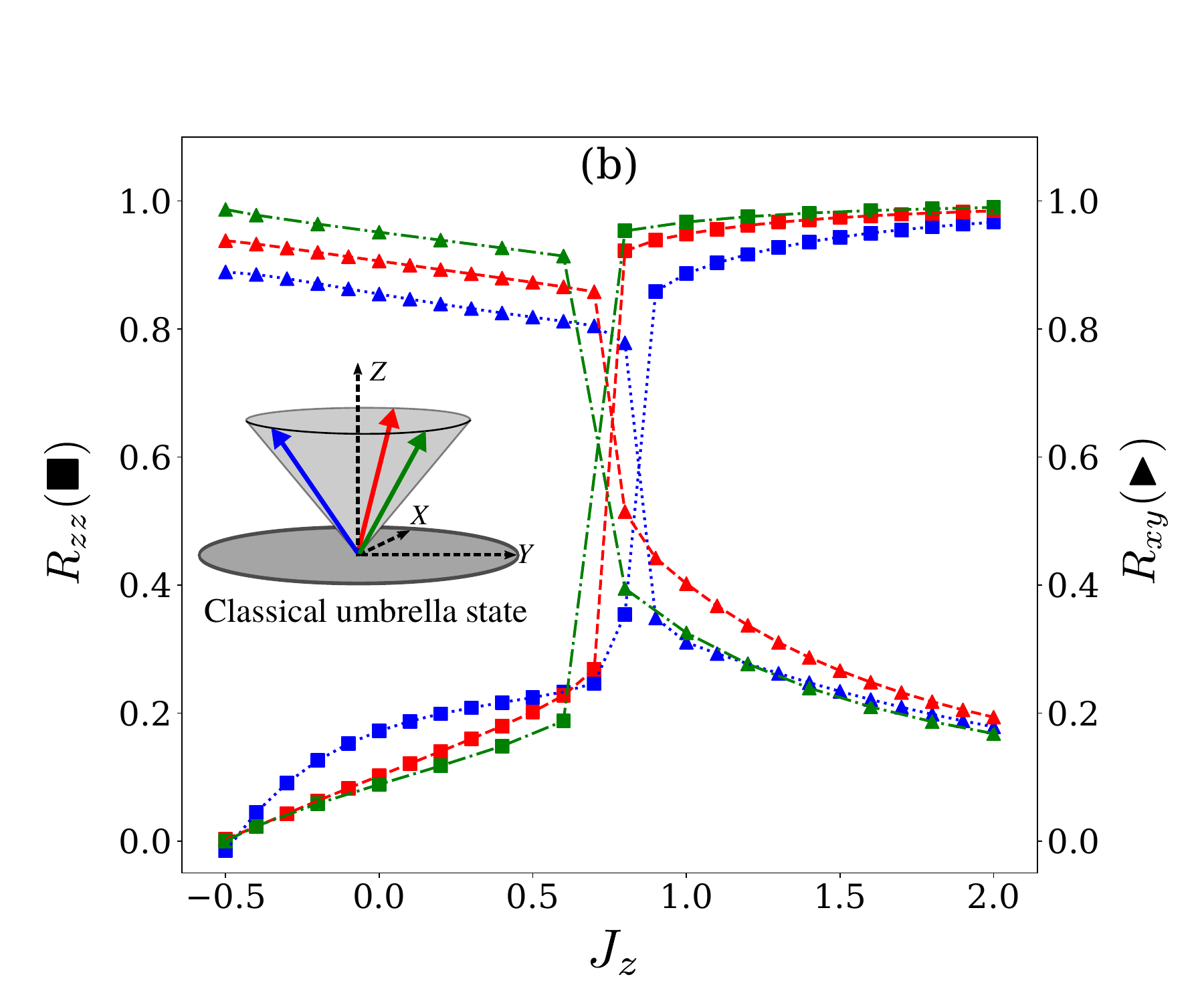}
  \caption{\label{fig:triangular_Sz=OT}
	For the triangular lattice $XXZ$ Hamiltonian in the $\frac{m}{m_s}=\frac{1}{3}$ sector, the left panel shows the ground state energy per bond ($E_0$) vs. $J_z$.
	The inset of the left panel shows the evolution of the structure factors 
	$\left( \frac{S_{zz}(\mathbf{q}_0)}{N} \text{ and } \frac{S_{xy}(\mathbf{q}_0)}{N} \right)$, 
	calculated at the ordering vector $\mathbf{q}_0=(4\pi/3,0)$.
  The right panel shows the correlation ratios ($R_{zz}$ and $R_{xy}$) as defined in the text (Eq.~\ref{eqn:CR}) vs. $J_z$  plot at the ordering wave-vector $\mathbf{q}_0$ for the representative case of
$\Delta \mathbf{q}=(\frac{2\pi}{L_x}, \frac{4\pi}{\sqrt{3}L_y} - \frac{2\pi}{\sqrt{3}L_x})$.
The inset of the right panel shows a schematic of classical spin configurations in the classical umbrella state 
	that correspond to the corresponding three-colored sites (Fig.~\ref{fig:cartoon}).
				  }
\end{figure*}

The ground state of the $\frac{m}{m_s}=\frac{1}{3}$ sector of the triangular Heisenberg AFM 
has been argued to be a magnetization
plateau state.~\cite{Chubukov_jpcm1991,alicea2009quantum} 
In this state, each triangle has two spin-ups and one spin-down
in a modulated pattern at the wave vector $\mathbf{q}_0=(4\pi/3,0)$
(the ``UUD" state) which is equivalent to $\frac{1}{3}$ filling of hardcore bosons 
ordering at the same wave vector.
A magnetization plateau state is an incompressible state with a 
gap to excitations that change magnetization. 
In contrast, coloring ground states are expected to be 
gapless with low energy Goldstone modes lying above it.
At the classical level for $\frac{m}{m_s}=\frac{1}{3}$, the ground state in the $XY$ regime ($0 < J_z < 1$), 
is an ``umbrella" state 
whose projection on to the $XY$ plane has $120^\circ$ correlations
(see a schematic in the inset of Fig.~\ref{fig:triangular_Sz=OT}).~\cite{yamamoto2014quantum}
This classical umbrella state in fact extends all the way to the $3c$ point.
Since the $3c$ point exists for any magnetization sector,
it is natural to ask
how the quantum counterpart of the classical umbrella state that emerges from the
$3c$ point eventually transitions
to the magnetization plateau state.

Starting from Eq.(\ref{eq:mat_elem_def}) in this 
$\frac{m}{m_s}=\frac{1}{3}$ sector, i.e. setting $S^*_z=N/6$, gives
%\textcolor{red}{
\begin{equation}
  \langle C^{(k)}_{\Sop^z=N/6} | C^{(l)}_{\Sop^z=N/6} 
	\rangle =\begin{cases}
    1, & \text{for $k=l$}.\\
    \frac{^{N/3}C_{2N/9}}{^{N}C_{2N/3}}, & \text{for $k\neq l$}.
  \end{cases}
\end{equation}
%}
%\begin{eqnarray}
%	\langle C^{(k)}_{\Sop^z=N/6} | C^{(l)}_{\Sop^z=N/6} 
%	\rangle &=& 1 ~~~\text{for $k=l$}\nonumber\\
% &=& \frac{^{N/3}C_{2N/9}}{^{N}C_{2N/3}}~~~\text{for $k\neq l$}
%\end{eqnarray}
and in the thermodynamic limit, the overlap between the two three-coloring
ground state again goes to zero. Similarly, we have
\begin{eqnarray}
	\langle C^{(k)}_{\Sop^z=N/6} | S^z_m S^z_n | C^{(k)}_{\Sop^z=N/6} \rangle &=& 
	-\frac{1}{4} \left[\frac{8}{9}\frac{N}{N-1}-1\right] \\
\label{EqnSzSz}
	\langle C^{(k)}_{\Sop^z=N/6} | S_m^xS_n^x |C^{(k)}_{\Sop^z=N/6} \rangle &=& 
	\langle C^{(k)}_{\Sop^z=N/6} | S_m^yS_n^y |C^{(k)}_{\Sop^z=N/6} \rangle \nonumber \\
	&=& \frac{\epsilon_{mn}}{9}\frac{N}{N-1} 
\end{eqnarray}
where again $\epsilon_{mn}=-1/2$ for a pair of sites $\{m,n\}$ having different colors,
while $\epsilon_{mn}=1$ for $\{m,n\}$ with same color. 
This again reflects $120^\circ$ sub-lattice LRO in the $XY$ plane in 
triangular lattice. 
As expected, $\langle S^z_mS^z_n\rangle $ is now finite as $N \to \infty$
in this nonzero magnetization sector with the thermodynamic
value of this correlator in complete agreement with $\frac{m}{m_s}=\frac{1}{3}$.
This along with $\langle  S_m^xS_n^x \rangle\propto \epsilon_{mn}$ tells us 
that the state at the solvable point in this magnetization sector is the
quantum counterpart of
umbrella state illustrated in Fig.~\ref{fig:triangular_Sz=OT}.

In Fig.~\ref{fig:triangular_Sz=OT}(a), we show the ground state energy (per bond)
for a wide range of $J_z$.
It shows a sharp kink at $J_{z}^{c} \approx 0.75$ on the XY side,
indicative of a first-order phase transition that occurs 
before the $SU(2)$-symmetric Heisenberg point.
In the range $-0.5<J_z < J_z^{c}$, 
$S_{xy}(\mathbf{q}_0)$ dominates over $S_{zz}(\mathbf{q}_0)$ at $\mathbf{q}_0=(4\pi/3,0)$ in accordance with an umbrella state.
Due to the net magnetization, $S_{zz}$ has a peak at the zero wavevector (not shown) for all $J_z$.
Once $J_z > J^c_z$,  $S_{zz}(\mathbf{q}_0)$ becomes the dominant order parameter, while
$S_{xy}(\mathbf{q}_0)$ is suppressed in accordance with the UUD state.

We confirm the first-order nature of the transition using the correlation ratio as
shown in Fig.~\ref{fig:triangular_Sz=OT}(b): 
To the left of $J^c_z$, $R_{xy}$ tends to unity while
$R_{zz}$ tends to zero.
To the right of $J^c_z$, $R_{zz}$ tends towards unity, while $R_{xy}$ tends towards zero. 
In this magnetization sector, the finite size trends of the order parameter
and the correlation ratio are clear-cut,
and we clearly see the first-order behavior
as sharp discontinuities in these quantities near $J^c_z$.
Our results for $J_z>0$ 
are in agreement with previous work on the triangular phase 
diagram~\cite{yamamoto2014quantum,sellmann2015phase} and extend it to the solvable point. 
Through this work, we realize that 
the umbrella state in the phase diagram as actually being
inherited from the $3c$ point, but quantum fluctuations eventually drive
a phase transition to the UUD plateau state. 

%===============================================================

\section{Colors and dimers in the anisotropic Majumdar-Ghosh chain}
\label{sec:dimers}

\begin{figure}[h]
\includegraphics[width=0.9\linewidth]{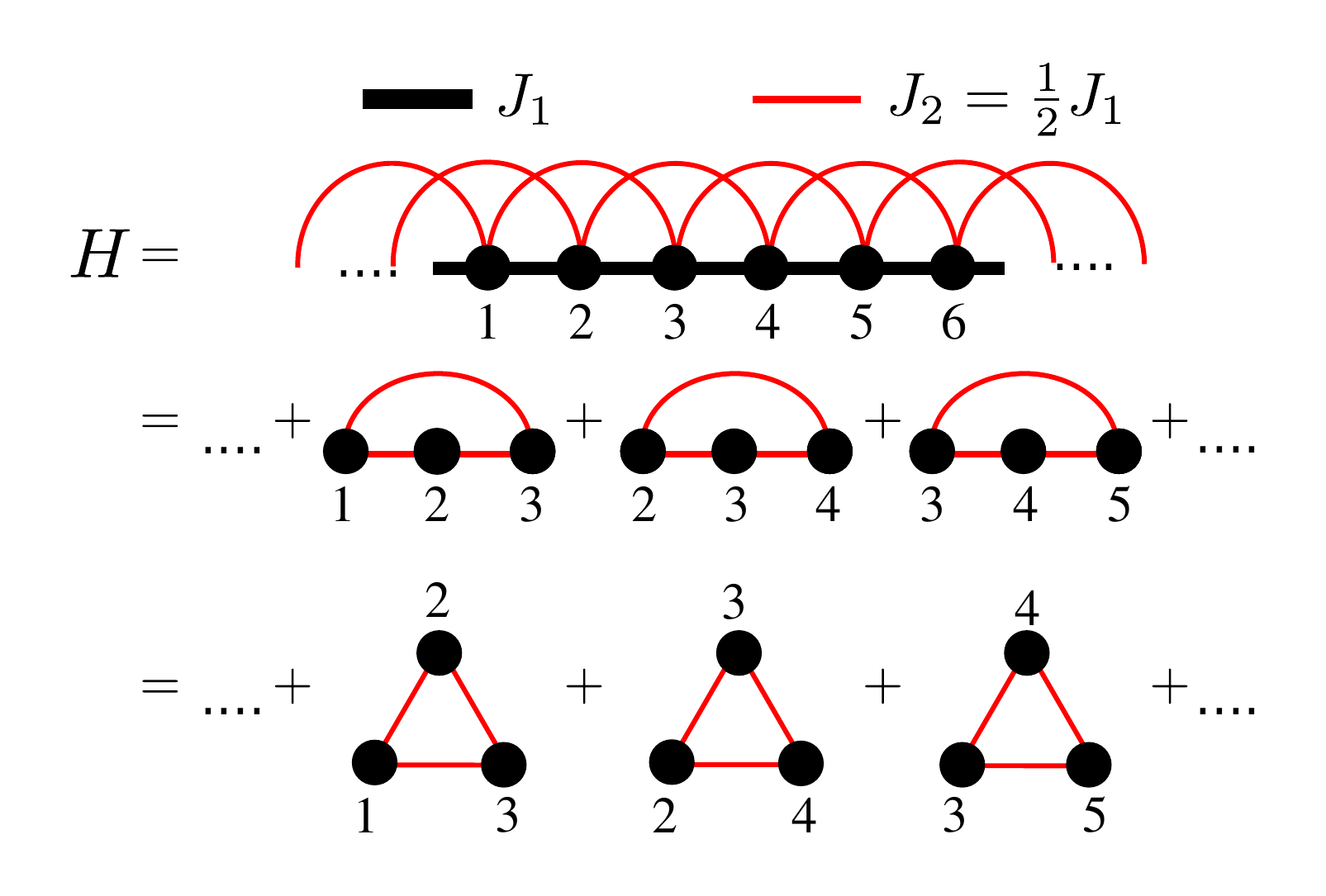}
\caption{Decomposition of the Hamiltonian for the Majumdar-Ghosh chain (and its anisotropic generalization), which can be visualized as 
overlapping three site contiguous chunks. Each three site motif has the $XXZ$ Hamiltonian on a triangular motif which, at $J_z=-\frac{1}{2}$ 
can be three-colored and dimer covered consistently, without creating any conflicts.}
\label{fig:MG_decomp}
\end{figure}

We now study a model which illuminates the competition between 
three-coloring states and other quantum 
ground states.
Our inspiration stems from the Majumdar-Ghosh (MG) model,~\cite{Majumdar_Ghosh}
one of the earliest known exactly solvable models of frustrated $1d$ quantum magnetism. 
The model has nearest neighbor $J_1$ and second neighbor $J_2$ 
isotropic Heisenberg interactions in the ratio $\frac{J_2}{J_1} = \frac{1}{2}$, 
which allows the Hamiltonian to be written as 
$H^{MG} = \frac{1}{2} \sum_{i=1}^{N} \Big( \vec{S}_{i-1} + \vec{S}_{i} + \vec{S}_{i+1} \Big)^2$
up to an innocuous constant for $N$ sites and periodic boundary conditions 
($i+1$ and $i-1$ are taken modulo $N$). 
Each term in this sum corresponds to the square of the total spin of three consecutive sites
schematically shown in Fig.~\ref{fig:MG_decomp}.

For even $N$, all terms can be simultaneously minimized, 
a property of ``frustration free" Hamiltonians, i.e. each three site motif can be brought in a total $S=\frac{1}{2}$ state in two different ways. 
These correspond to the two dimer coverings of the one dimensional chain and 
are referred to as the valence bond solid (VBS) states in the literature. 
Ref.~\onlinecite{Caspers_Emmett_Magnus_1984} rigorously proved that these are the only 
two exact ground states of the MG chain. 

\begin{figure}[]
\includegraphics[width=0.75\linewidth,trim=0 5mm 0 5mm,clip=true]{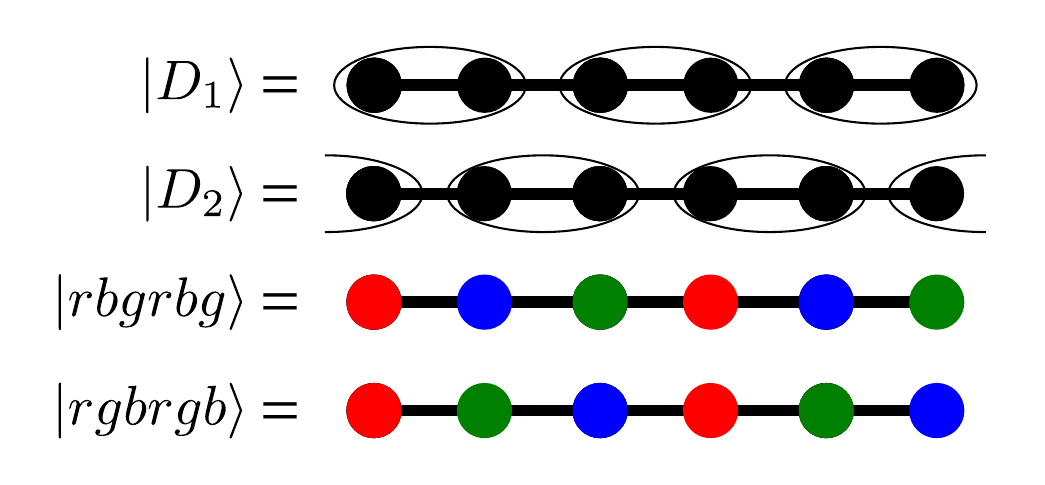}
\caption{Dimer and three-color solutions on a six site motif. These patterns repeat for larger systems. 
For $J_z=-\frac{1}{2}$ all four states are exact ground states, for periodic chains that have sizes which 
are multiples of six. (For exactly six sites, the four solutions are not linearly independent). 
For $|D_2 \rangle$, the sites on the boundary pair up into a dimer for periodic boundary conditions.}
\label{fig:MG_GS}
\end{figure}

We generalize the MG Hamiltonian to anisotropic interactions, 
\begin{align}
	H^{MG}_{XXZ} =& \:
J_1 \sum_{i=1}^{N} \left(  S^{x}_i S^{x}_{i+1} + S^{y}_i S^{y}_{i+1} + J_z \: S^{z}_i S^{z}_{i+1} \right) + \nonumber \\
& J_2 \sum_{i=1}^{N}  \left( S^{x}_i S^{x}_{i+2} + S^{y}_i S^{y}_{i+2}  + J_z \: S^{z}_i S^{z}_{i+2} \right)
\end{align}
with $\frac{J_2}{J_1} = \frac{1}{2}$ which we fix for the rest of this discussion, and $J_z$ is a
dimensionless parameter in this section. 
For this ratio of $\frac{J_2}{J_1} = \frac{1}{2}$, the entire Hamiltonian still remains
a sum of triangular pieces each of which has the $3c$ form for $J_z=-\frac{1}{2}$; this decomposition 
has been schematically depicted in Fig.~\ref{fig:MG_decomp}. 
The ground state of this Hamiltonian is thus locally a three-coloring state on each triangular piece.
As long as each of these three-site motifs can be 
three-colored consistently, without creating any ``color conflicts" 
(no neighboring sites have the same color, and each contiguous three site motif has three distinct colors), 
the resulting wavefunction is an exact ground state of the anisotropic MG chain. 
For chain lengths that are multiples of three, 
this can be done in precisely two ways - $rbgrbg...$ and $rgbrgb...$, 
as is shown in Fig. \ref{fig:MG_GS}. 
%Note that for chain lengths which are multiples of three, 
%any consecutive three sites have exactly one $r$, one $b$ and one $g$ degree of freedom.
For chain lengths that are also even, i.e. multiples of six, 
we may project the two three-colorings to the $S_z=0$ sector and, as mentioned earlier, 
this projection still preserves the property that it is an eigenstate. 

Additionally, the set of two dimer coverings are \emph{also}
exact ground states at the $3c$ point of the anisotropic MG chain. 
This is because on any three-site triangular motif, the
two linearly independent three-colorings (schematically $|rgb\rangle$ and $|rbg\rangle$)
may be linearly combined and then projected to $S_z=\pm \frac{1}{2}$
to make a dimer or valence bond and a free spin-$\frac{1}{2}$. Indeed, this is the
situation at the familiar $J_z=1$ MG point as well.
Requiring all three-site triangular motifs to have a dimer and a free spin-$\frac{1}{2}$
yields the two dimer covering states. 

For a six-site chain, the proposed set of four solutions (two three-colorings and two dimer coverings) 
are not linearly independent. We establish this with an explicit enumeration of the amplitudes of three-coloring 
and dimer wavefunctions for 
all 20 Ising configurations in the $S_z=0$ sector (see Table~\ref{tab:MG_six_sites} in App.~\ref{app:tables}). 
We obtain the relation,
\begin{equation}
	|D_1 \rangle + | D_2 \rangle = \frac{\sqrt{20}}{\omega^2 - \omega} P_{S_z=0} (|rgbrgb \rangle - |rbgrbg \rangle)	
\end{equation}
where $|D_1\rangle$ and $|D_2\rangle$ are depicted in Fig.~\ref{fig:MG_GS}, and in our notation, $P_{S_z=0} |... \rangle$ 
corresponds to a coloring wavefunction that has been projected and normalized. 
In defining our sign convention for the dimer solutions, 
we have used that any local dimer of sites $i$ and $i+1$ (modulo N) is 
$\frac{1}{\sqrt{2}} \Big( |\uparrow \rangle_{i} \otimes |\downarrow \rangle_{i+1} - |\downarrow \rangle_{i} \otimes |\uparrow \rangle_{i+1} \Big)$.
For chains that are higher multiples of six, there is no such linear dependence between
the four states. On larger system sizes $N = 12, 18, 24, 30$, 
we find the number of solutions to be four or greater. We have empirically observed the precise number 
to be $(\frac{N}{6} + 2)$ but do not have an explanation for the extra solutions.

\begin{figure}
\includegraphics[width=0.9\linewidth]{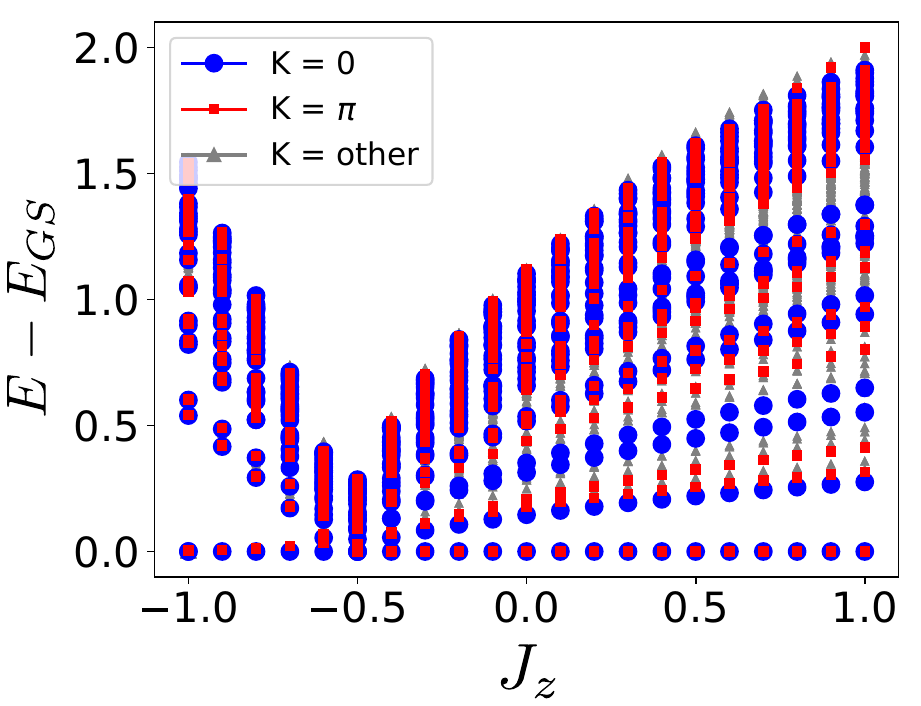}
\caption{Energy spectrum of the 18 site periodic anisotropic Majumdar-Ghosh chain as a function of $J_z$ ($J_{\perp}=1$) in the $S_z = 0$ sector. For a given $J_z$, 
only the lowest 30 energies in each momentum sector are plotted. The ground state energy is subtracted out 
at each value of $J_z$. There is a phase transition at $J_z=-\frac{1}{2}$. For $J_z > -\frac{1}{2}$ the ground state is exactly two fold degenerate (one state in $k=0$ and the other in $k=\pi$), 
which correspond to linear combinations of the two Majumdar-Ghosh dimer (valence bond) solutions. 
These ground states persist all the way to the Heisenberg point ($J_z=1$) and beyond (not shown), 
consistent with the analytic arguments.}
\label{fig:spectrum_MG_18}
\end{figure}

We now address the case of $J_z \geq -\frac{1}{2}$. We 
rewrite the anisotropic MG Hamiltonian (up to a constant) as
\begin{eqnarray}
	H^{MG}_{XXZ} &=&      H^{MG}_{3c} + \frac{(J_z+\frac{1}{2})}{2} \sum_{i=1}^{N} (S_{i-1}^z + S_{i}^z + S_{i+1}^z)^2 \nonumber \\
		   &\equiv& H^{MG}_{3c} + H_{ZZ}
\end{eqnarray}
As the second term involves the square of the sum over only the $S^z$ components, 
therefore for $(J_z + \frac{1}{2}) \geq 0$ this term is 
minimized for any state that satisfies 
$(S_{i-1}^z + S_{i}^z + S_{i+1}^z)^2 |\psi\rangle = (\frac{1}{2})^2 |\psi \rangle$ 
for any three consecutive sites $\{i-1,i,i+1\}$. 
While $H^{MG}_{3c}$ and $H_{ZZ}$ do not commute, 
any wavefunction that simultaneously minimizes their individual 
contributions is guaranteed to be a ground state of the anisotropic MG model. 
This condition is indeed achieved by the dimer VBS states since,
as discussed earlier, they respect the condition that any three-site triangular
motif is composed of a dimer and a free spin-$\frac{1}{2}$. 
Thus, they are indeed the lowest energy eigenstates of $H^{MG}_{3c}$ and $H_{ZZ}$
simultaneously and therefore of $H^{MG}_{XXZ}$.
This analytic result is confirmed with exact diagonalization, 
and demonstrated for the representative example of the 18 site periodic chain 
in Fig.~\ref{fig:spectrum_MG_18}. While the dimer solutions break translational 
invariance, appropriate linear combinations of them restore this symmetry, these linear combinations appear 
in exact diagonalization (with momentum symmetry). We observe two exactly degenerate states, 
one with momentum $k=0$ and the other $k=\pi$ that are selected from the degenerate manifold at $J_{z}=-\frac{1}{2}$, 
and stay degenerate for all $J_z>-\frac{1}{2}$, gapped out from the rest of the spectrum. 

The three-coloring states (projected or unprojected) possess LRO, and in accordance with the 
Mermin-Wagner theorem are not allowed to be the ground state of a Hamiltonian with continuous symmetry in one dimension. 
However, at precisely the $J_z=-\frac{1}{2}$ point which is a critical point in parameter space, 
both the short range 
ordered solutions (dimer VBS) and the three-coloring states coexist.~\cite{chertkov2018computational} This leads us to 
conclude that the presence of competing states at the solvable point 
can strongly influence the stability of the coloring ground state, and in this particular case, 
they immediately lose out to the VBS ground states for any $J_z>-\frac{1}{2}$.

%===============================================================

\section{Conclusion}

In this work, we have reported a ground state solvable point
$H_{2c}$ in the $XXZ$ phase diagram of lattice translationally invariant
bipartite quantum magnets in any magnetization sector.
The associated $U(1)$ symmetric
$XY$ N\'eel order in the zero magnetization sector 
is numerically demonstrated to be adiabatically connected
to the $SU(2)$ symmetric N\'eel order
at the Heisenberg point. This is unified with a similar
thread in the tripartite triangular lattice with $120^\circ$ AFM order
and associated solvable point $H_{3c}$ with finite number of three-colorings. 
For the case of the $\frac{m}{m_s} = \frac{1}{3}$
sector on the triangular lattice, we found that the umbrella state at $H_{3c}$ 
extends up to $J_z \sim 0.75$, after which the magnetization 
plateau UUD state is obtained. 
We also studied the anistropic generalization of the MG chain and 
found it to be ground state solvable.
Both long-range ordered colorings and valence bond ordered states coexist at the
$3c$ point,
while the latter  are the only ground states on moving towards the
$SU(2)$-symmetric point and beyond. This offers an interesting contrast to the previous results that we 
presented on magnetic LRO.

It is also interesting to ask whether 
the existence of the $2c$ point offers a natural explanation for the 
numerically observed existence of  LRO on diluted unfrustrated AFM
at their percolation threshold.~\cite{Yu_Quenched,Wang_Sandvik,Sandvik_Quenched,Todo_quenched,Changlani_Quenched,Ghosh_Quenched} 
This problem has seen several conflicting opinions, 
owing to the possible smallness of the order parameter (the staggered magnetization).
Parts of the system become dimer covered with dominant VBS correlations, 
and hence magnetically inert, yet LRO tenuously survives 
on such fractal clusters. LRO at the $2c$ point on such bipartite clusters is obviously
guaranteed (Eq.~\ref{eq:mat_elem_SxSx}), and one would anticipate that it
adiabatically persists to the Heisenberg point, but this remains to be firmly established. 

In comparison to the ordered cases presented here, that involved finite number of
colorings of the lattice, the highly-frustrated Kagome lattice
harbors a macroscopic degeneracy due to an exponential
number of three-colorings.~\cite{Changlani_PRL}
While it is not clear which state is stabilized as one moves towards the
Heisenberg point,
there is evidence of adiabaticity of the Heisenberg point
to $H_{3c}$ from DMRG computations.~\cite{He_2015,LauchliMoessner,changlani2019resonating}
Evidence for adiabaticity to the $XY$ point ($J_z=0$) 
was also observed previously in the context of 
chiral spin liquid on the Kagome lattice in the 
$m/m_s = 2/3$ magnetization sector.~\cite{KumarChanglani2016}
These findings suggest a unifying picture of the ground state
behavior in $XXZ$ models. A natural question to ask then
is what happens to the excited states and the associated dynamics
of coloring states on tuning the anisotropy. (The question of non-equilibrium
dynamics in the vicinity of the $3c$ point on the Kagome, 
as a function of anisotropy has been addressed recently~\cite{Lee_PRB_2020,Lee_arxiv_2020}).
Finally, for completeness, we note that our numerical evidence for 
adiabaticity from solvable points towards the isotropic regime
strictly applies to finite size systems,
and rigorously showing this
in the thermodynamic limit at the level of a mathematical theorem
is an open problem.

%===============================================================

\section{Acknowledgements}
We acknowledge useful discussions with Johannes Richter, Nandini Trivedi,
Subhro Bhattacharjee, Arnab Sen and Yasir Iqbal. SP acknowledges the support (17IRCCSG011) of IRCC, IIT Bombay and
SERB, DST, India (SRG/2019/001419).
HJC and PS acknowledge funds from Florida State University and the National High Magnetic Field Laboratory.
The National High Magnetic Field Laboratory is supported by the National Science Foundation through NSF/DMR-1644779 and the state of Florida.
We also thank the Research Computing Cluster (RCC) and the Planck cluster at Florida State University for computing resources.

%===============================================================

\bibliographystyle{apsrev4-1}
\bibliography{Notes}

\begin{thebibliography}{56}
\bibitem [{pro()}]{projector}%
  \BibitemOpen
  \href@noop {} {}\bibinfo {note} {For any operator $O$ that commutes with the
  Hamiltonian $[H,O]=0$, if $|\psi\rangle$ is an eigenstate, $H |\psi\rangle =
  E |\psi\rangle$, then it also follows that $H (O|\psi\rangle) = E
  (O|\psi\rangle)$ and thus $O|\psi\rangle$ is an eigenstate as well. In our
  case, $O$ is the projection operator $P_{S_z}$. The opposite however does not
  follow, i.e. $H (P_{S_z}|\psi\rangle) = E ({P_{S_z}}|\psi\rangle)$ does not
  imply $H |\psi\rangle = E |\psi\rangle$.}\BibitemShut {Stop}%
\end{thebibliography}

%===============================================================

\appendix

\begin{widetext}
\section{Two-point ground state correlators for $H_{2c}$}
\label{app:2c_mat_elem}

Here we calculate matrix elements  
for the unique two-coloring state $|C_{S^z=0}\rangle$ on any bipartite lattice
with equal number of A and B sublattice sites. 
We start with the overlap $\langle C_{S^z=0} |C_{S^z=0} \rangle$ as in Eq.~\ref{eq:mat_elem_def} 
to highlight the basic algebraic manipulations that will used throughout in these calculations.
We recall that
$|c_j\rangle$ can be either $|r\rangle \equiv \frac{1}{\sqrt{2}}(|\uparrow\rangle+|\downarrow\rangle)$ 
or $|b\rangle \equiv \frac{1}{\sqrt{2}}(|\uparrow\rangle-|\downarrow\rangle)$, and
$|s_j\rangle$ are the Ising states $|\uparrow\rangle$ or $|\downarrow\rangle$ on site $j$.
Terms of the form $\langle c_j|s_j\rangle$ follow from these definitions.
Taking into account the overall normalization of $\frac{2^N}{^N C_{N/2}}$ in the $S^*_z=0$ sector
as discussed in the main text, we have, 
%the overlap of the $|C_{S^z=0}\rangle$ is
\begin{eqnarray}
\langle C_{S^z=0}|C_{S^z=0}\rangle &=& \mathcal{N}\sum_p\prod_j \frac{1}{2}(e^{ip/2}+ e^{-ip/2}) \hspace*{3cm} (\text{where}~~ \mathcal{N}=\frac{1}{N+1}\frac{2^N}{^N C_{\frac{N}{2}}})\nonumber \\
&=& \frac{\mathcal{N}}{2^N}\sum_{k=0}^N(e^{ik\theta}+e^{-ik\theta})^N  \hspace*{2cm}
(\text{where} ~~p=\frac{2\pi k}{N+1}=2k\theta ~~\text{and}~~ k~\text{ is integer})\nonumber\\
&=& \frac{\mathcal{N}}{2^N}\sum_{k=0}^N\sum_{m=0}^{N}  {^NC_m}  (e^{ik\theta})^m(e^{-ik\theta})^{N-m} \hspace*{2cm}(\text{using the binomial expansion})\nonumber\\
&=& \frac{\mathcal{N}}{2^N}\sum_{m=0}^{N}  {^NC_m} (N+1)\delta_{N/2-m,0}\nonumber\\
&=& \mathcal{N}\frac{N+1}{2^N} {^NC_{\frac{N}{2}}}=1~.
	\label{eq:2c_overlap}
\end{eqnarray}
as expected.

Analogous to Eq.~\ref{eq:mat_elem_def}, the general expression for diagonal correlation function 
in the zero magnetization sector is
\begin{eqnarray}
&&\langle C_{S^z=0}|S^z_mS^z_n|C_{S^z=0}\rangle \label{zz-corr} \\
	&=&\mathcal{N}\sum_p \left[ \left( \sum_{s_m,s'_m,s_n,s'_n} 
	e^{ips_m}\langle c_m|s_m\rangle \langle s_m|S^z_m|s'_m\rangle \langle s'_m|c_m\rangle 
	\; e^{ips_n}\langle c_n|s_n\rangle \langle s_n|S^z_n|s'_n\rangle \langle s'_n|c_n\rangle \right)
	\left( \prod_{j\neq\{m,n\}}\sum_{s_j} e^{ips_j}  \langle c_j|s_j\rangle \langle s_j|c_j\rangle 
\right) \right] \nonumber .  
\end{eqnarray} 
Perfoming the $s_m,s'_m,s_n,s'_n,s_j$ sums, we get
\begin{eqnarray}
\langle C_{S^z=0}|S^z_mS^z_n|C_{S^z=0}\rangle 
&=& \mathcal{N}\sum_p \left[ \frac{1}{4} (e^{ip/2}-e^{-ip/2}) \frac{1}{4} (e^{ip/2}-e^{-ip/2})
	\prod_{j \neq \{m,n\}} \frac{1}{2}(e^{ip/2}+e^{-ip/2})\right] \nonumber \\
&=& \mathcal{N}\sum_p \left[ \frac{1}{4}\frac{(e^{ip/2}-e^{-ip/2})(e^{ip/2}-e^{-ip/2})}{(e^{ip/2}+e^{-ip/2})(e^{ip/2}+e^{-ip/2})}\prod_j\frac{1}{2}(e^{ip/2}+e^{-ip/2})\right] \nonumber \\
&=& -\frac{\mathcal{N}}{2^{N+2}}\sum_p \left[(e^{ip/2}+e^{-ip/2})^{N-2}-(e^{ip/2}+e^{-ip/2})^N \right]\nonumber \\
&=& -\frac{\mathcal{N}(N+1)}{2^{N+2}}[ 4\times ^{N-2}C_{\frac{N}{2}-1}-^NC_{\frac{N}{2}}]\nonumber \\
&=& -\frac{1}{4}\frac{1}{N-1}
\label{diagonal_corr}
\end{eqnarray}
where we use similar manipulations as in Eq.~\ref{eq:2c_overlap}. 
Similarly, the general expression for off-diagonal correlation function
in the zero magnetization sector is
\begin{eqnarray}
&&\langle C_{S^z=0}|S^\pm_mS^\mp_n|C_{S^z=0}\rangle \label{pm-corr} \\
	&=&\mathcal{N}\sum_p \left[ \left( \sum_{s_m,s'_m,s_n,s'_n} e^{ips_m}\langle c_m|s_m\rangle \langle s_m|S^\pm_m | s'_m\rangle \langle s'_m|c_m\rangle \; e^{ips_n}\langle c_n|s_n\rangle \langle s_n|S^\mp_n | s'_n\rangle \langle s'_n|c_n\rangle  \right) \left( \prod_{j \neq \{m,n\}} \sum_{s_j} e^{i p s_j}\langle c_j|s_j\rangle \langle s_j|c_j\rangle \right) \right]
	~. \nonumber
\end{eqnarray}
Again, perfoming the $s_m,s'_m,s_n,s'_n,s_j$ sums, we get
\begin{eqnarray}
\langle C_{S^z=0}|S^\pm_mS^\mp_n|C_{S^z=0}\rangle 
	&=& \mathcal{N}\sum_p \left[\frac{\epsilon_{mn} e^{\pm ip/2}e^{\mp ip/2}}{(e^{ip/2}+e^{-ip/2})(e^{ip/2}+e^{-ip/2})} \prod_j\frac{1}{2}(e^{ip/2}+e^{-ip/2}) \right] \nonumber \\
	&=& \epsilon_{mn} \frac{\mathcal{N}}{2^N}\sum_p (e^{ip/2}+e^{-ip/2})^{N-2}\nonumber \\
	&=& \frac{\epsilon_{mn}}{4}\frac{N}{N-1}~.
\label{ODE}
\end{eqnarray}
where we use similar manipulations as in Eq.~\ref{eq:2c_overlap},
and $\epsilon_{mn}=-1$ in Eq. \ref{ODE} when 
$\{ m,n \}$ belongs to sites with different colors, while $\epsilon_{mn}=1$ 
when sites $\{ m,n \}$ have the same color. 
Following Eq. \ref{ODE}, it is straightforward to get the form of Eq. \ref{eq:mat_elem_SxSx} in the main text.

%=========================================================
\section{Details of the gaplessness argument}
\label{app:gapless}
To show that the unprojected two-coloring state $|C\rangle$ is a gapless ground state
of $H_{2c}$,
we consider the following state $|C'\rangle$ built by modulating the
two-coloring of $|C\rangle$ as mentioned in the main text:
\begin{eqnarray}
|C'\rangle & \equiv &  \prod_{i \in A} \otimes_i e^{i \hat{S}^z_i \delta_i} 
|r\rangle_i \prod_{j \in B} \otimes_j 
e^{i \hat{S}^z_j \delta_j} |b\rangle_j \nonumber \\
&=& \prod_{i \in A} \left( \cos\left(\frac{\delta_i}{2}\right)|r\rangle 
+ i\sin\left(\frac{\delta_i}{2}\right)|b\rangle \right) \prod_{j \in B} \left( \cos\left(\frac{\delta_j}{2}\right)|b\rangle 
+ i\sin\left(\frac{\delta_j}{2}\right)|r\rangle \right)\nonumber \\
&=& \left( \prod_{i} \cos\left(\frac{\delta_i}{2}\right) \right) |C\rangle + \ldots \nonumber \\
& \equiv & \sqrt{\epsilon} |C\rangle + \ldots
\label{eq:ep_def}
\end{eqnarray}
where $\epsilon \equiv \prod_{i} \cos^2 \left(\frac{\delta_i}{2}\right)$,
and $\delta_i$ are to be small numbers $\rightarrow 0$.
Both $|C\rangle$ and $|C'\rangle$ are clearly normalized.
Now, for the variational excited state, we will consider a state $|\psi\rangle$ as
that part of $|C'\rangle$ which does not contain any component along
$|C\rangle$, i.e. $\langle \psi|C\rangle=0$. This is simply achieved by
\begin{eqnarray}
|\psi\rangle \equiv |C'\rangle - 
%\left(\prod_{i}  \cos\left(\frac{\delta_i}{2}\right) \right) 
\sqrt{\epsilon} |C\rangle.
\end{eqnarray}
This state has to be renormalized to respect normalization, i.e, presently
\begin{eqnarray}
\langle\psi|\psi\rangle 
& = & \langle C'|C'\rangle + 
%\prod_{i} \left(\cos\left(\frac{\delta_i}{2}\right)\right)^2
\epsilon \langle C|C\rangle 
- 
%\prod_{i} \cos\left(\frac{\delta_i}{2}\right) 
\sqrt{\epsilon} \left( \langle C'|C\rangle + \langle C|C'\rangle \right) \nonumber \\
&=& 1- \epsilon %\prod_{i} \left(\cos\left(\frac{\delta_i}{2}\right)\right)^2
\end{eqnarray}
In the above, we simply used $\langle C'|C\rangle= \langle C|C'\rangle=
\sqrt{\epsilon}$ as defined in Eq.~\ref{eq:ep_def}.
%$\equiv \prod_{i} \cos\left(\frac{\delta_i}{2}\right)$).
Now in the following, we establish %will aim to establish
a variational upper bound for the excitation gap
using $|\psi\rangle$ which being orthogonal $|C\rangle$ is a legitimate variational
excited state. The energy in the properly normalized state will be
\begin{eqnarray}
\frac{\langle\psi|H_{2c}|\psi\rangle}{\langle \psi | \psi \rangle}
&= & \frac{\langle C'|H_{2c}|C'\rangle + 
%\prod_{i} \left(\cos\left(\frac{\delta_i}{2}\right)\right)^2
\epsilon \langle C|H_{2c}|C\rangle - 
%\prod_{i} \cos\left(\frac{\delta_i}{2}\right) 
\sqrt{\epsilon} \left( \langle C'|H_{2c}|C\rangle 
+ \langle C|H_{2c}|C'\rangle \right)}{1 - \epsilon} \nonumber \\
%&=& \frac{\langle C'|H_{2c}|C'\rangle - \prod_{i} 
%\left(\cos\left(\frac{\delta_i}{2}\right)\right)^2\langle C|H_{2c}|C\rangle }{1-\prod_{i} \left(\cos\left(\frac{\delta_i}{2}\right)\right)^2} \nonumber\\
&=& \frac{\langle C'|H_{2c}|C'\rangle - \epsilon\langle C|H_{2c}|C\rangle }{1-\epsilon}
\end{eqnarray}
where we make use of the fact that $|C\rangle$ is the (ground) eigenstate
of $H_{2c}$, i.e. $H_{2c}|C\rangle = \langle C | H_{2c} | C \rangle |C\rangle$, and thereby
$\langle C|H_{2c}|C'\rangle=\langle C'|H_{2c}|C\rangle = 
%\prod_{i} \cos\left(\frac{\delta_i}{2}\right)
\sqrt{\epsilon} \langle C|H_{2c}|C\rangle$.
%and $\epsilon=\prod_{i} \left(\cos\left(\frac{\delta_i}{2}\right)\right)^2$ 
%(very small quantity).
Therefore, the variational estimate of the excitation energy is
\begin{equation}
\Delta E \equiv \frac{\langle\psi|H_{2c}|\psi\rangle}{\langle \psi | \psi \rangle} - \langle C|H_{2c}|C\rangle
= \frac{\langle C'|H_{2c}|C'\rangle - \langle C|H_{2c}|C\rangle }{1-\epsilon}
\label{eq:var_est}
\end{equation}
%===========================================
\par
We are primarily interested in the $N$ dependence or scaling of $\Delta E$ in the
arguments below.
For the numerator of $\Delta E$ in Eq.~\ref{eq:var_est}, for a single bond, the states on the
sites that are part of the bond are relevant, and therefore we have for the bond $\langle i,j \rangle$
(with $i \in A\text{ sublattice and }j \in B\text{ sublattice}$ without loss of generality):
\begin{eqnarray}
\langle C'|H_{ij}|C'\rangle - \langle C|H_{ij}|C\rangle 
&=& \left(e^{-i \hat{S}^z_i \delta_i} \langle r |_i \otimes e^{-i \hat{S}^z_j \delta_j} \langle b |_j \right) 
 H_{ij} 
\left(e^{i \hat{S}^z_i \delta_i} | r \rangle_i \otimes e^{i \hat{S}^z_j \delta_j} | b \rangle_j \right)
-
\left(\langle r |_i \otimes \langle b |_j \right) 
H_{ij}
\left( | r \rangle_i \otimes | b \rangle_j \right)
 \nonumber \\
&=& \left[-\frac{1}{4}\left(\cos\delta_i\cos\delta_j
+\sin\delta_i\sin\delta_j\right)+\frac{1}{4} \right] =
\frac{1-\cos(\delta_i-\delta_j)}{4} \nonumber \\
&\simeq & \frac{(\delta_i-\delta_j)^2}{8}
\end{eqnarray}
since $\delta_i \rightarrow 0 \;\forall \;i$, and it is understood
that $j$ in the above expressions are the nearest neighbor sites in the
unit cell to which $i$ belongs.

As described in the main text, let us choose the following modulation:
$\delta_i=\delta \sin(\mathbf{q}.\mathbf{r}_i)$ with 
%$\mathbf{q}=(\frac{2\pi}{N^{1/d}},0,0)$,
%$\mathbf{q}=(\frac{2\pi}{L_x},0,0)$ 
$\mathbf{q}=\frac{2\pi}{L_x} \hat{x}$ 
$\rightarrow \mathbf{0}$ as $L_x \rightarrow \infty$.
Let's recall that $L_x, L_y, \ldots$ are the linear dimensions, and the
number of sites $N = \prod_i L_i$ in $d$ dimensions. %$N=L_x \times L_y$.
We sum over all the bonds along the $x$-axis (since in other
directions, $\delta_i - \delta_j = 0$ identically in our choice
of modulation) to get
\begin{eqnarray}
\langle C'|H_{2c}|C'\rangle - \langle C|H_{2c}|C\rangle 
&=& \sum^N_{i=1}  \frac{\delta^2}{8} \left[\sin\left(\mathbf{q} \cdot \mathbf{r}_i\right)
-\sin\left(\mathbf{q} \cdot \left(\mathbf{r}_i+ \hat{x}\right)\right)\right]^2 \nonumber \\
&=& \sum^N_{i=1} \frac{\delta^2}{2} \left[ \sin \left(\mathbf{q} \cdot \frac{\hat{x}}{2} \right) 
\cos \left(\mathbf{q} \cdot \left(\mathbf{r}_i+\frac{ \hat{x}}{2}\right)\right)\right]^2 \nonumber \\
%& \simeq &  \frac{\delta^2 \pi^2 }{2 N^{2/d}}
& \simeq &  \frac{\delta^2 \pi^2 }{2 L^2_x}
\left[ \sum^N_{i=1} \cos^2\left(\mathbf{q} \cdot \left(\mathbf{r}_i+\frac{\hat{x}}{2}\right)\right) \right]
\end{eqnarray}
by using small angle approximation as 
%$\mathbf{q} \cdot \hat{x} \sim N^{-1/d}$.
$\mathbf{q} \cdot \hat{x} = 2\pi/L_x \sim N^{-1/d}$.
We also have
$\sum^N_{i=1} \cos^2(\mathbf{q}.(\mathbf{r}_i+\frac{\hat{x}}{2}))=\frac{1}{4}~^2C_{1}~N\sim N$
by using very similar steps for the power of cosine sums as in previous
sections.
Therefore, the numerator in Eq.~\ref{eq:var_est} for $\Delta E$ scales as
\begin{eqnarray}
\langle C'|H_{2c}|C'\rangle - \langle C|H_{2c}|C\rangle \sim  \delta^2 N^{1-2/d} 
\end{eqnarray}

Another way to see the above scaling is
by choosing $i$,$j$ such that $\mathbf{r}_i=0$, i.e.
$\delta_i=0$ and 
%$\delta_j=\delta \frac{2\pi}{N^{1/d}}$.
$\delta_j=\delta \frac{2\pi}{L_x}$.
For this choice,
one obtains the maximum value of $(\delta_i-\delta_j)$ over all bonds
(simply because for $f(x)=\sin x$, the variation or slope around $x=0$ is maximum).
This gives an upper bound for $\langle C'|H_{2c}|C'\rangle - \langle C|H_{2c}|C\rangle$
which leads to the same scaling as before, i.e. $N \; \text{max}[(\delta_i - \delta_j)^2]
\sim \delta^2 N^{1 - 2/d}$.

If $\delta$ scales as $\delta \sim N^{\alpha}$,
then the numerator of $\Delta E$ in Eq.~\ref{eq:var_est} scales as
\begin{equation}
\langle C'|H_{2c}|C'\rangle - \langle C|H_{2c}|C\rangle \sim N^{1+2\alpha-\frac{2}{d}},
\label{eq:numerator_scaling}
\end{equation}
which will $\rightarrow 0$ (as is the goal of this appendix) if $\alpha<0$ (for $d=2$).
This is consistent with our initial assumption above that the modulations are small,
i.e. $\delta_i \ll 1 \; \forall \; i$.
However, to complete the argument, it
remains to analyse the scaling of the denominator of Eq.~\ref{eq:var_est}
as well to make sure that $\Delta E$ indeed scales to zero.
We note here that the denominator $1-\epsilon$ is 
%essentially 
directly related to the
overlap of $|C\rangle$ and $|C'\rangle$. Going ahead,
\begin{eqnarray}
\; \; \; \; \epsilon & = &  \prod^N_{i=1} \cos^2 \left( \frac{\delta_i}{2} \right) \simeq 
\prod^N_{i=1}  \left( 1 - \frac{\delta^2_i}{4} \right) \nonumber \\
\implies & \log (\epsilon) \simeq & \sum^N_{i=1} \log \left( 1-\frac{\delta_i^2}{4} \right) 
\simeq \sum_i \left( -\frac{\delta_i^2}{4} \right)\notag \\
\implies & \epsilon \; \; \simeq & \exp \left[ -\sum^N_{i=1} \frac{\delta_i^2}{4} \right]
%=e^{-\sum^N_{i=1} \delta^2 \sin^2(\frac{2\pi i}{N^{1/d}})}
=\exp \left[-\frac{\delta^2}{4}\sum^N_{i=1} \sin^2(\mathbf{q}\cdot \mathbf{r}_i) \right]
\end{eqnarray}
%where $\delta_i=\delta\sin(\frac{2\pi i}{N^{1/d}})$. 
Now we again use a power of sines sum identity to arrive at
%\begin{eqnarray}
%\sum_k \left(\sin(\frac{2\pi k}{2(N+1)})\right)^M
%=\frac{N+1}{2^M}~ ^MC_{M/2}
%\end{eqnarray}
%This gives
%\begin{eqnarray}
%$\sum^N_{i=1} \sin^2(\frac{2\pi i}{N^{1/d}})
$\sum^N_{i=1} \sin^2(\mathbf{q}\cdot \mathbf{r}_i)
=\frac{N}{2^2}~^2C_1 \sim N$. 
%\end{eqnarray}
Therefore for $\delta\sim N^\alpha$, $\epsilon$ behaves as
\begin{eqnarray}
\epsilon \sim e^{-N.N^{2\alpha}}=e^{-N^{1+2\alpha}}
\label{eq:overlap_scale}
\end{eqnarray}
In order to ensure gaplessness, i.e. $\Delta E \rightarrow 0$ as $N \rightarrow \infty$,
we need to ensure that the denominator $1 - \epsilon$ remain finite and $\emph{not}$
scale to zero simultaneously. Given Eq.~\ref{eq:overlap_scale},
this is clearly ensured by
%\begin{eqnarray}
$1+2\alpha >0 \Rightarrow -1/2<\alpha$
%\end{eqnarray}
Thus, we have arrived at the desired scaling choice for $\delta$ such that
the variational estimate for the excitation energy $\Delta E$ scales to zero
when 
\begin{equation}
-1/2<\alpha< 0
\end{equation}
which implies gaplessness for the spectrum at the solvable point $H_{2c}$
as is to be expected for a $U(1)$-symmetry broken N\'eel state. This completes
our proof.

%Finally we note in passing 
Finally, it is instructive to consider
how the above gaplessness argument fails when $\alpha$ is not in the desired
range stated above. E.g. when $\alpha$ is below the range, say $\alpha=-1$,
then the numerator of $\Delta E$ indeed still scales to zero as desired,
however the denominator now \emph{also} scales to zero! This tells us that
the modulation magnitude can not be too small either on a \emph{finite} lattice,
otherwise the overlap does not scale to zero fast enough to make the
gaplessness argument work, inspite of the naive expectation that $\epsilon$
is simply the product of $N$ factors each being
less than one (of the form $\cos^2(\delta_i/2$)).
On the other side, when
$\alpha$ is above the range, say $\alpha=0$, then the denominator
does scale to a finite value (one) as desired, but now the numerator
of $\Delta E$ \emph{does not} scale to zero thus again 
%spoiling 
invalidating the gaplessness argument.

%=========================================================

\section{Two-point ground state correlators for the $m=0$ sector of $H_{3c}$}
	\label{app:3c_zeromag_mat_elem}

In this section, we calculate matrix elements for triangular lattice where 
the coloring ground states is two-fold degenerate (see Sec.~\ref{sec:triangle_zeromag}). 
We recall that $|c_j\rangle$ on site $j$ can be 
$|r\rangle \equiv \frac{1}{\sqrt{2}}(|\uparrow\rangle+|\downarrow\rangle)$,
$|b\rangle \equiv \frac{1}{\sqrt{2}}(|\uparrow\rangle+\omega|\downarrow\rangle)$
or $|g\rangle \equiv \frac{1}{\sqrt{2}}(|\uparrow\rangle+\omega^2|\downarrow\rangle)$
corresponding to the colors on the three sublattices of the triangular lattice.
$\omega=e^{i 2\pi/3}$ and $\omega^2=\omega^*$ are the cube roots of unity.
Therefore, if we associate integers 0, 1, 2 to $c_j$ for $|r\rangle,|b\rangle,|g\rangle$ respectively, 
it follows that  
$\langle c_j|s_j\rangle=\omega^{(c_j-2c_js_j)}/\sqrt{2}$. 
Taking into account the overall normalization of $\frac{2^N}{^N C_{N/2}}$ in the $S^*_z=0$ sector, 
the overlap in general can be written as
\begin{eqnarray}
\langle C_{S^z=0}|C_{S^z=0}'\rangle = \mathcal{N}\sum_p\prod_j \frac{1}{2}(e^{ip/2}+e^{i2\pi \lambda_j/3} e^{-ip/2})
\end{eqnarray}
	where $\lambda_j=(2c_j+c'_j)\text{ (mod 3)}$.
Therefore for the two three-coloring states, we get 
$\langle C^{(1)}_{S^z=0}|C^{(1)}_{S^z=0}\rangle=\langle C^{(2)}_{S^z=0}|C^{(2)}_{S^z=0}\rangle = 1$
	as expected using the very same steps as in Eq.~\ref{eq:2c_overlap}.
	
For the overlap between the two three-coloring states, we have
\begin{eqnarray}
\langle C^{(1)}_{S^z=0}|C^{(2)}_{S^z=0}\rangle = \langle C^{(2)}_{S^z=0}|C^{(1)}_{S^z=0}\rangle 
&=& \mathcal{N}\sum_p \left[\frac{1}{2^3}(e^{ip/2}+ e^{-ip/2})(e^{ip/2}+ \omega e^{-ip/2})(e^{ip/2}+\omega^2 e^{-ip/2}) \right]^{N/3}\nonumber \\
&=& \frac{\mathcal{N}}{2^{N}}\sum_p (e^{i3p/2}+ e^{-i3p/2})^{N/3} \nonumber\\
&=& \frac{^{N/3}C_{N/6}}{^{N}C_{N/2}} 
\end{eqnarray}
This overlap vanishes in the thermodynamic limit, i.e., $N\rightarrow\infty$.

%--------------------------------------------------------------------------------

For the spin-spin correlations, we will make use of the following identities: 
\begin{eqnarray}
\sum_{s_m,s'_m} e^{ips_m}\langle c_m|s_m\rangle \langle s_m|S^z_m|s'_m\rangle \langle s'_m|c_m\rangle &=& \frac{1}{4}(e^{ip/2}-e^{i2\pi \lambda_m/3} e^{-ip/2})\nonumber\\
\sum_{s_m,s'_m} e^{ips_m}\langle c_m|s_m\rangle \langle s_m|S^+_m|s'_m\rangle \langle s'_m|c_m\rangle &=& \frac{1}{2}(e^{ip/2}~e^{i2\pi c_m/3} )\\
\sum_{s_m,s'_m} e^{ips_m}\langle c_m|s_m\rangle \langle s_m|S^-_m|s'_m\rangle \langle s'_m|c_m\rangle &=& \frac{1}{2}(e^{-ip/2}~e^{i4\pi c_m/3} )\nonumber
\label{EqnSet}
\end{eqnarray} 
Then, starting from  the analogs of 
Eq.\ref{zz-corr} and Eq.\ref{pm-corr} for the three-coloring case, we have 
\begin{eqnarray}
\langle C^{(l)}_{S^z=0}|S^z_mS^z_n|C^{(l)}_{S^z=0}\rangle &=& \mathcal{N}\sum_{p} \left[ \frac{1}{4}\frac{(e^{ip/2}-e^{-ip/2}e^{i2\pi\lambda_{m}/3})(e^{ip/2}-e^{-ip/2}e^{i2\pi\lambda_{n}/3})}{(e^{ip/2}+e^{-ip/2}e^{i2\pi\lambda_{m}/3})(e^{ip/2}+e^{-ip/2}e^{i2\pi\lambda_{n}/3})}\prod_{j}\frac{1}{2}(e^{ip/2}+e^{-ip/2}e^{i2\pi\lambda_{j}/3}) \right] \nonumber\\
&=& -\frac{\mathcal{N}}{2^{N+2}}\sum_p \left[(e^{ip/2}+e^{-ip/2})^{N-2}-(e^{ip/2}+e^{-ip/2})^N \right] \hspace*{1cm} (\text{for}~ c_j=c'_j~,~\lambda_{m/n/j}=0~\text{mod}~ 3)\nonumber \\
&=& -\frac{\mathcal{N}(N+1)}{2^{N+2}} \left[ 4\times ^{N-2}C_{\frac{N}{2}-1}-^NC_{\frac{N}{2}} \right]\nonumber \\
&=& -\frac{1}{4}\frac{1}{N-1}~,
\label{EqnSzSz1}
\end{eqnarray}

\begin{eqnarray}
\langle C^{(l)}_{S^z=0}|S^+_mS^-_n|C^{(l)}_{S^z=0}\rangle &=& \mathcal{N}\sum_{p} \left[ \frac{(\frac{1}{2}e^{i2\pi c_m/3} e^{ip/2})~(\frac{1}{2}e^{i4\pi c_n/3} e^{-ip/2})}{\frac{1}{2}(e^{ip/2}+e^{-ip/2}e^{i2\pi\lambda_{m}/3})\frac{1}{2}(e^{ip/2}+e^{-ip/2}e^{i2\pi\lambda_{n}/3})}\prod_j \frac{1}{2}(e^{ip/2}+e^{-ip/2}e^{i2\pi\lambda_{j}/3}) \right] \nonumber\\
&=& \frac{\mathcal{N}}{2^N}e^{i\frac{2\pi}{3}(c_m+2c_n)}\sum_{p}(e^{ip/2}+e^{-ip/2})^{N-2} \nonumber \\
	&=&\frac{N}{4(N-1)}e^{i\frac{2\pi}{3}(c_m+2c_n)}
\label{eq:B5}
\end{eqnarray}
and similarly
\begin{eqnarray}
\langle C^{(l)}_{S^z=0}|S^-_mS^+_n|C^{(l)}_{S^z=0}\rangle &=& \mathcal{N}\sum_{p} \left[ \frac{(\frac{1}{2}e^{i4\pi c_m/3} e^{ip/2})~(\frac{1}{2}e^{i2\pi c_n/3} e^{-ip/2})}{\frac{1}{2}(e^{ip/2}+e^{-ip/2}e^{i2\pi\lambda_{m}/3})\frac{1}{2}(e^{ip/2}+e^{-ip/2}e^{i2\pi\lambda_{n}/3})}\prod_j \frac{1}{2}(e^{ip/2}+e^{-ip/2}e^{i2\pi\lambda_{j}/3}) \right] \nonumber\\
&=&\frac{\mathcal{N}}{2^N}e^{i\frac{2\pi}{3}(2c_m+c_n)}\sum_{p}(e^{ip/2}+e^{-ip/2})^{N-2} \nonumber \\
&=& \frac{N}{4(N-1)}e^{i\frac{2\pi}{3}(2c_m+c_n)}
\label{eq:B6}
\end{eqnarray}
where $l\in (1,2)$.  
Since we made the choice $(c_r,c_b,c_g)=(0,1,2)$ above, 
thus for sites $\{m, n\}$ that have different colors,
we obtain
\begin{eqnarray}
\langle C^{(l)}_{S^z=0} | S_r^xS_b^x+S_r^yS_b^y| C^{(l)}_{S^z=0} \rangle &=& \frac{N}{8(N-1)}\omega(1+\omega)=-\frac{N}{8(N-1)}\nonumber \\
\langle C^{(l)}_{S^z=0} | S_r^xS_g^x+S_r^yS_g^y| C^{(l)}_{S^z=0} \rangle &=& \frac{N}{8(N-1)}\omega^2(1+\omega^2)=-\frac{N}{8(N-1)}\nonumber \\
\langle C^{(l)}_{S^z=0} | S_b^xS_g^x+S_b^yS_g^y| C^{(l)}_{S^z=0} \rangle &=& \frac{N}{8(N-1)}1.(\omega+\omega^2)=-\frac{N}{8(N-1)}~.
\end{eqnarray}
In the above equations, we have used the identity $1+\omega+\omega^2=0$.  
$U(1)$ symmetry implies $\langle C^{(l)}_{S^z=0} | S_m^xS_n^x| C^{(l)}_{S^z=0} \rangle = \langle C^{(l)}_{S^z=0} | S_m^yS_n^y| C^{(l)}_{S^z=0} \rangle$, and therefore
$\langle C^{(l)}_{S^z=0} | S_m^xS_n^x| C^{(l)}_{S^z=0} \rangle = \langle C^{(l)}_{S^z=0} | S_m^yS_n^y| C^{(l)}_{S^z=0} \rangle =-\frac{1}{2}(\frac{1}{8}\frac{N}{N-1})$
for sites $\{m,n\}$ that have different colors.
For sites $\{m, n\}$ that have the same color, putting 
$c_m+2c_n=2c_m+c_n=0 \mod 3$ in Eqs.~\ref{eq:B5} and \ref{eq:B6},
we obtain
$\langle C^{(l)}_{S^z=0} | S_m^xS_n^x| C^{(l)}_{S^z=0} \rangle = \langle C^{(l)}_{S^z=0} | S_m^yS_n^y| C^{(l)}_{S^z=0} \rangle =\frac{1}{8}\frac{N}{N-1}$.
In general, we may write
\begin{eqnarray}
\langle C^{(l)}_{S^z=0} | S_m^xS_n^x| C^{(l)}_{S^z=0} \rangle &=& \langle C^{(l)}_{S^z=0} | S_m^yS_n^y| C^{(l)}_{S^z=0} \rangle =\epsilon_{mn}\frac{1}{8}\frac{N}{N-1} 
\end{eqnarray}
where $\epsilon_{mn}=1$ and $-1/2$ for sites $\{m, n\}$ 
with same and different colors respectively. 
As $\cos(120^\circ)=\cos(240^\circ)=-1/2$ and $\cos(0^\circ)=1$, this is often
called as $120^\circ$ or three sub-lattice order (in the $XY$ plane).

%================================================================

\section{Two-point ground state correlators for the $\frac{m}{m_s}=\frac{1}{3}$ sector of $H_{3c}$}
	\label{app:3c_onebythree_mat_elem}

For the $\frac{m}{m_s}=\frac{1}{3}$ sector, the calculation steps are 
similar to the $m=0$ sector shown in the previous section
with the only difference being $S_z^*=0$ gets replaced by $S_z^*=N/6$ 
and the overall normalization factor thus becomes 
$\frac{1}{N+1} \frac{2^N}{^N C_{2N/3}}=\bar{ \mathcal{N}}$. Therefore,
\begin{eqnarray}
\langle C^{(1)}_{S^z=N/6}|C^{(1)}_{S^z=N/6}\rangle=\langle C^{(2)}_{S^z=N/6}|C^{(2)}_{S^z=N/6}\rangle 
&=& \frac{\bar{\mathcal{N}}}{2^N}\sum_p (e^{ip/2}+ e^{-ip/2})^N e^{-ipN/6}\nonumber \\
&=& \frac{\bar{\mathcal{N}}}{2^N}\sum_{k=0}^N\sum_{m=0}^{N}  {^NC_m}  (e^{ik\theta})^m(e^{-ik\theta})^{N-m} e^{-ikN\theta/3}~~
(\text{where} ~~p=\frac{2\pi k}{N+1}=2k\theta) 
\nonumber\\
&=& \frac{\bar{\mathcal{N}}}{2^N}\sum_{m=0}^{N}  {^NC_m} (N+1)\delta_{2N/3-m,0}\nonumber\\
&=& \frac{\bar{\mathcal{N}}}{2^N}~(N+1)~~ ^NC_{\frac{2N}{3}}=1~. 
\label{EqnCC_1/3}
\end{eqnarray}
Similarly, we have
\begin{eqnarray}
\langle C^{(1)}_{S^z=N/6}|C^{(2)}_{S^z=N/6}\rangle = \langle C^{(2)}_{S^z=N/6}|C^{(1)}_{S^z=N/6}\rangle 
&=& \bar{\mathcal{N}}\sum_p \left[ \frac{1}{2^3}(e^{ip/2}+ e^{-ip/2})(e^{ip/2}+ \omega e^{-ip/2})(e^{ip/2}+\omega^2 e^{-ip/2}) \right]^{N/3}e^{-ipN/6}\nonumber \\
&=& \frac{\bar{\mathcal{N}}}{2^{N}}\sum_p (e^{i3p/2}+ e^{-i3p/2})^{N/3}e^{-ipN/6} \nonumber\\
&=& \frac{^{N/3}C_{2N/9}}{^{N}C_{2N/3}} 
\label{EqnCC'_1/3}
\end{eqnarray}
In the thermodynamic limit, the right hand side of Eq. \ref{EqnCC'_1/3} vanishes and the two 
three-coloring states become orthogonal to each other similar to the $m=0$ sector.
The expression for diagonal correlation function in this sector is
\begin{eqnarray}
\langle C^{(l)}_{S^z=N/6}|S^z_mS^z_n|C^{(l)}_{S^z=N/6}\rangle 
&=& -\frac{\bar{\mathcal{N}}}{2^{N+2}}\sum_p \left[(e^{ip/2}+e^{-ip/2})^{N-2}-(e^{ip/2}+e^{-ip/2})^N \right]~e^{-ipN/6} \nonumber \\
&=& -\frac{\bar{\mathcal{N}}(N+1)}{2^{N+2}} \left[ 4\times ^{N-2}C_{\frac{2N}{3}-1}-^NC_{\frac{2N}{3}} \right]\nonumber \\
&=& -\frac{1}{4} \left[\frac{8}{9} \frac{N}{N-1}-1 \right]~,
\label{EqnSzSz_1/3}
\end{eqnarray}
whereas the off-diagonal correlation function has the form
\begin{eqnarray}
\langle C^{(l)}_{S^z=N/6}|S^+_mS^-_n|C^{(l)}_{S^z=N/6}\rangle &=& \frac{\bar{\mathcal{N}}}{2^N}e^{i\frac{2\pi}{3}(c_m+2c_n)}\sum_{p}(e^{ip/2}+e^{-ip/2})^{N-2}~e^{-ipN/6}=\frac{2N}{9(N-1)}e^{i\frac{2\pi}{3}(c_m+2c_n)}
\label{off1/3_pm}
\end{eqnarray}
and 
\begin{eqnarray}
\langle C^{(l)}_{S^z=N/6}|S^-_mS^+_n|C^{(l)}_{S^z=N/6}\rangle &=&\frac{\bar{\mathcal{N}}}{2^N}e^{i\frac{2\pi}{3}(2c_m+c_n)}\sum_{p}(e^{ip/2}+e^{-ip/2})^{N-2}~e^{-ipN/6}=\frac{2N}{9(N-1)}e^{i\frac{2\pi}{3}(2c_m+c_n)}~.
\label{off1/3_mp}
\end{eqnarray}
Combining Eq. \ref{off1/3_pm} with Eq. \ref{off1/3_mp} and following the same steps as for the $m=0$ sector, we have 
\begin{eqnarray}
\langle C^{(l)}_{S^z=N/6} | S_m^xS_n^x| C^{(l)}_{S^z=N/6} \rangle &=& \langle C^{(l)}_{S^z=N/6} | S_m^yS_n^y| C^{(l)}_{S^z=N/6} \rangle = \epsilon_{mn}\frac{1}{9}\frac{N}{N-1} 
\end{eqnarray}
with $\epsilon_{mn}$ as defined in the previous section.

%===================================================================

\section{Ground State structure factors for $H_{2c}$ and $H_{3c}$}
\label{app:exact_struc_fac}

Here we compute the exact structure factors 
of the two-coloring and two three-coloring states for the square and 
triangular lattice in the zero magnetization sector respectively. 
The calculations follow directly
from the exact expressions of real space correlation functions derived
previously: 
a) $\langle C_{S^z=0} | S^z_mS^z_n| C_{S^z=0} \rangle = -\frac{1}{4}\frac{1}{N-1} \text{ for } m\neq n$
and $0.25 \text{ for } m=n$,
and b)
$\langle C_{S^z=0} | S^x_m S^x_n| C_{S^z=0} \rangle = 
\langle C_{S^z=0} | S^y_m S^y_n| C_{S^z=0} \rangle = 
\frac{\epsilon_{mn}}{8}\frac{1}{N-1} \text{ for } m\neq n$ and $0.25 \text{ for } m=n$ 
for all the coloring states with appropriate definitions of $\epsilon_{mn}$ for
the square and triangular cases as noted in Apps.~\ref{app:2c_mat_elem} and \ref{app:3c_zeromag_mat_elem}. 
For both cases, the diagonal structure factor has the form

\begin{eqnarray}
S_{zz}(\textbf{q}) &\equiv& \frac{1}{N}\sum_{m,n}e^{-i\textbf{q}\cdot(\textbf{r}_m-\textbf{r}_n)}\langle C^{(l)}_{S^z=0} | S^z_mS^z_n| C^{(l)}_{S^z=0} \rangle \nonumber \\
&=& \frac{1}{N}\left[0.25N - \frac{1}{4}\frac{1}{N-1}\sum_{m\neq n}e^{-i\textbf{q}\cdot(\textbf{r}_m-\textbf{r}_n)} \right] \nonumber\\
&=& 0.25-\frac{1}{4}\frac{1}{N(N-1)}\left[\sum_{m,n}e^{-i\textbf{q}\cdot(\textbf{r}_m-\textbf{r}_n)}-N\right] \nonumber\\
&=& 0.25-\frac{1}{4(N-1)}\left[N\delta_{\textbf{q},0}-1 \right]
\end{eqnarray}
and therefore $S_{zz}(\textbf{q})=0$ for Brillouin zone center and $S_{zz}(\textbf{q})=0.25+\frac{1}{4(N-1)}$ for other points.
The off-diagonal structure factors has the form
\begin{eqnarray}
S_{xy}(\textbf{q}) &=& \frac{1}{N}\sum_{m,n}e^{-i\textbf{q}\cdot(\textbf{r}_m-\textbf{r}_n)}\langle C^{(l)}_{S^z=0} | S^x_mS^x_n| C^{(l)}_{S^z=0} \rangle \nonumber \\
&=& \frac{1}{N}\left[0.25N + \frac{1}{8}\frac{N}{N-1}\sum_{m\neq n}\epsilon_{mn}~ e^{-i\textbf{q}\cdot(\textbf{r}_m-\textbf{r}_n)}\right] 
\end{eqnarray}

For the $2c$ case, $\epsilon_{mn} = e^{i \mathbf{q}_0 \cdot (\textbf{r}_m-\textbf{r}_n)}$
with $\mathbf{q}_0 = (\pi,\pi)$, 
Therefore,
\begin{eqnarray}
	S_{xy}(\textbf{q}) &=& 0.25+\frac{1}{8}\frac{1}{(N-1)} \left[\sum_{m,n} e^{i (\mathbf{q} - \mathbf{q}_0)
	\cdot (\textbf{r}_m-\textbf{r}_n)} - \sum_{m} 1\right] \nonumber\\
	&=& 0.25+\frac{1}{8}\frac{1}{(N-1)} \left[ N^2 \delta_{\mathbf{q},\mathbf{q}_0} - N\right]
\label{Eqn:S(q)xx_2c}
\end{eqnarray}
For the $3c$ case, 
$\epsilon_{mn} = \frac{e^{i \mathbf{q}_0 \cdot (\textbf{r}_m-\textbf{r}_n)} + e^{-i \mathbf{q}_0 \cdot (\textbf{r}_m-\textbf{r}_n)}}{2}$
with $\mathbf{q}_0 = (\frac{4\pi}{3},0) \text{ or } (-\frac{4\pi}{3},0)$, and therefore
\begin{eqnarray}
	S_{xy}(\textbf{q}) &=& 0.25+\frac{1}{8}\frac{1}{(N-1)} \left[ \frac{N^2}{2} 
	\left( \delta_{\mathbf{q},\mathbf{q}_0} + \delta_{\mathbf{q},-\mathbf{q}_0} \right) - N\right] 
	\label{Eqn:S(q)xx_3c}
\end{eqnarray}
These values are observed in ED and DMRG at the solvable points (Fig.~\ref{fig:square_ED} and 
\ref{fig:triangular_Sz=0}) as expected.

%====================================================================================

\section{Real space spin correlations}
\label{app:real_space_corr}
In the main text, we discussed the evolution of features in the static spin structure factor of the triangular and square 
lattice antiferromagnet as a function of the anisotropy $J_z$, in the zero magnetization $m=0$ ($S_z = 0$) sector. 
Here we present the ground state real space spin correlation functions on the $12 \times 6$ cylinder for the triangular lattice, and $8 \times 8$ cylinder for the square lattice. 
We plot $\frac{1}{2} \langle S^{x}_i S^{x}_c + S^{y}_i S^{y}_c \rangle$ and  $\langle S^{z}_i S^{z}_c \rangle$, with respect to a site $c$ located in the bulk of the cylinder, 
for various representative $J_z$ values. 

In Fig.~\ref{fig:RSC_Sz=0} we discuss our results for the triangular case. At $J_z=-1.0$, the system spontaneously forms two (equal sized) ferromagnetic domains, one with spins pointing in the $z$ direction and the other with spins pointing in the $-z$ direction, consistent with the $S_z=0$ constraint imposed in the DMRG calculation. 
Due to the choice of the cylindrical geometry (length being bigger than the width) the two domains are placed horizontally, to minimize
the energy cost of having a domain wall. The transverse ($XY$ plane) correlations exist only along the domain wall. 

At the exactly solvable point $J_z=-\frac{1}{2}$, the correlation functions are consistent with the exact formulae derived for the projected coloring 
wavefunction. The $\langle S_i^zS_c^z \rangle$ correlator is constant, independent of sublattice. The transverse correlations 
show correlations consistent with $120^\circ$ order, and do not depend on the distance between sites, but only on which sublattice they belong to.

On moving away from the solvable point towards the Heisenberg point i.e. for $J_z>-\frac{1}{2}$, next nearest neighbor ferromagnetic correlations gradually 
begin to develop in the $z$ direction. The in-plane correlations qualitatively resemble the pattern seen at $J_z=-\frac{1}{2}$, 
but the long range order is weakened, as is evidenced from the fall off of the size of the circles (see caption). At the Heisenberg point, 
both patterns evolve to be identical (as they must) owing to the full rotational symmetry of the Hamiltonian at $J_z=1$, 
and given that the ground state is non-degenerate. 
At $J_z=3$, evidence of ordering in both channels is seen, at least on the finite size system 
studied here. 
This is the co-existence of diagonal and off-diagonal ordering, discussed in the main text.

For completeness, we also show the case of the square lattice in Fig.~\ref{fig:RSC_Square_Sz=0}. 
The ordering wavevector of N\'eel order 
is now $(\pi,\pi)$ and the critical points in the $XXZ$ phase diagram are at $J_z=-1$ and $J_z=1$. 

\begin{figure*}[htpb]
  \centering
	\includegraphics[width=\linewidth]{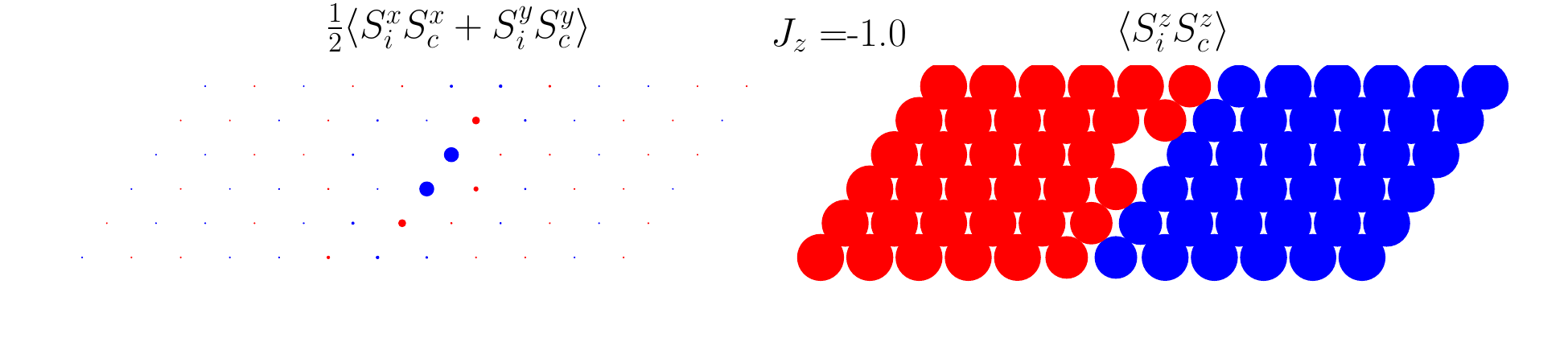}
	\includegraphics[width=\linewidth]{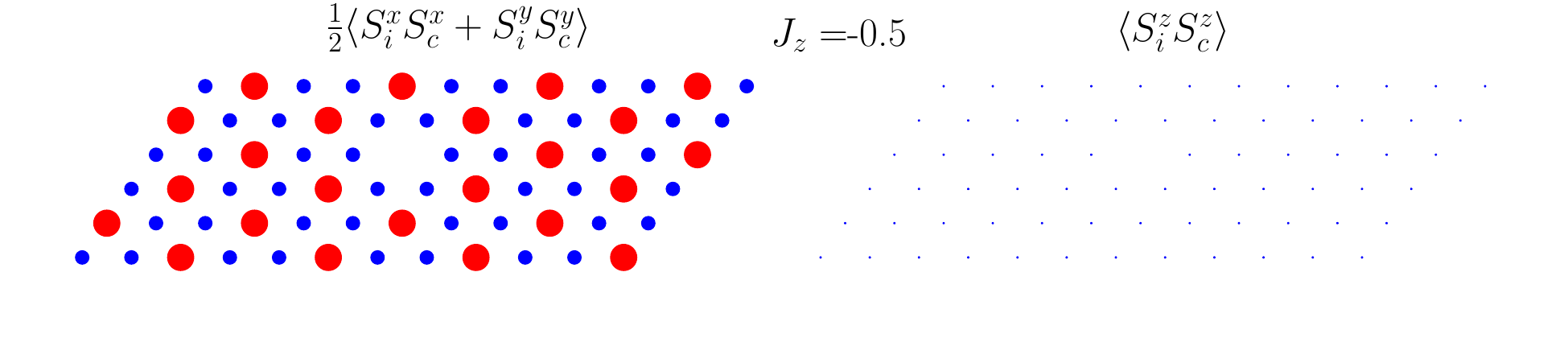}
	\includegraphics[width=\linewidth]{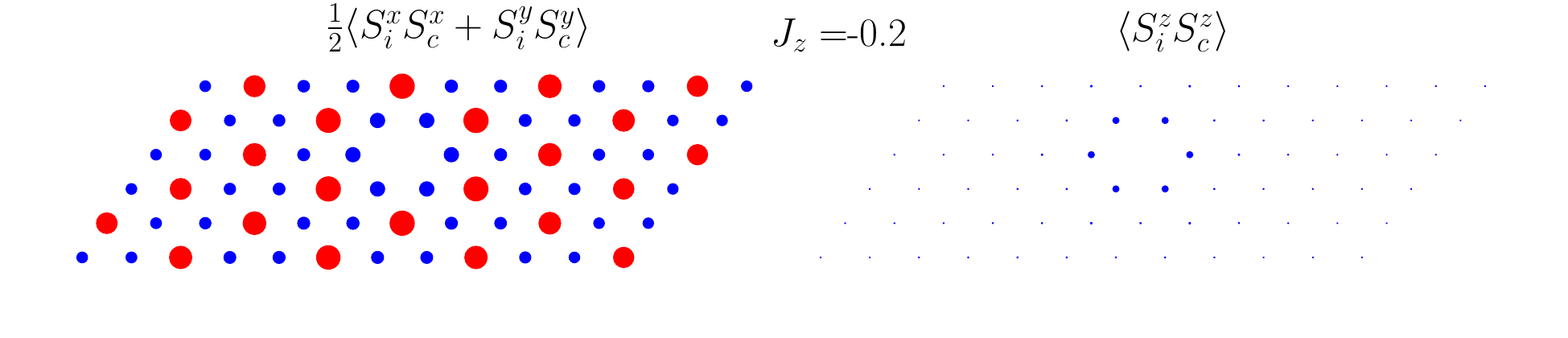}
	\includegraphics[width=\linewidth]{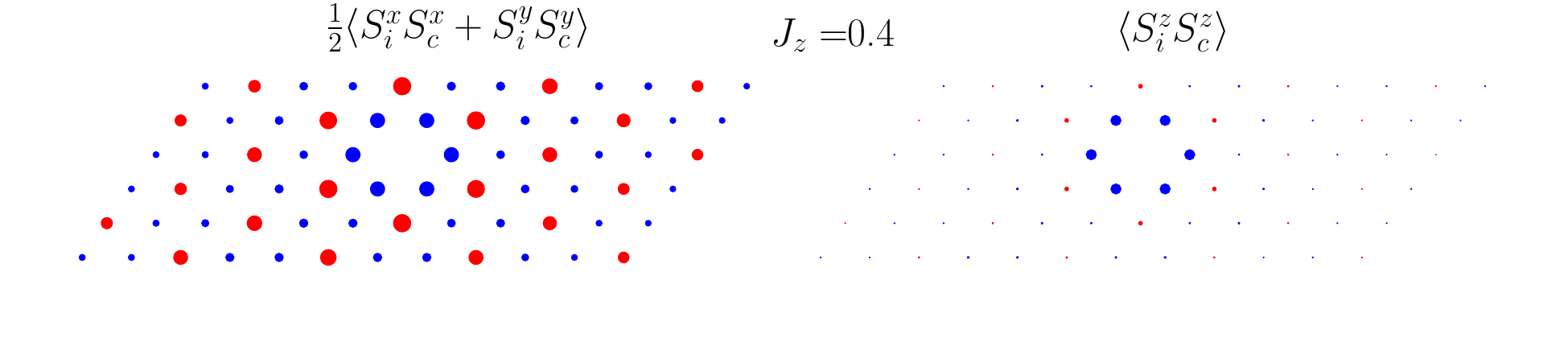}
	\includegraphics[width=\linewidth]{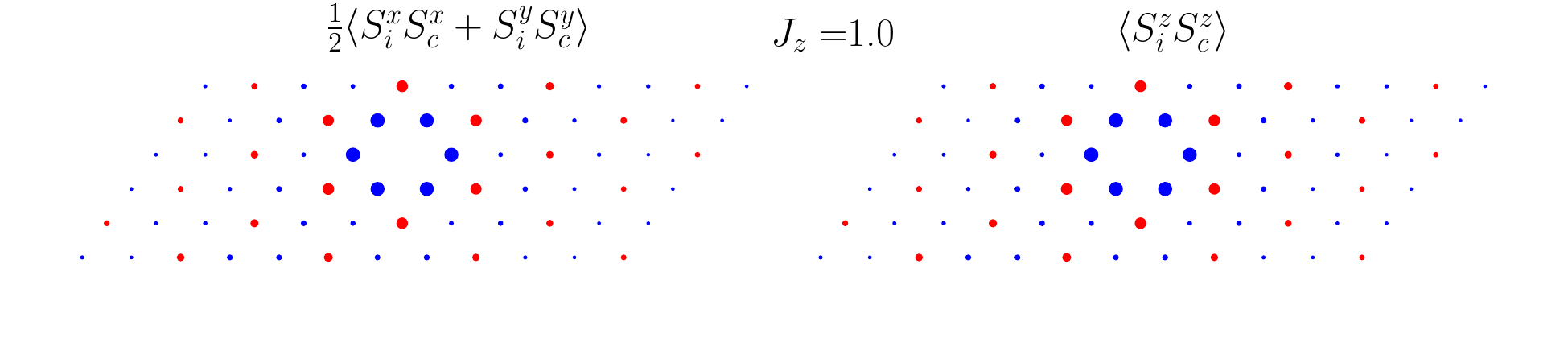}
	\includegraphics[width=\linewidth]{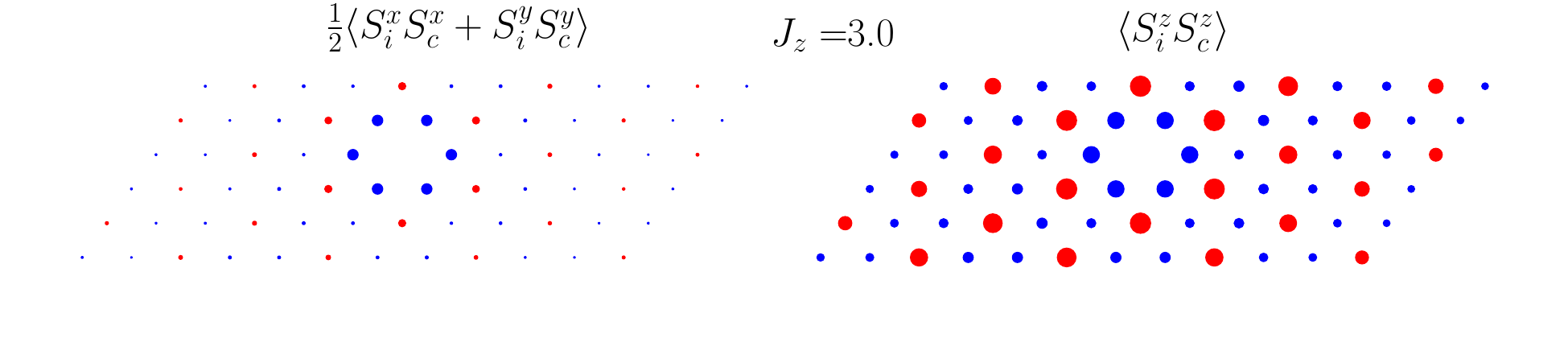}
	\caption{\label{fig:RSC_Sz=0} Real space correlation functions (measured with respect to a central site $c$) $\frac{1}{2} \langle S^{x}_i S^{x}_c + S^{y}_i S^{y}_c \rangle$ and
	$\langle S^{z}_i S^{z}_c \rangle$ for the triangular lattice in the $m=0$ ($S_z=0$) magnetization sector, at various representative $J_z$. The correlation function of the spin at a site with itself is not plotted, and is left empty. The size of the circles indicates the magnitude of the correlator and the color indicates the sign, blue being negative and red being positive.
 }
\end{figure*}

\begin{figure*}[htpb]
  \centering
		\includegraphics[width=0.49\linewidth,trim= 3cm 2cm 3cm 2cm ,clip=true]{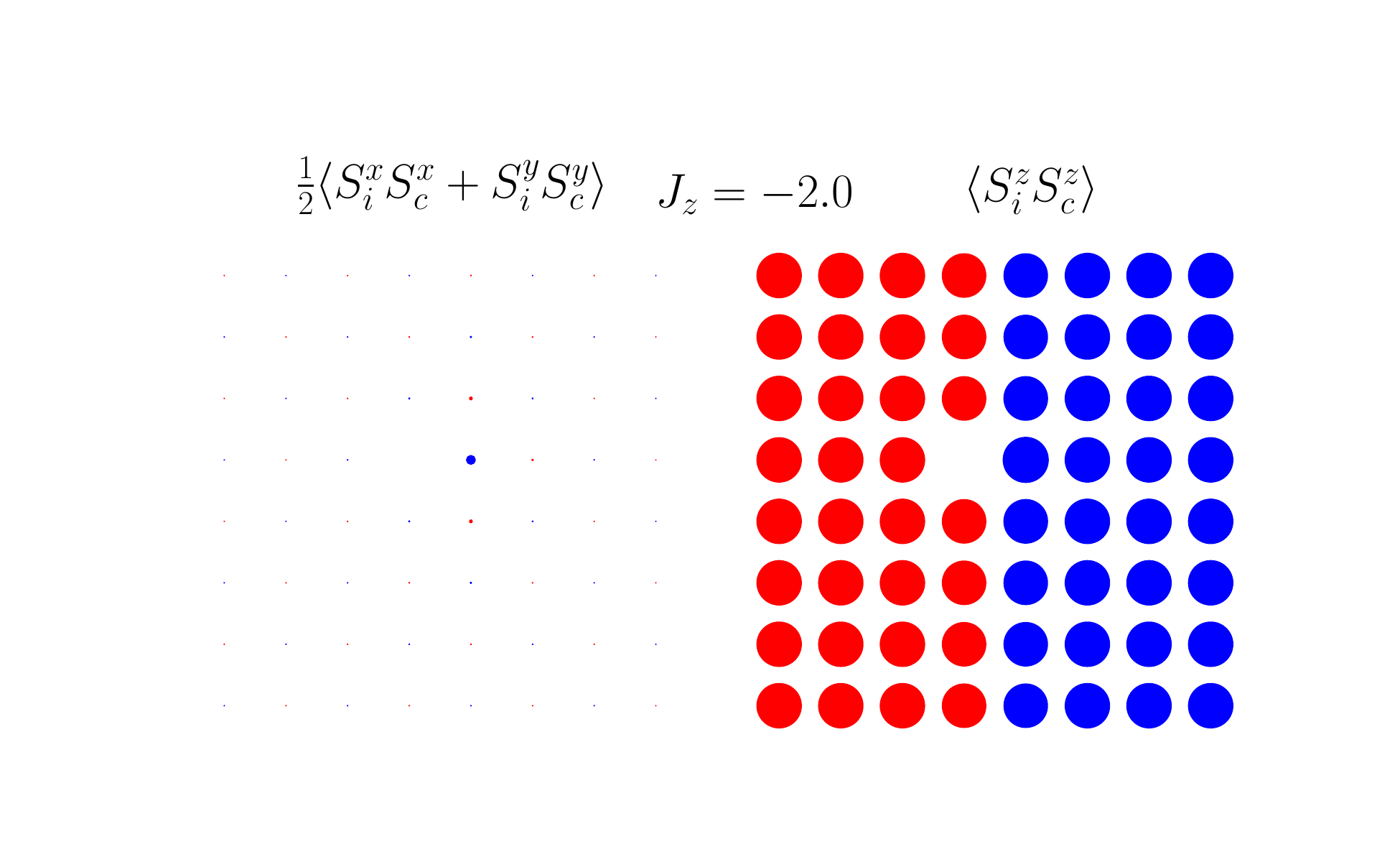}
	\includegraphics[width=0.49\linewidth,trim= 3cm 2cm 3cm 2cm ,clip=true]{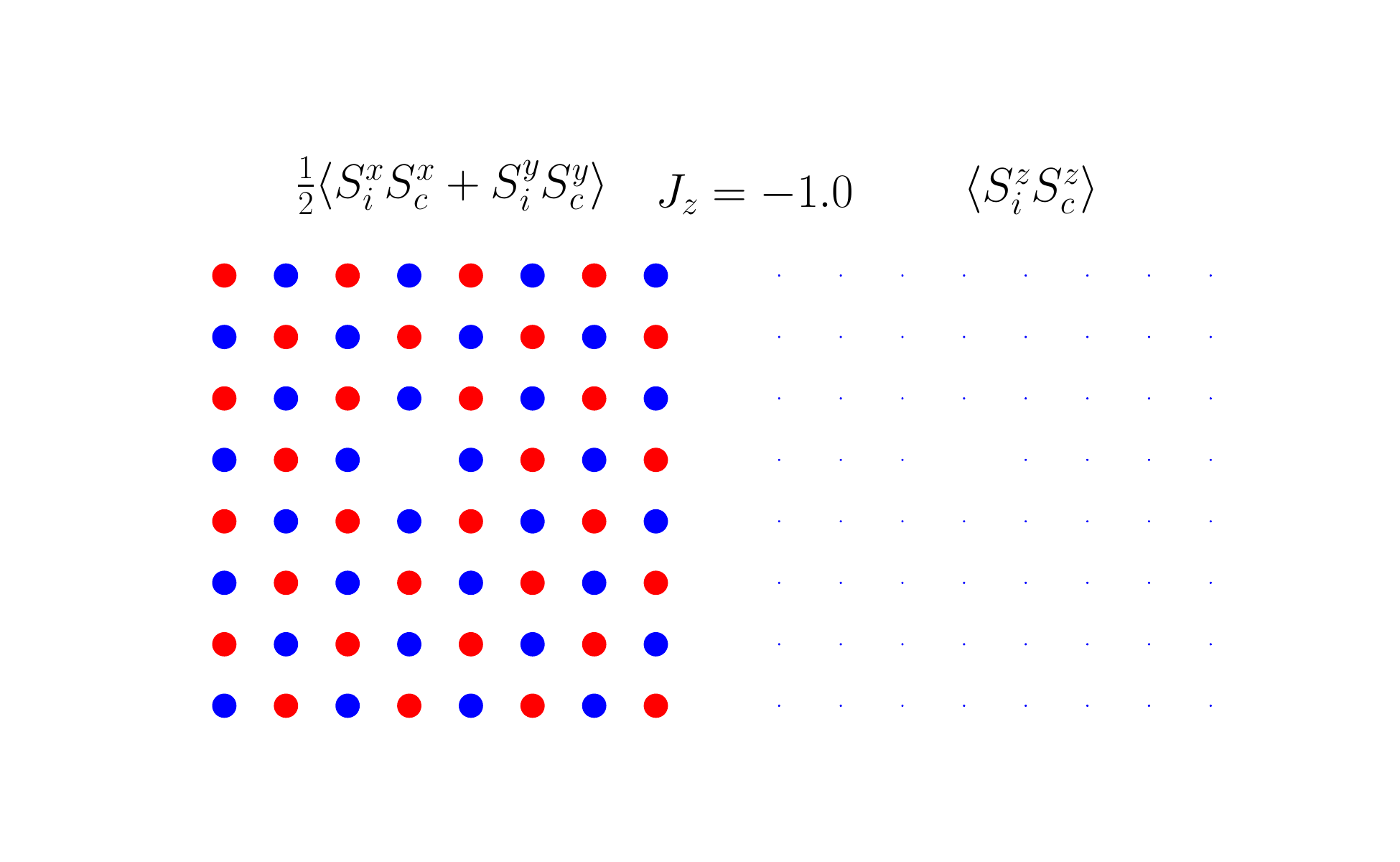}
	\includegraphics[width=0.49\linewidth,trim= 3cm 2cm 3cm 2cm ,clip=true]{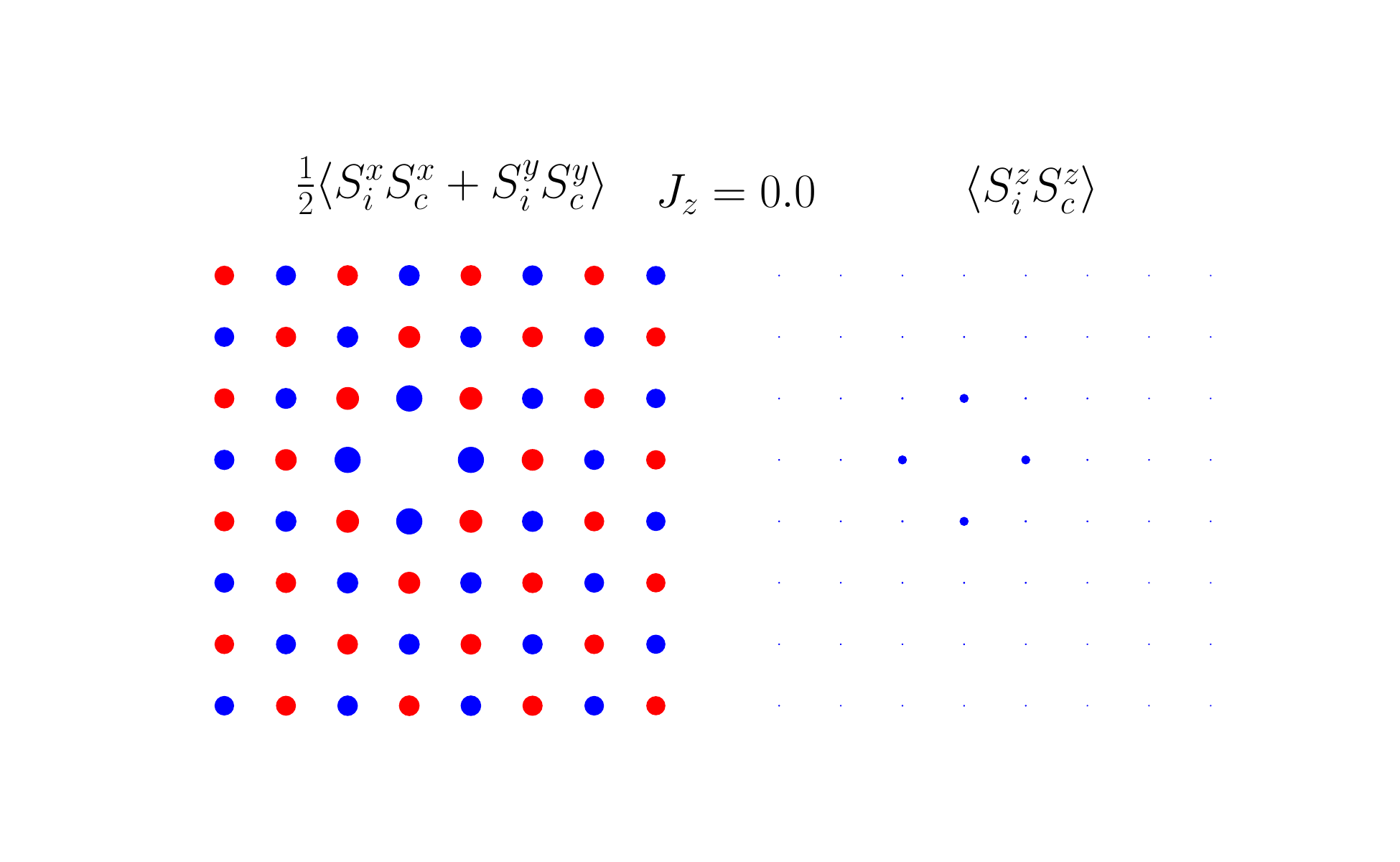}
	\includegraphics[width=0.49\linewidth,trim= 3cm 2cm 3cm 2cm ,clip=true]{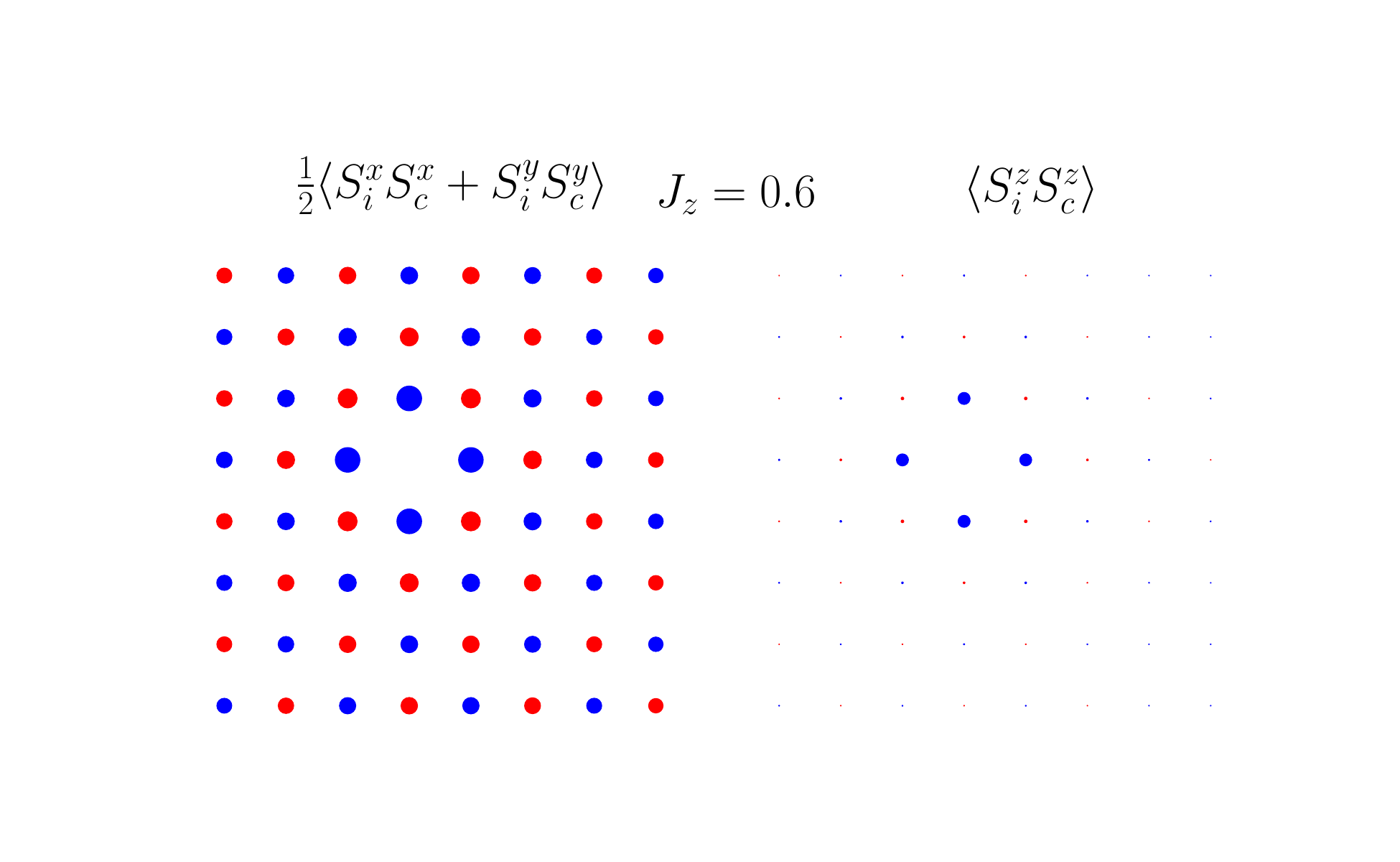}
	\includegraphics[width=0.49\linewidth,trim= 3cm 2cm 3cm 2cm ,clip=true]{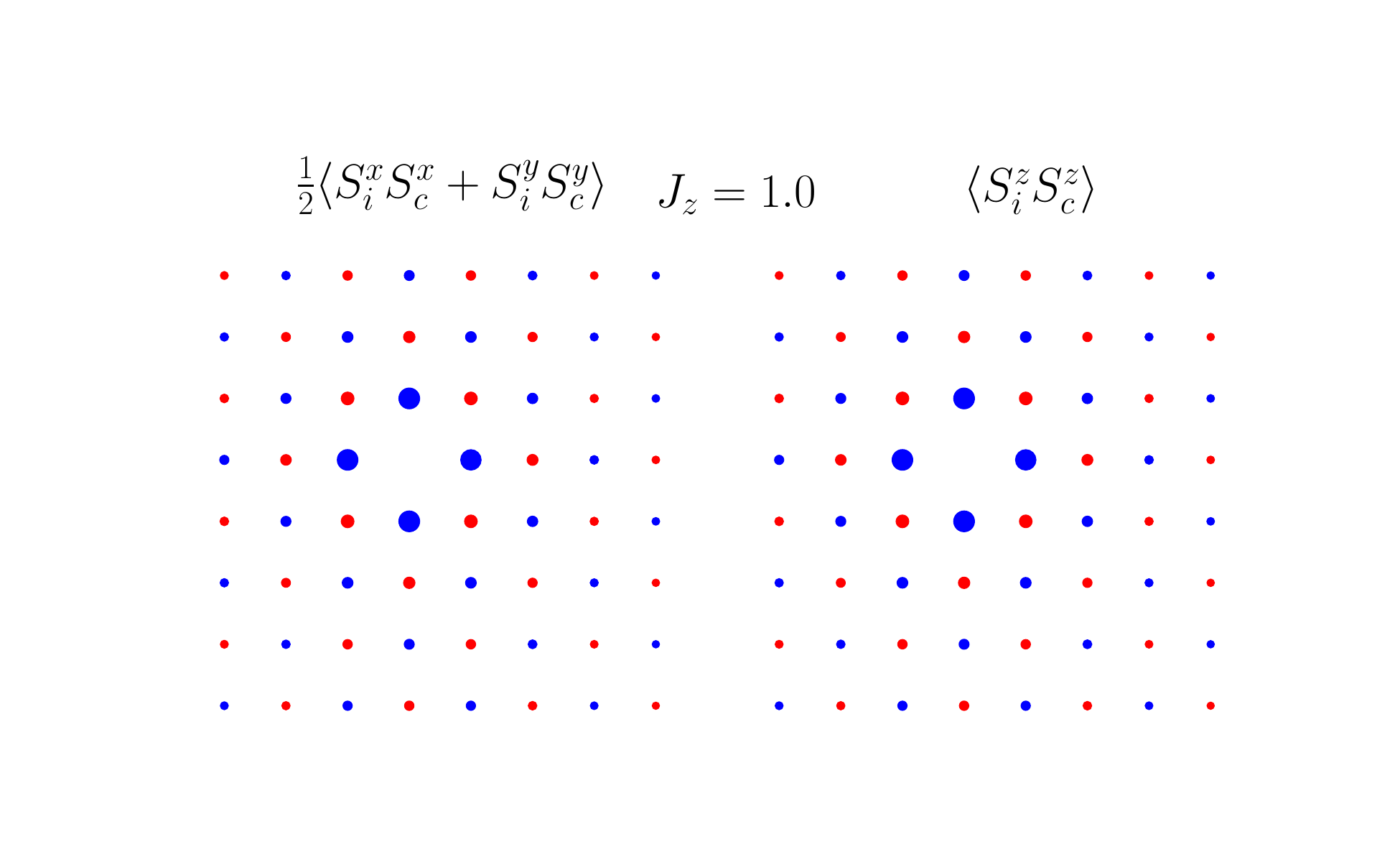}
	\includegraphics[width=0.49\linewidth,trim= 3cm 2cm 3cm 2cm ,clip=true]{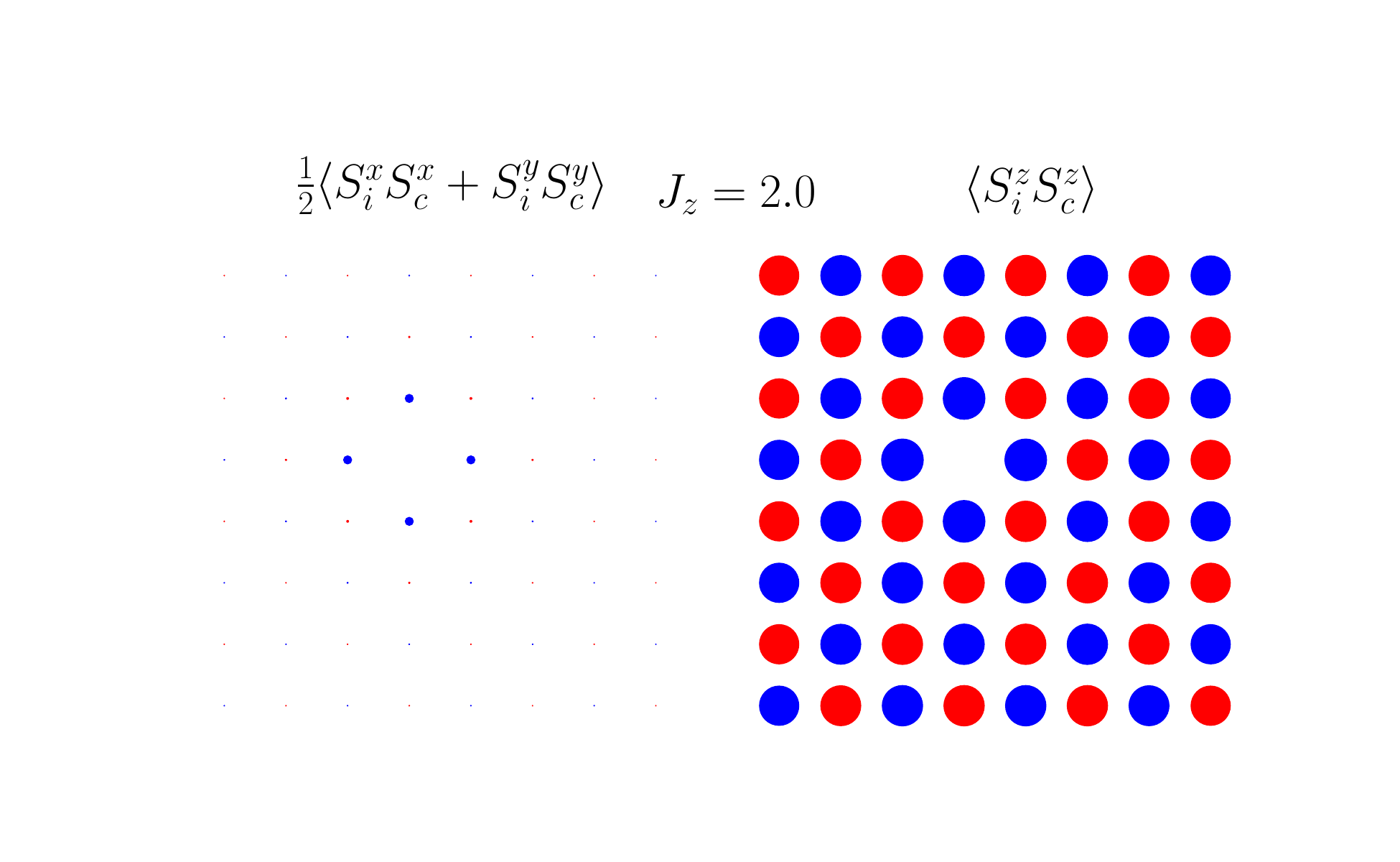}	
	\caption{\label{fig:RSC_Square_Sz=0} Real space correlation functions (measured with respect to a central site $c$) $\frac{1}{2} \langle S^{x}_i S^{x}_c + S^{y}_i S^{y}_c \rangle$ and
	$\langle S^{z}_i S^{z}_c \rangle$ for the $8 \times 8$ square lattice cylinder in the $m=0$ ($S_z=0$) magnetization sector, at various representative $J_z$. The correlation function of the spin at a site with itself is not plotted, and is left empty. The size of the circles indicates the magnitude of the correlator and the color indicates the sign, blue being negative and red being positive.
 }
\end{figure*}

%=========================================================

\section{Tables}
\label{app:tables}

In this section, we list the tables that have been referred to in the 
main text in Sec.~\ref{sec:square_zeromag} and ~\ref{sec:dimers}.

\begin{table*}[h]
\scriptsize
\begin{center}
\begin{tabular}{|C{1.8cm}|C{1.8cm}|C{1.8cm}|C{1.5cm}|C{1cm}|C{1.8cm}|C{1.8cm}|C{1.8cm}|C{1.8cm}|}

\hline
$(L_x,L_y)$ & $\frac{E_{|C\rangle}}{\#\text{bonds}}$ &  $\frac{E_{ED}}{ \#\text{bonds}}$ &  $\#_{|C\rangle}$ $(N+1)$ & $\#_{ED}$ & $\langle \hat{S}^z_i\hat{S}^z_j\rangle_{|C\rangle}$ $(-\frac{1}{4}\frac{1}{N-1})$ & $\langle \hat{S}^z_i\hat{S}^z_j\rangle_{ED}$ & $\epsilon_{ij}\langle \hat{S}^x_i\hat{S}^x_j\rangle_{|C\rangle}$ $(\frac{1}{8}\frac{N}{N-1})$ & $\epsilon_{ij}\langle \hat{S}^x_i\hat{S}^x_j\rangle_{ED}$ \\ 
\hline
\hline
 (2,2)/(4,1)& -0.25  & -0.250000..& 5 & 5 & -1/12 & -0.083333.. & 1/6 & 0.166666..\\ 
\hline
(4,2) & -0.25  & -0.250000..& 9 & 9 & -1/28 & -0.035714.. & 1/7 & 0.142857.. \\ 
\hline
(6,2) & -0.25  & -0.250000..& 13 & 13 & -1/44 & -0.022727.. & 3/22 & 0.136364.. \\ 
\hline
(4,4) & -0.25  & -0.250000..& 17 & 17 & -1/60 & -0.016666.. & 2/15 & 0.133333.. \\ 
\hline
 (6,4) & -0.25  & -0.250000..& 25 & 25 & -1/92 & -0.0108696 & 3/23 & 0.130435..\\ 
\hline
(8,4) & -0.25  & -0.250000..& 33 & 33 & -1/124 & -0.0080645 & 4/31 & 0.129032..\\ 
\hline
\end{tabular}
\end{center}
\caption{\label{tab:ED_comp} 
Comparison of exact analytic results with exact diagonalization computations for the case of two-coloring. The results hold for both periodic and open boundary conditions.}
\end{table*}

\begin{table*}[h]
\normalsize
\begin{center}
\begin{tabular}{|c|c|c|c|c|c|c|}
\hline
	Configuration & $2^{3/2} |D_1\rangle$ & $2^{3/2} |D_2 \rangle$ &  $2^{3/2} \Big( |D_1\rangle + |D_2\rangle \Big) $ & $\sqrt{20}P_{S_z=0}|rbgrbg\rangle$ & $\sqrt{20} P_{S_z=0}|rgbrgb \rangle$ & $\frac{\sqrt{20}}{\omega^2-\omega}\Big(|rgbrgb\rangle-|rbgrbg\rangle \Big)$ \\ 
\hline
\hline
 $|\uparrow \uparrow \uparrow \downarrow \downarrow \downarrow \rangle $   &  0   &  0     & 0     & 1            & 1            & 0 \\ 
\hline
 $|\uparrow \uparrow \downarrow \downarrow \downarrow \uparrow \rangle $   &  0   &  0     & 0     & 1            & 1            & 0 \\ 
\hline
 $|\uparrow \downarrow \uparrow \downarrow \downarrow \uparrow \rangle $   & $-1$ &  0     & $-1$  & $\omega^2$   & $\omega$     & $-1$ \\ 
\hline
 $|\downarrow \uparrow \uparrow \downarrow \downarrow \uparrow \rangle $   & $+1$ &  0     & $+1$  & $\omega$     & $\omega^2$   & $+1$ \\ 
\hline
 $|\uparrow \uparrow \downarrow \downarrow \uparrow \downarrow \rangle $   &  0   & $+1$   & $+1$  & $\omega$     & $\omega^2$   & $+1$ \\ 
\hline
 $|\uparrow \downarrow \uparrow \downarrow \uparrow \downarrow \rangle $   & $+1$ & $-1$   &  0    &  1           &       1      & 0 \\ 
\hline
 $|\downarrow \uparrow \uparrow \downarrow \uparrow \downarrow \rangle $   & $-1$ &  0     & $-1$  & $\omega^2$   & $\omega$     & $-1$ \\ 
\hline
 $|\uparrow \uparrow \downarrow \uparrow \downarrow \downarrow \rangle $   &  0   & $-1$   & $-1$  & $\omega^2$   & $\omega$     & $-1$ \\ 
\hline
 $|\uparrow \downarrow \uparrow \uparrow \downarrow \downarrow \rangle $   &  0   & $+1$   & $+1$  & $\omega$     & $\omega^2$   & $+1$ \\ 
\hline
 $|\downarrow \uparrow \uparrow \uparrow \downarrow \downarrow \rangle $   &  0   &  0     & 0     & 1            & 1            & 0 \\ 
\hline
 $|\downarrow \downarrow \downarrow \uparrow \uparrow \uparrow \rangle $   &  0   &  0     & 0     & 1            & 1            & 0 \\ 
\hline
 $|\downarrow \downarrow \uparrow \uparrow \uparrow \downarrow \rangle $   &  0   &  0     & 0     & 1            & 1            & 0 \\ 
\hline
 $|\downarrow \uparrow \downarrow \uparrow \uparrow \downarrow \rangle $   & $+1$ &  0     & $+1$  & $\omega$     & $\omega^2$   & $+1$ \\ 
\hline
 $|\uparrow \downarrow \downarrow \uparrow \uparrow \downarrow \rangle $   & $-1$ &  0     & $-1$  & $\omega^2$   & $\omega$     & $-1$ \\ 
\hline
 $|\downarrow \downarrow \uparrow \uparrow \downarrow \uparrow \rangle $   &  0   & $-1$   & $-1$  & $\omega^2$   & $\omega$     & $-1$ \\ 
\hline
 $|\downarrow \uparrow \downarrow \uparrow \downarrow \uparrow \rangle $   & $-1$ & $+1$   &  0    &  1           & 1            & 0 \\ 
\hline
 $|\uparrow \downarrow \downarrow \uparrow \downarrow \uparrow \rangle $   & $+1$ &  0     & $+1$  & $\omega$     & $\omega^2$   & $+1$ \\ 
\hline
 $|\downarrow \downarrow \uparrow \downarrow \uparrow \uparrow \rangle $   &  0   & $+1$   & $+1$  & $\omega$     & $\omega^2$   & $+1$ \\ 
\hline
 $|\downarrow \uparrow \downarrow \downarrow \uparrow \uparrow \rangle $   &  0   & $-1$   & $-1$  & $\omega^2$   & $\omega$     & $-1$ \\ 
\hline
 $|\uparrow \downarrow \downarrow \downarrow \uparrow \uparrow \rangle $   &  0   &  0     & 0     &  1           & 1            & 0 \\ 
\hline
\end{tabular}
\end{center}
\caption{\label{tab:MG_six_sites} 
	Amplitudes of dimer and three-coloring wavefunctions (and linear combinations) for all 20 Ising configurations in the $S_z=0$ sector for the six site chain with periodic boundary conditions. $\omega \equiv \exp(\frac{i2\pi}{3})$ is the cube root of unity.}
\end{table*}
\end{widetext}

%===============================================================

\end{document}